\documentclass[fleqn,usenatbib]{mnras}

\usepackage{newtxtext,newtxmath}
\usepackage{longtable}
\usepackage{threeparttablex}
\usepackage{booktabs}
\usepackage{caption}

\usepackage[T1]{fontenc}

\DeclareRobustCommand{\VAN}[3]{#2}
\let\VANthebibliography\thebibliography
\def\thebibliography{\DeclareRobustCommand{\VAN}[3]{##3}\VANthebibliography}

\usepackage{graphicx}	
\usepackage{amsmath}	

\newcommand{\fermi}{{\it Fermi}-LAT}
\newcommand{\gray}{$\gamma$-ray}
\newcommand{\grays}{$\gamma$-rays}

\title[Broadband emission from high redshift blazars]{Broadband Study of Gamma-Ray Blazars at Redshifts $z=2.0-2.5$}

\author[N. Sahakyan et al.]{
N. Sahakyan$^{1}$, \thanks{E-mail: narek@icra.it}
G. Harutyunyan$^{1}$,
S. Gasparyan$^{1}$,
D. Israyelyan$^{1}$\\
$^{1}$ICRANet-Armenia, Marshall Baghramian Avenue 24a, Yerevan 0019, Armenia
}

\date{Accepted XXX. Received YYY; in original form ZZZ}
\pubyear{2022}

\begin{document}
\label{firstpage}
\pagerange{\pageref{firstpage}--\pageref{lastpage}}
\maketitle

\begin{abstract}
High redshift blazars are among the most powerful non-explosive sources in the Universe and play a crucial role in understanding the evolution of relativistic jets. To understand these bright objects, we performed a detailed investigation of the multiwavelength properties of 79 \(\gamma\)-ray blazars with redshifts ranging from z = 2.0 to 2.5, using data from Fermi LAT, Swift XRT/UVOT, and NuSTAR observations. In the \gray\ band, the spectral analysis revealed a wide range of flux and photon indices, from \(5.32 \times 10^{-10}\) to \(3.40 \times 10^{-7}\) photons cm\(^{-2}\) s\(^{-1}\) and from 1.66 to 3.15, respectively, highlighting the diverse nature of these sources. The detailed temporal analysis showed that flaring activities were observed in 31 sources. Sources such as 4C+71.07, PKS 1329-049, and 4C+01.02, demonstrated significant increase in the \gray\ luminosity and flux variations, reaching peak luminosity exceeding \(10^{50}\) erg s\(^{-1}\). The temporal analysis extended to X-ray and optical/UV bands, showed  clear flux changes in some sources in different observations. The time-averaged properties of high redshift blazars were derived through modeling the spectral energy distributions with a one-zone leptonic scenario, assuming the emission region is within the broad-line region (BLR) and the X-ray and \gray\ emissions are due to inverse Compton scattering of synchrotron and BLR-reflected photons. This modeling allowed us to constrain the emitting particle distribution, estimate the magnetic field inside the jet, and evaluate the jet luminosity, which is discussed in comparison with the disk luminosity derived from fitting the excess in the UV band.
\end{abstract}




\section{Introduction}
Blazars are jetted active galactic nuclei (AGN) and are among the most powerful persistent sources of electromagnetic radiation in the universe. Current unification theories assumes that blazars are a subtype of AGNs with jets oriented at a small angle relative to the observer's line of sight \citep{1995PASP..107..803U}. Their emission is believed to be primarily powered by the accretion of matter into supermassive black holes \citep{1977MNRAS.179..433B} and exhibit extreme characteristics, such as high-amplitude and short-timescale variability, core dominance, superluminal motion, and significant optical polarization. Blazars are typically grouped into two main groups based on the presence of emission lines in their spectra \citep{1995PASP..107..803U}: BL Lacertae objects (BL Lacs), which have weak or absent emission lines, and flat spectrum radio quasars (FSRQs), which display strong emission lines.

The nonthermal emission from the jets of blazars is observable across almost all accessible bands of the electromagnetic spectrum \citep{2017A&ARv..25....2P} up to high energy (HE; $>100$ MeV) and (VHE; $>100$ MeV) \gray\ bands. Their broadband spectral energy distribution typically shows two broad humps. The lower-energy component, from the radio to the optical/X-ray band, is generally attributed to the synchrotron emission of electrons. The frequency of the synchrotron peak (\(\nu_{\rm p}\)) serves as a criterion to further classify blazars: as low synchrotron peaked sources (LSPs or LBLs), intermediate synchrotron peaked sources (ISPs or IBLs), or high synchrotron peaked sources (HSPs or HBLs) when \(\nu_{\rm p} < 10^{14}\) Hz, \(10^{14}\) Hz < \(\nu_{\rm p} < 10^{15}\) Hz, and \(\nu_{\rm p} > 10^{15}\) Hz, respectively \citep{Padovani1995,Abdo_2010}. The origin of the second component, however, remains a subject of debate. In leptonic scenarios, this second peak is interpreted as the result of inverse Compton scattering of low-energy photons, which may be of internal or external origin. In the synchrotron self-Compton (SSC) models \citep{1985A&A...146..204G, 1992ApJ...397L...5M, 1996ApJ...461..657B}, it is the synchrotron photons that are upscattered through inverse Compton processes. On the other hand, external inverse Compton (EIC) scenarios \citep[e.g.,][]{1994ApJ...421..153S} suggest that the photons originate outside the jet, coming directly from the accretion disk \citep{1992A&A...256L..27D, 1994ApJS...90..945D}, or they may be reprocessed by the broad-line region (BLR) \citep{1994ApJ...421..153S}, or emitted from the torus \citep{2000ApJ...545..107B}. In \citet{2023arXiv231102979B}, a novel approach for fitting the blazar Spectral Energy Distribution (SED) utilizing convolutional neural networks is introduced which allows self-consistent modeling, enabling a more detailed interpretation of the observed results.

In alternative models such as the hadronic or lepto-hadronic scenarios, the second spectral bump is assumed to be from the synchrotron emission of ultra-HE protons or from the decay of secondary particles produced during hadronic interactions \citep{1993A&A...269...67M, 1989A&A...221..211M, 2001APh....15..121M, mucke2, 2013ApJ...768...54B, 2015MNRAS.447...36P, 2022MNRAS.509.2102G}. Interest in these models has increased, particularly after establishing a potential link between blazars and VHE neutrinos, following the association of TXS 0506+056 with the IceCube-170922A neutrino event \citep{2018Sci...361..147I, 2018Sci...361.1378I, 2018MNRAS.480..192P}, and the detection of multiple neutrino events concurrent with the active phase of PKS 0735+178 in optical/UV, X-ray, and \gray\ bands \citep{2023MNRAS.519.1396S}. These multimessenger observations have started extensive discussions, with various models being applied to explain the multimessenger observations of blazars \citep{2018ApJ...863L..10A,2018ApJ...864...84K, 2018ApJ...865..124M, 2018MNRAS.480..192P, 2018ApJ...866..109S, 2019MNRAS.484.2067R,2019MNRAS.483L..12C, 2019A&A...622A.144S, 2019NatAs...3...88G, 2022MNRAS.509.2102G}.

The emission from blazars is highly beamed, and their bolometric luminosity can exceed \(10^{48}\:{\rm erg\:s^{-1}}\), allowing them to be observed even at very high redshifts \citep[e.g., see][]{2017ApJ...837L...5A, 2020MNRAS.498.2594S}. These distant blazars are particularly interesting, as their study offers insights into the formation and evolution of supermassive black holes, relativistic jets, and the connections between accretion disks and jets. Moreover, their \gray\ emission is important for probing the early universe; \gray\ emission from distant blazars undergoes attenuation via \(\gamma\gamma\) absorption when interacting with extragalactic background light (EBL) photons, thereby enabling observations that can constrain the EBL's density.
\begin{figure}
	\includegraphics[width=0.48\textwidth]{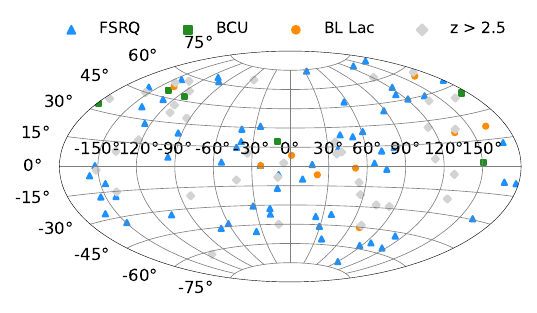}
    \caption{Hammer-Aitoff projection in Galactic coordinates showing the distribution of \(\gamma\)-ray blazars with redshifts above \(z>2.0\). BL Lacs within the redshift range \(2.0 \leq z \leq 2.5\) are represented by orange symbols, FSRQs are depicted in blue, BCUs in green, and blazars with redshifts \(z>2.5\) are shown as grey diamonds.}
    \label{proj}
\end{figure}
The multiwavelength properties of distant blazars have been extensively studied in a number of publications \citep[e.g., see][]{2023MNRAS.521.1013S, 2020MNRAS.498.2594S, 2020ApJ...900...72L, 2019ApJ...879L...9L, 2017ApJ...839...96M, 2016Galax...4...26O, 2016MNRAS.455.1881D, 2015ApJ...803..112P, 2012MNRAS.425.2015P, 2015ApJ...804...74P,2016ApJ...825...74P, 2017ApJ...851...33P, 2019ApJ...871..211P}. In \citet{2020MNRAS.498.2594S}, the origin of emission from the most distant blazars detected in the HE \gray\ band (\( z > 2.5 \)) was investigated analyzing data in the optical/UV, X-ray, and \gray\ bands. From the temporal evolution of emission in these bands, flaring periods were identified when the luminosity substantially increased. Their broadband emission was modeled using one-zone SSC and EIC models, assuming that the external photons are infrared (IR) photons from the dusty torus (see \citet{2018A&A...616A..63A} for a discussion on the contribution of different external fields in shaping the HE emission of LSPs). As a result, the parameters that characterize the particle emission and jet in these distant blazars were estimated.\\
\indent In this work, we expand the study by \citet{2020MNRAS.498.2594S} to \gray\ emitting blazars which have estimated redshifts that are in the range of $z=2.0-2.5$. A prolonged observation period of $\sim14$ years enables a comprehensive investigation of emissions from these sources across optical/UV, X-ray, and \gray\ bands as well as to perform detailed spectral and temporal analysis. Using the analyzed data broadband SEDs for a substantial number of sources were constrained and modeled which allows for a systematic comparison of blazar emission parameters at varying distances, which, in turn, could enhance our understanding of these brightest objects. 

The structure of the paper is as follows: Section \ref{sample} introduces the sample of sources under consideration; Section \ref{dat:anal} details the data analysis methodology; Section \ref{res:data_anal} presents the results of the data analysis; Section \ref{theormodel} discusses the modeling of broadband SEDs; and Section \ref{res} outlines the modeling results. Finally, the conclusions are summarized in Section \ref{concl}. 

\section{Source sample}\label{sample}
The fourth catalogue of AGNs detected by the Fermi Large Area Telescope (\fermi) \citep[][]{2022ApJS..263...24A} contains 3,814 blazars, among which 792 are FSRQs, 1,458 are BL Lacs, and 1,493 are BCUs. The most distant blazar, GB 1508+5714, is at \( z=4.31 \). A small fraction of the blazars, 79 (2.07\%), have a redshift between \( 2.0 < z < 2.5 \), and these are the ones selected for the current study (the source sample). This group includes 64 FSRQs, 9 BL Lacs, and 6 BCUs. The BL Lacs generally have redshifts lower than \( z=2.1 \), with the exception of SDSS J145059.99+520111.7, which is at \( z=2.47 \). Among the BL Lacs, there are 3 LBLs, 4 IBLs, and only 2 HBLs. The FSRQs have a more homogeneous redshift distribution, being observed at almost all redshifts; the most distant FSRQ in the sample is S5 1053+70 at \( z=2.49 \). BCUs, which exhibit characteristics similar to blazars but lack reliable optical associations, have been observed across a range of redshifts. For instance, 4FGL J1139.0+4033 (CRATES J113903+403303) has been observed at \( z=2.36 \), while 4FGL J1003.4+0205 (SDSS J100326.63+020455.6) is located at \( z=2.08 \). In \citet{2022MNRAS.tmp.3466S} BCUs were classified by training machine learning algorithms on the \gray\ properties of FSRQs and BL Lacs. According to those criteria, four BCUs from our sample show probability similar to FSRQs (CRATES J113903+403303, MG4 J162750+4802, TXS 2315+189 and SDSS J120542.82+332146.9), one to BL Lacs (SDSS J100326.63+020455.6) and one (SDSS J120542.82+332146.9) can not be classified.

In summary, the source sample considered in this study comprises 79 objects. In Table \ref{tab:param}, the three leftmost columns list these sources along with their 4FGL names and classes. The spatial distribution of these blazars in Galactic coordinates is depicted in the Hammer–Aitoff projection shown in Fig. \ref{proj}. For completeness, this figure also includes blazars with redshifts greater than \( z>2.5 \), as reported by \citet{2020MNRAS.498.2594S}.
\begin{table*}
    \centering
\caption{The table shows the source sample, detailing the outcomes of the \gray\ analysis. For each source, the name, associated 4FGL name, and class are provided. $\alpha_{\rm p}$ represents the photon index when the source spectrum is best modeled with a power-law model, whereas $\alpha$ and $\beta$ denote the slope and curvature, respectively, when the spectrum is modeled with a log-parabola. The flux is reported in units of \(10^{-8}\) photons cm\(^{-2}\) s\(^{-1}\), and the luminosity is expressed in \(10^{47}\) ergs s\(^{-1}\). The redshift $z$ is given in last column.}
\label{tab:param}
\begin{tabular}{lclccclc}
\toprule
Object &  4FGL name & Class &  $\alpha_{\rm p}/\alpha$   & $\beta$  & Flux  & Luminosity & z \\
\midrule
S5 1053+70 & 4FGL J1056.8+7012 & FSRQ & 2.70 $\pm$ 0.08 & 0.23 $\pm$ 0.08 & 1.92 $\pm$ 0.55 & 6.51 $\pm$ 0.46 & 2.492 \\
PMN J1344-1723 & 4FGL J1344.2-1723 & FSRQ & 2.03 $\pm$ 0.04 & 0.14 $\pm$ 0.02 & 1.68 $\pm$ 0.14 & 9.79 $\pm$ 0.5 & 2.490 \\
SDSS J145059.99+520111.7 & 4FGL J1450.8+5201 & BLL & 2.11 $\pm$ 0.08 & 0.09 $\pm$ 0.04 & 0.64 $\pm$ 0.16 & 3.79 $\pm$ 0.44 & 2.471 \\
PKS 1915-458 & 4FGL J1919.4-4550 & FSRQ & 3.15 $\pm$ 0.12 & - & 1.47 $\pm$ 0.18 & 2.19 $\pm$ 0.26 & 2.470 \\
PKS 0226-559 & 4FGL J0228.3-5547 & FSRQ & 2.24 $\pm$ 0.02 & 0.11 $\pm$ 0.01 & 5.04 $\pm$ 0.18 & 22.1 $\pm$ 0.49 & 2.464 \\
S3 2214+30 & 4FGL J2216.8+3103 & FSRQ & 2.66 $\pm$ 0.28 & 0.33 $\pm$ 0.19 & 0.26 $\pm$ 0.12 & 0.89 $\pm$ 0.28 & 2.462 \\
PKS 2315-172 & 4FGL J2318.6-1657 & FSRQ & 2.21 $\pm$ 0.01 & - & 0.20 $\pm$ 0.00 & 0.76 $\pm$ 0.02 & 2.462 \\
PKS 0601-70 & 4FGL J0601.1-7035 & FSRQ & 2.27 $\pm$ 0.03 & 0.09 $\pm$ 0.02 & 3.51 $\pm$ 0.17 & 11.22 $\pm$ 0.39 & 2.409 \\
B2 1436+37B & 4FGL J1438.9+3710 & FSRQ & 2.33 $\pm$ 0.05 & 0.10 $\pm$ 0.03 & 1.88 $\pm$ 0.14 & 5.24 $\pm$ 0.27 & 2.401 \\
2MASS J16561677-3302127$^1$ & 4FGL J1656.3-3301 & FSRQ & 2.94 $\pm$ 0.11 & 0.32 $\pm$ 0.09 & 3.47 $\pm$ 0.72 & 6.68 $\pm$ 0.70 & 2.400 \\
MG1 J173624+0632 & 4FGL J1736.6+0628 & FSRQ & 2.67 $\pm$ 0.08 & - & 1.58 $\pm$ 0.21 & 2.90 $\pm$ 0.39 & 2.387 \\
TXS 1645+635 & 4FGL J1645.6+6329 & FSRQ & 2.46 $\pm$ 0.06 & 0.13 $\pm$ 0.04 & 1.41 $\pm$ 0.12 & 3.28 $\pm$ 0.2 & 2.379 \\
PKS B1149-084 & 4FGL J1152.3-0839 & FSRQ & 2.33 $\pm$ 0.04 & 0.11 $\pm$ 0.02 & 2.79 $\pm$ 0.19 & 7.9 $\pm$ 0.38 & 2.370 \\
S5 0212+73 & 4FGL J0217.4+7352 & FSRQ & 2.93 $\pm$ 0.04 & 0.15 $\pm$ 0.03 & 3.52 $\pm$ 0.17 & 5.56 $\pm$ 0.2 & 2.367 \\
B2 0552+39A & 4FGL J0555.6+3947 & FSRQ & 2.67 $\pm$ 0.00 & 0.15 $\pm$ 0.00 & 3.98 $\pm$ 0.02 & 7.5 $\pm$ 0.02 & 2.365 \\
CRATES J113903+403303 & 4FGL J1139.0+4033 & BCU & 2.73 $\pm$ 0.09 & - & 1.13 $\pm$ 0.18 & 1.90 $\pm$ 0.30 & 2.360 \\
B2 2112+28B & 4FGL J2114.8+2831 & FSRQ & 2.50 $\pm$ 0.21 & 0.42 $\pm$ 0.18 & 0.33 $\pm$ 0.15 & 1.54 $\pm$ 0.36 & 2.345 \\
PKS 2149-306 & 4FGL J2151.8-3027 & FSRQ & 2.86 $\pm$ 0.03 & 0.17 $\pm$ 0.03 & 7.01 $\pm$ 0.21 & 10.85 $\pm$ 0.3 & 2.345 \\
PKS 1430-178 & 4FGL J1433.0-1801 & FSRQ & 2.71 $\pm$ 0.02 & 0.30 $\pm$ 0.01 & 0.60 $\pm$ 0.03 & 1.53 $\pm$ 0.1 & 2.331 \\
MG4 J162750+4802 & 4FGL J1627.3+4758 & BCU & 2.64 $\pm$ 0.16 & 0.11 $\pm$ 0.11 & 0.46 $\pm$ 0.16 & 0.96 $\pm$ 0.21 & 2.326 \\
PMN J0743-5619 & 4FGL J0743.0-5622 & FSRQ & 2.90 $\pm$ 0.19 & - & 1.01 $\pm$ 0.22 & 1.45 $\pm$ 0.32 & 2.319 \\
GB6 J0742+4900 & 4FGL J0742.1+4902 & FSRQ & 2.24 $\pm$ 0.09 & - & 0.46 $\pm$ 0.10 & 1.38 $\pm$ 0.30 & 2.312 \\
S3 0458-02 & 4FGL J0501.2-0158 & FSRQ & 2.29 $\pm$ 0.01 & 0.09 $\pm$ 0.01 & 10.80 $\pm$ 0.19 & 27.57 $\pm$ 0.4 & 2.291 \\
PMN J0157-4614 & 4FGL J0157.7-4614 & FSRQ & 2.26 $\pm$ 0.13 & 0.26 $\pm$ 0.09 & 0.55 $\pm$ 0.15 & 2.08 $\pm$ 0.24 & 2.287 \\
PKS 0726-476 & 4FGL J0728.0-4740 & FSRQ & 2.36 $\pm$ 0.24 & - & 0.45 $\pm$ 0.32 & 1.06 $\pm$ 0.75 & 2.282 \\
PKS 0420+022 & 4FGL J0422.8+0225 & FSRQ & 2.63 $\pm$ 0.19 & 0.37 $\pm$ 0.17 & 1.11 $\pm$ 0.24 & 1.95 $\pm$ 0.28 & 2.277 \\
PKS 2245-328 & 4FGL J2248.7-3235 & FSRQ & 2.67 $\pm$ 0.05 & - & 2.01 $\pm$ 0.14 & 3.23 $\pm$ 0.22 & 2.268 \\
PKS B2224+006 & 4FGL J2226.8+0051 & FSRQ & 2.98 $\pm$ 0.24 & 0.31 $\pm$ 0.10 & 0.92 $\pm$ 0.17 & 1.81 $\pm$ 0.32 & 2.262 \\
TXS 1322+479 & 4FGL J1324.9+4748 & FSRQ & 2.67 $\pm$ 0.07 & - & 1.09 $\pm$ 0.13 & 1.74 $\pm$ 0.21 & 2.260 \\
PKS 2244-37 & 4FGL J2247.5-3700 & FSRQ & 2.40 $\pm$ 0.20 & - & 0.19 $\pm$ 0.09 & 0.41 $\pm$ 0.19 & 2.252 \\
B2 0242+23 & 4FGL J0245.4+2408 & FSRQ & 2.67 $\pm$ 0.05 & - & 1.98 $\pm$ 0.16 & 3.09 $\pm$ 0.25 & 2.243 \\
4C +71.07 & 4FGL J0841.3+7053 & FSRQ & 2.81 $\pm$ 0.02 & 0.21 $\pm$ 0.02 & 10.68 $\pm$ 0.17 & 14.76 $\pm$ 0.23 & 2.218 \\
PKS 2022+031 & 4FGL J2025.2+0317 & FSRQ & 2.07 $\pm$ 0.06 & 0.09 $\pm$ 0.03 & 0.92 $\pm$ 0.13 & 4.03 $\pm$ 0.37 & 2.210 \\
MG2 J174753+2323 & 4FGL J1747.4+2330 & FSRQ & 2.79 $\pm$ 0.03 & 0.39 $\pm$ 0.03 & 0.80 $\pm$ 0.03 & 1.2 $\pm$ 0.03 & 2.203 \\
MG2 J153938+2744 & 4FGL J1539.6+2743 & FSRQ & 2.23 $\pm$ 0.06 & 0.05 $\pm$ 0.03 & 1.18 $\pm$ 0.13 & 3.12 $\pm$ 0.23 & 2.196 \\
S4 0917+44 & 4FGL J0920.9+4441 & FSRQ & 2.35 $\pm$ 0.02 & 0.15 $\pm$ 0.02 & 5.88 $\pm$ 0.35 & 17.97 $\pm$ 0.43 & 2.186 \\
PMN J2135-5006 & 4FGL J2135.3-5006 & FSRQ & 2.40 $\pm$ 0.05 & 0.10 $\pm$ 0.03 & 1.99 $\pm$ 0.13 & 3.92 $\pm$ 0.2 & 2.181 \\
OX 131 & 4FGL J2121.0+1901 & FSRQ & 2.14 $\pm$ 0.03 & 0.05 $\pm$ 0.01 & 2.68 $\pm$ 0.17 & 8.77 $\pm$ 0.38 & 2.180 \\
PMN J1959-4246 & 4FGL J1959.1-4247 & FSRQ & 2.16 $\pm$ 0.10 & 0.19 $\pm$ 0.06 & 0.88 $\pm$ 0.15 & 2.76 $\pm$ 0.26 & 2.174 \\
B3 1520+437 & 4FGL J1521.8+4338 & FSRQ & 3.00 $\pm$ 0.10 & - & 1.14 $\pm$ 0.15 & 1.32 $\pm$ 0.17 & 2.168 \\
TXS 2315+189 & 4FGL J2318.2+1915 & BCU & 2.60 $\pm$ 0.01 & - & 1.60 $\pm$ 0.03 & 2.43 $\pm$ 0.05 & 2.163 \\
PKS 0446+11 & 4FGL J0449.1+1121 & FSRQ & 2.37 $\pm$ 0.03 & 0.13 $\pm$ 0.02 & 6.25 $\pm$ 0.21 & 12.12 $\pm$ 0.34 & 2.153 \\
PKS 1329-049 & 4FGL J1332.0-0509 & FSRQ & 2.41 $\pm$ 0.02 & 0.16 $\pm$ 0.02 & 6.33 $\pm$ 0.18 & 11.57 $\pm$ 0.32 & 2.150 \\
PMN J2227+0037 & 4FGL J2227.9+0036 & BLL & 1.77 $\pm$ 0.11 & 0.15 $\pm$ 0.05 & 0.31 $\pm$ 0.10 & 3.79 $\pm$ 0.54 & 2.145 \\
TXS 2321-065 & 4FGL J2323.6-0617 & FSRQ & 2.32 $\pm$ 0.10 & 0.18 $\pm$ 0.06 & 0.87 $\pm$ 0.16 & 2.14 $\pm$ 0.24 & 2.144 \\
PMN J1402-3334 & 4FGL J1402.6-3330 & FSRQ & 3.08 $\pm$ 0.28 & 1.30 $\pm$ 0.37 & 0.61 $\pm$ 0.13 & 1.44 $\pm$ 0.21 & 2.140 \\
PMN J0134-3843 & 4FGL J0134.3-3842 & FSRQ & 2.59 $\pm$ 0.08 & - & 0.72 $\pm$ 0.10 & 1.07 $\pm$ 0.15 & 2.140 \\
87GB 080551.6+535010 & 4FGL J0809.5+5341 & FSRQ & 2.19 $\pm$ 0.04 & 0.10 $\pm$ 0.02 & 1.71 $\pm$ 0.14 & 5.03 $\pm$ 0.26 & 2.133 \\
87GB 142651.1+564919 & 4FGL J1428.3+5635 & FSRQ & 2.69 $\pm$ 0.03 & - & 0.44 $\pm$ 0.03 & 0.60 $\pm$ 0.04 & 2.129 \\
PKS B1043-291 & 4FGL J1045.8-2928 & FSRQ & 2.61 $\pm$ 0.07 & - & 1.24 $\pm$ 0.15 & 1.80 $\pm$ 0.22 & 2.128 \\
OM 127 & 4FGL J1119.0+1235 & FSRQ & 2.29 $\pm$ 0.09 & 0.24 $\pm$ 0.06 & 1.21 $\pm$ 0.13 & 2.39 $\pm$ 0.19 & 2.126 \\
SDSS J120542.82+332146.9 & 4FGL J1205.8+3321 & BCU & 2.43 $\pm$ 0.17 & - & 0.29 $\pm$ 0.11 & 0.52 $\pm$ 0.19 & 2.125 \\
PMN J0124-0624 & 4FGL J0124.8-0625 & BLL & 2.21 $\pm$ 0.12 & - & 0.31 $\pm$ 0.08 & 0.78 $\pm$ 0.22 & 2.117 \\
PKS 0227-369 & 4FGL J0229.5-3644 & FSRQ & 2.45 $\pm$ 0.05 & 0.17 $\pm$ 0.04 & 2.19 $\pm$ 0.13 & 3.48 $\pm$ 0.16 & 2.115 \\
OF 200 & 4FGL J0403.3+2601 & FSRQ & 2.45 $\pm$ 0.03 & 0.71 $\pm$ 0.02 & 0.19 $\pm$ 0.01 & 0.8 $\pm$ 0.02 & 2.109 \\
B3 0803+452 & 4FGL J0806.5+4503 & FSRQ & 2.72 $\pm$ 0.11 & - & 0.79 $\pm$ 0.13 & 1.01 $\pm$ 0.16 & 2.102 \\
TXS 0036-099 & 4FGL J0039.0-0946 & FSRQ & 2.83 $\pm$ 0.08 & - & 1.31 $\pm$ 0.15 & 1.54 $\pm$ 0.17 & 2.102 \\
4C +01.02 & 4FGL J0108.6+0134 & FSRQ & 2.29 $\pm$ 0.01 & 0.11 $\pm$ 0.01 & 30.88 $\pm$ 0.36 & 66.23 $\pm$ 0.49 & 2.099 \\
87GB 145232.0+493854 & 4FGL J1454.0+4927 & BCU & 2.62 $\pm$ 0.18 & - & 0.27 $\pm$ 0.10 & 0.71 $\pm$ 0.22 & 2.085 \\
SDSS J105707.47+551032.2 & 4FGL J1057.2+5510 & BLL & 2.02 $\pm$ 0.11 & - & 0.18 $\pm$ 0.05 & 0.37 $\pm$ 0.14 & 2.085 \\
\bottomrule
\multicolumn{8}{l}{%
 \begin{minipage}{14.5cm}%
    $^1$ The result for this object is listed from the catalogue, as its ROI contains 10 extended sources, which complicates the analysis.
  \end{minipage}%
}
\end{tabular}
\end{table*}

\begin{table*}
    \centering
\caption*{Table \thetable{} (Continued)}
\begin{tabular}{lclccclc}
\toprule
Object &  4FGL name & Class &  $\alpha_{\rm p}/\alpha$   & $\beta$  & Flux  & Luminosity & z \\

\midrule
PKS 1348+007 & 4FGL J1351.0+0029 & FSRQ & 2.43 $\pm$ 0.11 & - & 0.47 $\pm$ 0.11 & 0.80 $\pm$ 0.18 & 2.084 \\
PKS B1112-080 & 4FGL J1114.5-0819 & FSRQ & 2.72 $\pm$ 0.02 & 0.12 $\pm$ 0.01 & 1.49 $\pm$ 0.04 & 1.95 $\pm$ 0.21 & 2.078 \\
1RXS J032342.6-011131 & 4FGL J0323.7-0111 & BLL & 1.82 $\pm$ 0.07 & 0.08 $\pm$ 0.03 & 0.38 $\pm$ 0.08 & 3.19 $\pm$ 0.36 & 2.075 \\
SDSS J100326.63+020455.6 & 4FGL J1003.4+0205 & BCU & 1.66 $\pm$ 0.13 & - & 0.05 $\pm$ 0.02 & 0.75 $\pm$ 0.32 & 2.075 \\
PKS 0528+134 & 4FGL J0530.9+1332 & FSRQ & 2.50 $\pm$ 0.03 & 0.23 $\pm$ 0.02 & 4.04 $\pm$ 0.28 & 7.72 $\pm$ 0.42 & 2.069 \\
TXS 0322+222 & 4FGL J0325.7+2225 & FSRQ & 2.52 $\pm$ 0.03 & 0.19 $\pm$ 0.03 & 4.78 $\pm$ 0.23 & 7.37 $\pm$ 0.26 & 2.066 \\
4C +13.14 & 4FGL J0231.8+1322 & FSRQ & 2.68 $\pm$ 0.01 & - & 2.05 $\pm$ 0.03 & 2.58 $\pm$ 0.03 & 2.065 \\
GB6 J1722+6105 & 4FGL J1722.6+6104 & FSRQ & 2.82 $\pm$ 0.15 & 0.09 $\pm$ 0.10 & 0.66 $\pm$ 0.17 & 0.83 $\pm$ 0.16 & 2.058 \\
SDSS J000359.23+084138.1 & 4FGL J0004.0+0840 & BLL & 1.79 $\pm$ 0.34 & 0.36 $\pm$ 0.24 & 0.03 $\pm$ 0.02 & 0.55 $\pm$ 0.22 & 2.057 \\
87GB 105148.6+222705 & 4FGL J1054.5+2211 & BLL & 2.16 $\pm$ 0.03 & - & 1.76 $\pm$ 0.13 & 4.72 $\pm$ 0.34 & 2.055 \\
NVSS J090226+205045 & 4FGL J0902.4+2051 & BLL & 2.06 $\pm$ 0.05 & 0.04 $\pm$ 0.02 & 1.43 $\pm$ 0.15 & 4.72 $\pm$ 0.34 & 2.055 \\
PMN J0625-5438 & 4FGL J0625.8-5441 & FSRQ & 2.70 $\pm$ 0.08 & - & 1.20 $\pm$ 0.13 & 1.46 $\pm$ 0.16 & 2.051 \\
IVS B0343+485 & 4FGL J0347.0+4844 & FSRQ & 2.47 $\pm$ 0.01 & - & 0.92 $\pm$ 0.03 & 1.40 $\pm$ 0.05 & 2.043 \\
GB1 1155+486 & 4FGL J1158.5+4824 & FSRQ & 2.49 $\pm$ 0.05 & - & 1.24 $\pm$ 0.12 & 1.82 $\pm$ 0.17 & 2.028 \\
PKS 1318-263 & 4FGL J1321.3-2641 & FSRQ & 2.61 $\pm$ 0.16 & - & 0.62 $\pm$ 0.18 & 0.80 $\pm$ 0.23 & 2.027 \\
OX 110 & 4FGL J2108.5+1434 & FSRQ & 2.66 $\pm$ 0.09 & 0.14 $\pm$ 0.08 & 1.26 $\pm$ 0.18 & 1.7 $\pm$ 0.19 & 2.017 \\
PKS 0437-454 & 4FGL J0438.9-4521 & BLL & 2.25 $\pm$ 0.06 & 0.11 $\pm$ 0.04 & 1.40 $\pm$ 0.17 & 3.07 $\pm$ 0.21 & 2.017 \\
PKS B1412-096 & 4FGL J1415.9-1002 & FSRQ & 0.91 $\pm$ 0.77 & 3.29 $\pm$ 1.20 & 0.06 $\pm$ 0.02 & 0.8 $\pm$ 0.22 & 2.001 \\
PKS 0549-575 & 4FGL J0550.3-5733 & FSRQ & 2.23 $\pm$ 0.10 & - & 0.34 $\pm$ 0.08 & 0.73 $\pm$ 0.17 & 2.001 \\
\bottomrule
\end{tabular}
\end{table*}

\section{Multiwavelength observations of considered blazars} \label{dat:anal}
In order to investigate the multiwavelength properties of the selected sources, data collected by the \fermi, Nuclear Spectroscopic Telescope Array (NuSTAR), Neil Gehrels Swift Observatory (hereafter Swift) X-Ray Telescope (XRT), and Ultraviolet/Optical Telescope (Swift UVOT) were downloaded and analyzed.
\subsection{\fermi\ data}
The Large Area Telescope (LAT), onboard the Fermi Gamma-ray Space Telescope, is a HE instrument that uses the pair-production technique to detect \grays\ in the energy range between 20 MeV and \(>300\) GeV. By default, it operates in scanning mode, continuously monitoring the \gray\ sky since its launch in 2008 \citep{2009ApJ...697.1071A}.

The \fermi\ PASS8 data collected from August 4, 2008, to December 4, 2022 ($\sim14.5$ years), were considered to study the properties of all 79 blazars under consideration. The standard data-reduction procedure was performed following the recommendations from the \fermi\ science team\footnote{\url{https://fermi.gsfc.nasa.gov/ssc/data/analysis/documentation/}}. For each blazar, events in the energy range between 100 MeV and 500 GeV from a region of interest (ROI) of $12^\circ$—reduced to $10^\circ$ for several sources to better represent the ROI—centered on the \gray\ position of the sources, were downloaded and analyzed. The Fermi ScienceTools version 2.0.8 and the P8R3\_SOURCE\_V3 instrument response function were used. To reduce contamination from the Earth's limb, a zenith angle cut of $90^\circ$ was applied. Events with a higher probability of being photons were selected using the filter {\it evclass = 128 and evtype = 3}, and the good time intervals were chosen with the expression \({\rm (DATA\_QUAL > 0) \&\& (LAT\_CONFIG == 1)}\). The analysis model file was created based on the \textit{Fermi} Fourth Source Catalog \citep[4FGL-DR3][]{2022ApJS..260...53A} and includes all sources within an ROI radius plus an additional $5^\circ$. During the likelihood fitting, the spectral parameters of all sources within the ROI were allowed to vary, while those of sources outside the ROI were fixed to their 4FGL values. The model file also includes the Galactic diffuse emission model gll\_iem\_v07 and the isotropic component iso\_P8R3\_SOURCE\_V3\_v1. The spectral parameters of the sources, along with the normalization of both background models, were optimized by applying a binned likelihood analysis using the {\it fermiPy} tool \citep{2017ICRC...35..824W}. The detection significance of the sources is quantified by the likelihood test statistic (TS), defined as \(TS = 2 \times (\log L - \log L_0)\), where \(L\) is the likelihood with the source at the position of interest included, and \(L_0\) is the likelihood without the source.

The \gray\ variability of the considered sources was investigated by generating light curves in two distinct manners. Initially, for all sources, the 14-year period was divided into equal intervals (e.g., 5, 7, 10 days, depending on the source's overall detection significance) to ensure that the light curves did not contain a significant number of upper limits. Within these intervals, the flux and photon index were estimated by applying the unbinned likelihood analysis method. This approach provides a general overview of the flux changes over time, but since the fluxes are averaged over several days, any potential short-scale flux variations are likely to be smoothed out. For a more detailed examination of flux evolution over time, light curves were also generated using the adaptive binning method \citep{2012A&A...544A...6L}, which is applied when photon statistics are sufficient. This technique allows for flexible time bin widths that are determined by assuming a constant uncertainty in the flux estimation. Consequently, brighter source states yield shorter bins, whereas longer bins are used during lower and/or average source states. Light curves produced by this method have been extensively utilized to study short-timescale flux variations in blazar emissions \citep[e.g., see][]{2013A&A...557A..71R, 2016ApJ...830..162B, 2017MNRAS.470.2861S, 2017A&A...608A..37Z,2017ApJ...848..111B,  2018ApJ...863..114G, 2018A&A...614A...6S, 2021MNRAS.504.5074S, 2022MNRAS.517.2757S, 2022MNRAS.513.4645S}.
\subsection{NuSTAR}
NuSTAR \citep[][]{2013ApJ...770..103H} is a hard X-ray telescope operating in the 3-79 keV range, equipped with two focal plane modules: FPMA and FPMB. Among the sources considered, PKS 0528+134, S3 0458-02, PKS 0446+11, PKS 1329-049, 87GB 080551.6+535010, and TXS 0322+222 were observed by NuSTAR once; PKS 2149-306, S5 0212+73, and PKS 0227-369 were observed twice; and 4C +71.07 was observed three times. In total, these amount to 15 observations that provide critical information on the hard X-ray band emission of the sources.

The analysis of all NuSTAR data was performed using the {\it NuSTAR\_Spectra} pipeline, a shell script built upon the NuSTAR Data Analysis Software \citep{2022MNRAS.514.3179M}. This script streamlines the process by autonomously retrieving calibrated and filtered event files, employing Ximage for accurate source positioning, and utilizing the {\it nuproducts} command to retrive science ready products. Source counts are extracted from a predefined circular region, while background counts are from an annular region, with inner and outer radii sizes dynamically determined by the source's count rate. Post data selection, the spectra are binned to ensure a minimum of one count per bin, and spectral fitting is executed within the XSPEC framework \citep{1996ASPC..101...17A}, applying Cash statistics \citep{1979ApJ...228..939C} for the energy range from 3 keV to the upper energy limit of detectable signal, which varies between 20 and 79 keV. The fitting is performed using both power-law and log-parabola models, the observed flux in the 3-10 and 10-30 keV bands is estimated and the corresponding SED is computed using the best-fit parameters. Details describing {\it NuSTAR\_Spectra} pipeline can be found in \citet{2022MNRAS.514.3179M}.   
\subsection{Swift XRT}
The high-redshift blazars selected for this study were also frequently monitored by the Swift XRT in the 0.3-10 keV energy range. Of the 79 considered sources, 60 were observed at least once by the Swift telescope. The source PKS 0528+134 was the most frequently observed—141 times—while 4C +71.07, PKS 2149-306, PKS 0226-559, S3 0458-02, 4C +01.02, and PMN J1344-1723 each were observed more than 20 times.

All the Swift XRT data accumulated from the observations of selected sources was retrieved and analyzed using {\it swift\_xrtproc} script. This automated tool accesses both PC and WT mode observations from Swift XRT and executes the standard analysis procedure which includes the creation of exposure maps, calibration of observational data. The source  spectral files are obtained by estimating the source counts within a 20-pixel radius circular region and the background from a surrounding annular region centered on the source. Additionally, it performs corrections for pile-up effects and conducts spectral fitting employing both power-law and log-parabola models within the XSPEC framework. Subsequently, it computes the SED spectral points using the optimal spectral model, computes the flux across specified energy intervals, and estimates the photon index for the selected energy range. For more details on the {\it swift\_xrtproc} tool see \citet{2021MNRAS.507.5690G}.

The majority of these sources considered here exhibit no significant variability in the X-ray band, as discussed in the next section, so to enhance the photon statistics and refine the estimation of the X-ray flux, if for a source multiple observations are available, they  were combined and analyzed using the tool provided by the UK Swift Science Data Centre \citep{2009MNRAS.397.1177E}. 
\subsection{Swift UVOT}
Together with the XRT, the sources were also observed by Swift UVOT, which provided data in three optical filters (V, B, and U) and three UV filters (W1, M2, and W2). All available Swift UVOT data from the observations of the considered sources were downloaded and analyzed.

The data were analyzed using the {\it uvotsource} task included in the HEAsoft package, version 6.29. Source counts were extracted from a circular region with a radius of 5'' centered on the source, while background counts were obtained from a larger circular region with a radius of 20'', located in a nearby source-free area. The observations of all sources were individually checked to ensure the accuracy of source and background region selection. For all sources, the magnitudes were derived using the {\it uvotsource} tool and corrected for reddening and Galactic extinction using the reddening coefficient \(E(B - V)\) obtained from the Infrared Science Archive\footnote{http://irsa.ipac.caltech.edu/applications/DUST/}. The corrected fluxes measured for each filter were used to construct the light curves and SEDs. As the considered sources are at high redshift, their optical/UV fluxes could be affected by absorption from neutral hydrogen in intervening $Lyman-\alpha$ absorption systems. These effects were corrected for in the SEDs during theoretical modeling, following the procedures described in \citet{2011MNRAS.411..901G} and \citet{2010MNRAS.405..387G}.

\subsection{Archival data}
To construct the most comprehensive multiwavelength SEDs possible for the considered sources, archival data were also extracted and analyzed in addition to the data discussed herein. This was accomplished using the VOU-Blazar tool \citep{2020A&C....3000350C} through Markarian Multiwavelength data center\footnote{\url{https://mmdc.am}}, which retrieves multiwavelength data from 71 catalogs and spectral databases through various online services.  
\section{Results of data analyses}\label{res:data_anal}
\begin{figure*}
     \centering
     \includegraphics[width=0.9\textwidth]{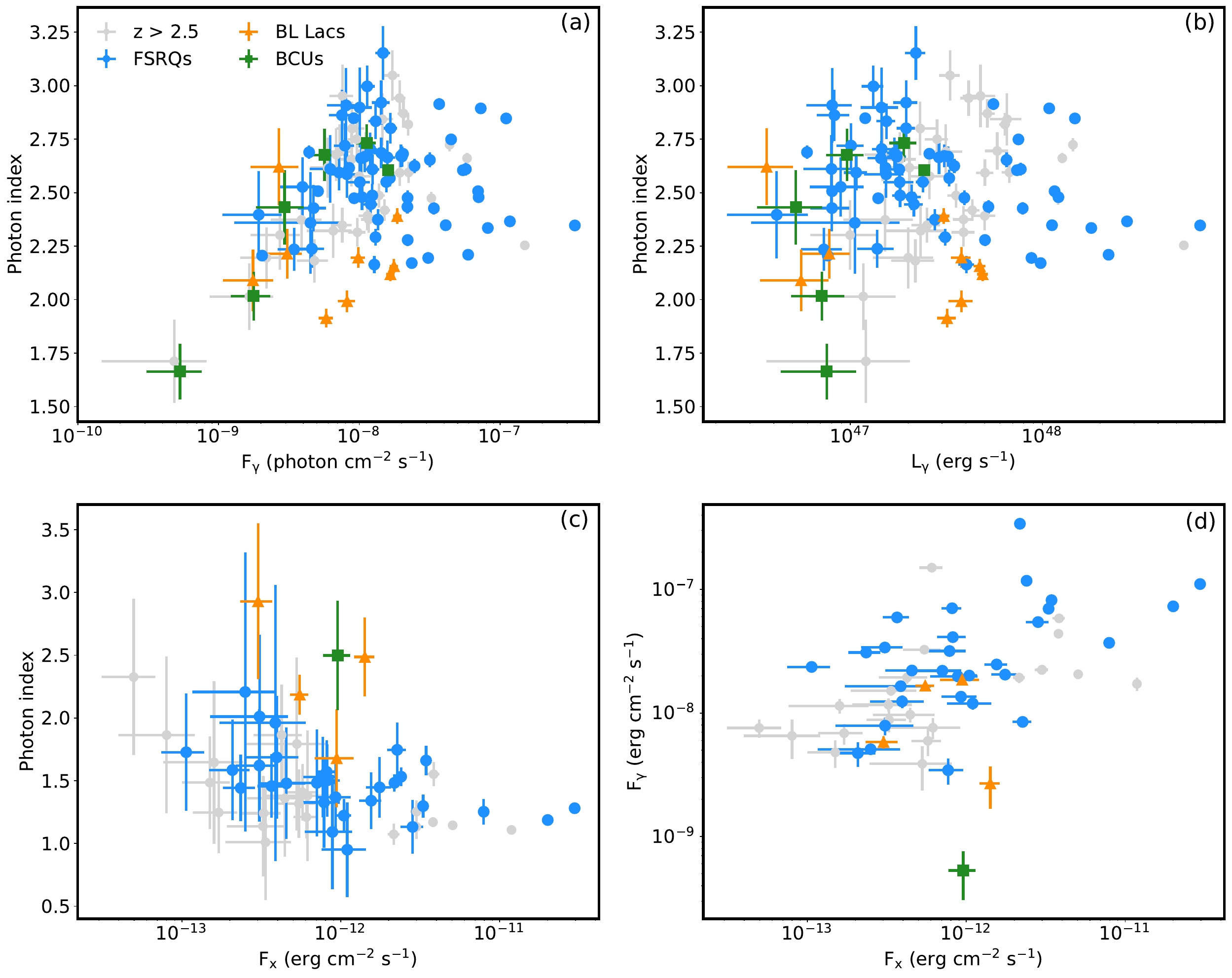}

     \caption{{\it Panels (a) and (b):} The \gray\ flux ($>100$ MeV) and luminosity versus the photon index. {\it Panel (c)} The X-ray flux versus the photon index. {\it Panel (d)} The X-ray flux versus the \gray\ flux.}
     \label{spectrum}
 \end{figure*}
In this section, a comprehensive spectral and temporal analysis of the sources included in our sample are performed. The outcomes of the \gray\ data analysis are provided in Table \ref{tab:param}. For each source, the power-law index ($\alpha_{\rm p}$), or (\(\alpha\)) together with the curvature parameter (\(\beta\)) when the data are best described using a low-parabola model, as well as the flux and luminosity (computed from the power-law fit), both with their respective uncertainties and the redshift of each source is reported. When the source spectrum is described by low-parabola model in 4FGL an additional analysis is performed using power-law model. 

The analysis results are presented in panel a of Fig. \ref{spectrum}, which depicts the flux and photon index derived from a power-law fit. The FSRQs are shown by blue markers, BL Lacs by orange, and BCUs by green. The \gray\ flux of sources within our sample ranges from $(5.32 \pm 2.25) \times 10^{-10}$ to $(3.40 \pm 0.02) \times 10^{-7}$ photon cm\(^{-2}\) s\(^{-1}\), with the lowest value corresponding to SDSS J100326.63+020455.6 and the highest to 4C +01.02. The mean flux is at $2.48 \times 10^{-8}$ photon cm\(^{-2}\) s\(^{-1}\). The photon index span from $1.66 \pm 0.12$ to $3.15 \pm 0.12$, with the lowest and highest indices observed for SDSS J100326.63+020455.6 and PKS 1915-458, respectively. The FSRQs, which are the most numerous in our sample, largely define these observed ranges. On the other hand, BCUs and BL Lacs exhibit narrower distributions in both \gray\ flux and photon index. Specifically, BL Lacs range between $(0.18-1.86) \times 10^{-8}$ photon cm\(^{-2}\) s\(^{-1}\) for flux and $1.91-2.62$ for photon index, while BCUs between $(0.05-1.60) \times 10^{-8}$ photon cm\(^{-2}\) s\(^{-1}\) for flux and $1.66-2.73$ for photon index. For comparison, blazars with a redshift exceeding $2.5$ from \citet{2020MNRAS.498.2594S} are depicted in light gray in panel a of Fig. \ref{spectrum}. This visualization shows that blazars with a redshift beyond $2.5$, as well as those included in the current sample, exhibit similar features, occupying similar regions on the photon index versus \( F_{\gamma} \) plane.

Panel b of Fig. \ref{spectrum} presents the \gray\ luminosity \( L_{\gamma} \) versus the photon index. This contrasts with the flux, luminosity, on the other hand, accounts for the total energy emitted by the source per unit time, so showing the intrinsic power of the sources. The luminosity of the sources under consideration spans from \( (3.67 \pm 1.37) \times 10^{46} \) erg s\(^{-1}\) to \( (6.62 \pm 0.05) \times 10^{48} \) erg s\(^{-1}\), with the lowest estimated for SDSS J105707.47+551032.2 and the highest for 4C +01.02. Notably, the luminosity of 4C +01.02 also exceeds that of B3 1343+451, which is the most luminous in the sample of sources with a redshift beyond $2.5$. The range of luminosities for the new sample is slightly shifted towards a higher luminosity range. For instance, the luminosities of PKS 0226-559 (\(2.21 \times 10^{48} \) erg s\(^{-1}\)), PKS 0601-70 (\(1.12 \times 10^{48} \) erg s\(^{-1}\)), PKS 2149-306 (\(1.09 \times 10^{48} \) erg s\(^{-1}\)), S3 0458-02 (\( 2.76 \times 10^{48} \) erg s\(^{-1}\)), 4C +71.07 (\( 1.48 \times 10^{48} \) erg s\(^{-1}\)), S4 0917+44 (\( 1.80 \times 10^{48} \) erg s\(^{-1}\)), PKS 0446+11 (\( 1.21 \times 10^{48} \) erg s\(^{-1}\)), PKS 1329-049 (\( 1.16 \times 10^{48} \) erg s\(^{-1}\)), and 4C +01.02 (\( 6.62 \times 10^{48} \) erg s\(^{-1}\)) are exceeding luminosity of \( 10^{48} \) erg s\(^{-1}\), making them among the luminous blazars detected in the \gray\ band.

Panel c of Fig. \ref{spectrum} shows the relationship between the X-ray photon index and the flux of the considered sources. The X-ray flux for the sources studied range from \( (1.06 \pm 0.32) \times 10^{-13} \) erg cm\(^{-2}\)s\(^{-1}\) for PMN J1344-1723 to \( (2.96 \pm 0.02) \times 10^{-11} \) erg cm\(^{-2}\)s\(^{-1}\) for 4C +71.07. The X-ray photon index is predominantly soft (less than 2.0) for the majority of the sources, suggesting that the X-ray emissions are likely dominated by the rising part of inverse Compton component. Notably, the brightest sources in the sample, such as 4C +71.07 and PKS 2149-306, have fluxes of \( (2.96 \pm 0.02) \times 10^{-11} \) erg cm\(^{-2}\)s\(^{-1}\) and \( (2.01 \pm 0.04) \times 10^{-11} \) erg cm\(^{-2}\)s\(^{-1}\), with photon indices of \( 1.28 \pm 0.01 \) and \( 1.19 \pm 0.03 \), respectively. These indices indicate their X-ray spectra are particularly hard compared to the rest of the sample. Correspondingly, these two sources also show the highest X-ray luminosities, being \( (1.13 \pm 0.11) \times 10^{48} \) erg s\(^{-1}\) for 4C +71.07, and \( (8.79 \pm 1.83) \times 10^{47} \) erg s\(^{-1}\) for PKS 2149-306.

In panel d of Fig. \ref{spectrum}, a comparison of the \(\gamma\)-ray and X-ray fluxes for the selected sources is shown. The wide spread observed in the data suggests that there is no direct or obvious correlation between the \(\gamma\)-ray and X-ray fluxes when considering time-averaged measurements. It should be noted, however, that these are average values and that during shorter time-scale events, such as flares, a correlation may appear. Interestingly,  the two bright in the X-ray band sources, 4C +71.07 and PKS 2149-306, also have notably high \(\gamma\)-ray fluxes of \( (1.11 \pm 0.02) \times 10^{-7} \) photon cm\(^{-2}\) s\(^{-1}\) and \( (7.30 \pm 0.20) \times 10^{-8} \) photon cm\(^{-2}\) s\(^{-1}\), respectively. Conversely, the \(\gamma\)-ray bright source 4C +01.02,  has only a moderate X-ray flux of \( (2.18 \pm 0.11) \times 10^{-12} \) erg cm\(^{-2}\)s\(^{-1}\), indicating that a high \(\gamma\)-ray flux does not necessarily imply a correspondingly high X-ray flux. This discrepancy shows the complexity of the emission mechanisms and the potential influence of other factors such as beaming, the environment of the source, or the presence of different emission processes at different wavelengths.

The NuSTAR analysis results are presented in Table \ref{nust}, where for each source the observation sequence, observation time, and flux in the $3-10$ keV and $10-30$ keV ranges, along with the photon index are provided. Three NuSTAR observations (60160099002 for S5 0212+73, 60002045002 for 4C +71.07, and 60367002002 for PKS 0227-369) were relatively short (on the order of a few hundred seconds), and hence no spectral analysis was conducted. In the $3-10$ keV range, the highest flux of \((2.02\pm0.01)\times10^{-11}\:\text{erg\:cm}^{-2}\:\text{s}^{-1}\) was observed for 4C +71.07 on MJD 56675.22, while the lowest flux of \((1.95\pm0.27)\times10^{-13}\:\text{erg\:cm}^{-2}\:\text{s}^{-1}\) was observed for 87GB 080551.6+535010. The photon index for all considered sources is hard, ranging from $1.09$ to $1.67$, suggesting that the hard X-ray component corresponds to the rising part of the inverse Compton component. The variability of the $3-10$ keV and $10-30$ keV fluxes could only be investigated for 4C +71.07 and PKS 2149-306, as multiple observations in different periods are available; however, the flux remained relatively stable.

\begin{table*}
\caption{NuSTAR analysis results. The source name, NuSTAR sequence ID, observation time in MJD, the logarithm of the 3-10 keV ($\log F_{3-10}$) and 10-30 keV ($\log F_{10-30}$) bands flux and the photon index for each observation are given.}
\label{nust}
\begin{tabular}{lccccc}
\toprule
Source & Sequence ID & MJD & $\log F_{3-10}$ & $\log F_{10-30}$ & photon index \\
\midrule
4C +71.07 & 60002045004 & 56675.22 & $-10.70 \pm 0.003$ & $-10.55 \pm 0.005$ & $1.63 \pm 0.01$ \\
4C +71.07 & 60002045002 & 56641.65 & $-10.88 \pm 0.004$ & $-10.75 \pm 0.007$ & $1.67 \pm 0.02$ \\
PKS 2149-306 & 60001099002 & 56643.22 & $-10.71 \pm 0.003$ & $-10.43 \pm 0.003$ & $1.35 \pm 0.01$ \\
PKS 2149-306 & 60001099004 & 56765.64 & $-10.79 \pm 0.003$ & $-10.55 \pm 0.004$ & $1.45 \pm 0.01$ \\
S3 0458-02 & 60367003001 & 58234.31 & $-11.65 \pm 0.014$ & $-11.51 \pm 0.020$ & $1.64 \pm 0.08$ \\
87GB 080551.6+535010 & 80001004002 & 56785.45 & $-12.71 \pm 0.061$ & $-12.30 \pm 0.087$ & $1.09 \pm 0.28$ \\
PKS 0446+11 & 60101078002 & 57358.10 & $-12.53 \pm 0.059$ & $-12.37 \pm 0.088$ & $1.60 \pm 0.29$ \\
TXS 0322+222& 60101079002 & 57334.11 & $-11.88 \pm 0.017$ & $-11.54 \pm 0.027$ & $1.24 \pm 0.09$ \\
PKS 0528+134& 60160238002 & 58509.08 & $-12.09 \pm 0.034$ & $-11.94 \pm 0.040$ & $1.61 \pm 0.16$ \\
PKS 0227-369 & 60367002002 & 57975.49 & $-12.56 \pm 0.055$ & $-12.27 \pm 0.077$ & $1.34 \pm 0.29$ \\
PKS 1329-049 & 60160541002 & 57902.35 & $-12.27 \pm 0.043$ & $-12.03 \pm 0.050$ & $1.44 \pm 0.19$ \\
S5 0212+73 & 60160099002 & 57442.46 & $-11.24 \pm 0.006$ & $-11.03 \pm 0.009$ & $1.52 \pm 0.03$ \\
\bottomrule
\end{tabular}
\end{table*}
\begin{figure*}
     \centering
     \includegraphics[ width=0.98\textwidth]{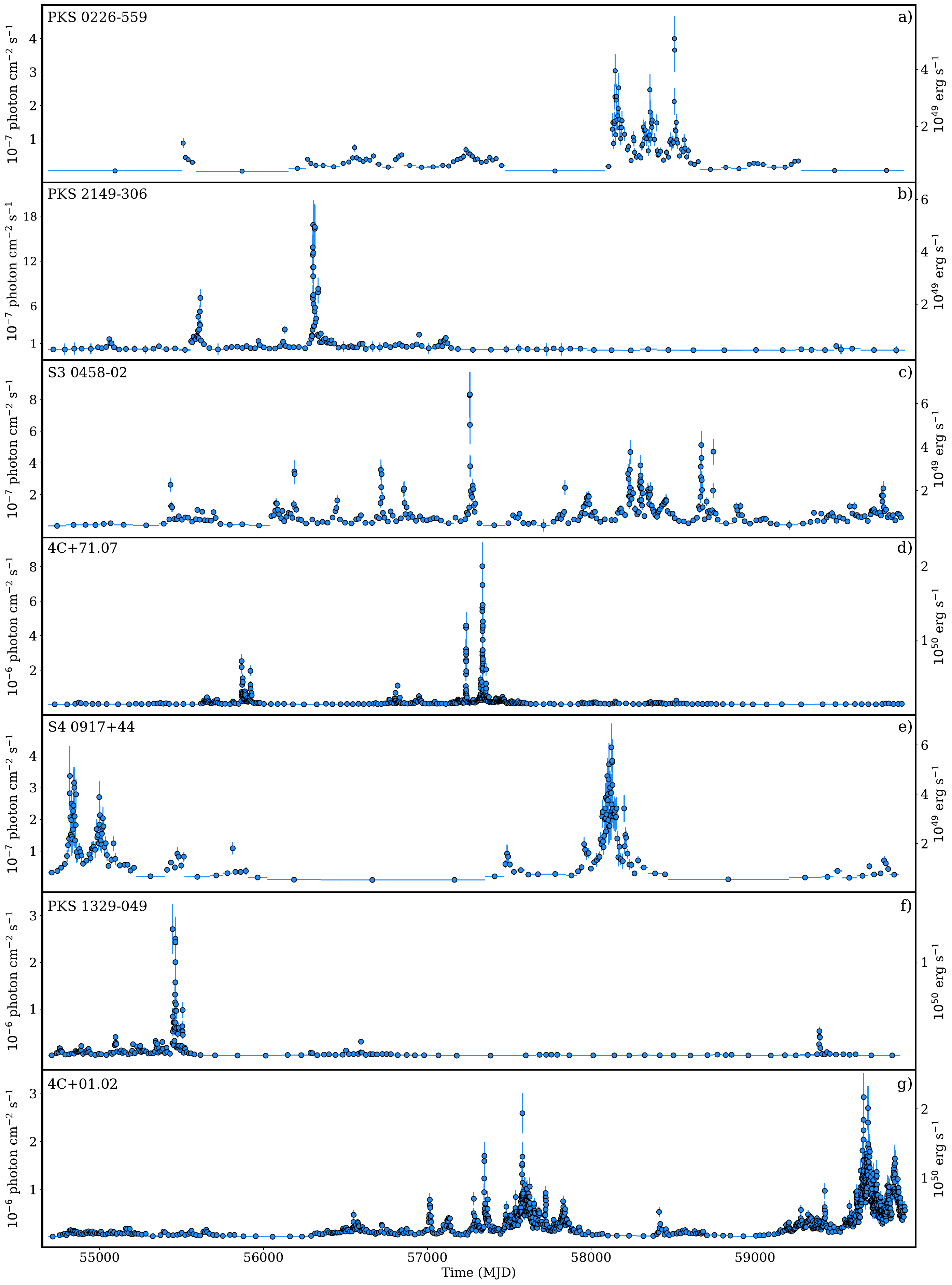}
     \caption{The \gray\ light curve of the sources considered in this study, which show a high amplitude \gray\ flux increase during the flares. The left axis shows the variation of flux over time, while the right axis displays the luminosity.}
     \label{strong_gray}
 \end{figure*}
\subsection{\gray\ variability}
The use of adaptive binning methods for computing light curves allowed detailed investigation of the \gray\ flux variation of considered sources. Flux variations, characterized by several times increases from average levels, have been observed in 31 sources. The most variable sources in the \gray\ band, where multiple flaring activities have been observed, are PKS 0226-559, PKS 2149-306, S3 0458-02, 4C+71.07, S4 0917+44, PKS 1329-049 and 4C+01.02. For example, the adaptively binned \gray\ light curve of PKS 0226-559, the most distant source in our sample at a redshift \(z=2.464\) showing variability, computed for energies above \(239.66\) MeV, is displayed in panel a) of Fig. \ref{strong_gray}. From the start of \fermi\ observations until MJD 55504 (November 04, 2010), the \gray\ emission of this source was in a low state being \((4.89 \pm 0.71) \times 10^{-9}\: \text{photon\:cm}^{-2}\:\text{s}^{-1}\). Subsequent periods of flaring activity occurred between MJD 55504.30-55584.50, 56153.18-57472.15, and MJD 58084.93-58664.52. During its most active state, between MJD 58084.93 and 58664.52, the maximum flux was \((3.99 \pm 0.67) \times 10^{-7}\: \text{photon\:cm}^{-2}\:\text{s}^{-1}\) observed on MJD 58507.69. Also, the luminosity of the source significantly increased during these flaring periods. While the long-term averaged luminosity was \((1.76 \pm 0.04) \times 10^{48}\: \text{erg\:s}^{-1}\), the luminosity during flares exceeded \(10^{49}\: \text{erg\:s}^{-1}\). For example, the source luminosity was \(>10^{49}\: \text{erg\:s}^{-1}\) $42$ times, with the peak luminosity of \((5.33 \pm 0.92) \times 10^{49}\: \text{erg\:s}^{-1}\) observed on MJD 58167.46. 

Among the flaring sources, significant increases in \gray\ emission have been observed for PKS 2149-306, 4C+71.07, PKS 1329-049, and 4C+01.02. For example, the time-averaged \gray\ flux of PKS 2149-306 (\(z=2.345\)) is \((2.01 \pm 0.47) \times 10^{-7}\: \text{photon\:cm}^{-2}\:\text{s}^{-1}\), but there was a notable period of elevated \gray\ emission between MJD 56288.92 and 56444.75 (Fig. \ref{strong_gray} panel b). During this interval, the flux above \(144.85\) MeV exceeded \(10^{-6}\: \text{photon\:cm}^{-2}\:\text{s}^{-1}\) on eight time intervals. The peak flux during this period was \((1.68 \pm 0.33) \times 10^{-6}\: \text{photon\:cm}^{-2}\:\text{s}^{-1}\) observed on MJD 56302.61. Additionally, the source exhibited another active \gray\ emission state between MJD 55554.39 and 55650.67. In this period, the \gray\ flux consistently exceeded \(10^{-7}\: \text{photon\:cm}^{-2}\:\text{s}^{-1}\), with a maximum of \((7.08 \pm 1.22) \times 10^{-7}\: \text{photon\:cm}^{-2}\:\text{s}^{-1}\) observed on MJD 55613.98. During multiple flaring periods of 4C+71.07 (Fig. \ref{strong_gray} panel d), which is at a redshift of \(z=2.218\), the \gray\ emission notably exceeded its time-averaged flux of \((1.06 \pm 0.01) \times 10^{-7}\: \text{photon\:cm}^{-2}\:\text{s}^{-1}\). Specifically, in three distinct flaring periods—MJD 55862.01-55932.97, MJD 57227-57245.79, and MJD 57296.92-57338.47—the flux above \(138.16\) MeV exceeded \(10^{-6}\: \text{photon\:cm}^{-2}\:\text{s}^{-1}\). The peak flux during these flaring events was at \((8.02 \pm 1.40) \times 10^{-6}\: \text{photon\:cm}^{-2}\:\text{s}^{-1}\), and was observed on MJD 57335.21. The \gray\ emission of PKS 1329-049, with a redshift of \(z=2.15\), is predominantly in its average emission state, as depicted in Fig. \ref{strong_gray} panel f). However, it exhibited an elevated emission state during the period from MJD 55442.09 to 55507.87 when the peak flux reached \((2.71 \pm 0.52) \times 10^{-6}\: \text{photon\:cm}^{-2}\:\text{s}^{-1}\) observed on MJD 55445.32. In contrast, 4C+01.02, with a redshift of \(z=2.099\), experiences alternating periods of flaring activity, with the most intense \gray\ flares observed after MJD 57000. The source entered an active \gray\ emission state starting from MJD 59615.91, when the peak \gray\ flux, measured above \(171.79\) MeV, reached \((2.92 \pm 0.50) \times 10^{-6}\: \text{photon\:cm}^{-2}\:\text{s}^{-1}\) on MJD 59663.19. The other two sources, depicted in panels c) and e) of Fig. \ref{strong_gray}, also exhibit multiple flaring periods but with only modest increases in their \gray\ flux. Specifically, the peak \gray\ flux of S3 0458-02, measured above \(202.55\) MeV, was \((8.32 \pm 1.41) \times 10^{-7}\: \text{photon\:cm}^{-2}\:\text{s}^{-1}\) observed on MJD 57258.92. Similarly, for S4 0917+44, it was  \((4.27 \pm 0.75) \times 10^{-7}\: \text{photon\:cm}^{-2}\:\text{s}^{-1}\) observed on MJD 57946.70.

Especially profound is the luminosity increase in 4C+71.07, PKS 1329-049, and 4C+01.02 (see Fig. \ref{strong_gray}). For 4C+71.07, during the flaring periods in MJD 57227-57245.79 and MJD 57296.92-57338.47, the source luminosity exceeded $10^{50}\:{\rm erg\:s^{-1}}$ 11 times. The highest luminosity of $(2.03\pm 0.36)\times10^{50}\:{\rm erg\:s^{-1}}$ was observed on MJD 57335.21. PKS 1329-049 was in an extreme bright state on MJD 55445.23 and MJD 55468.18, with luminosities of $(1.07\pm 0.21)\times10^{50}\:{\rm erg\:s^{-1}}$ and $(1.39\pm 0.26)\times10^{50}\:{\rm erg\:s^{-1}}$, respectively. Similarly, 4C+01.02 was in an extreme bright state between MJD 59662.03-59663.55, during which in 6 consecutive bins the flux exceeded $10^{50}\:{\rm erg\:s^{-1}}$, with the highest flux of $(2.08\pm 0.48)\times10^{50}\:{\rm erg\:s^{-1}}$ observed on MJD 59662.62. Because of such elevated luminosity, 4C+71.07, PKS 1329-049, and 4C+01.02 rank among the sources with the highest luminosity in the \gray\ band.

\begin{figure*}
     \centering
     \includegraphics[height=0.98\textheight, width=0.98\textwidth]{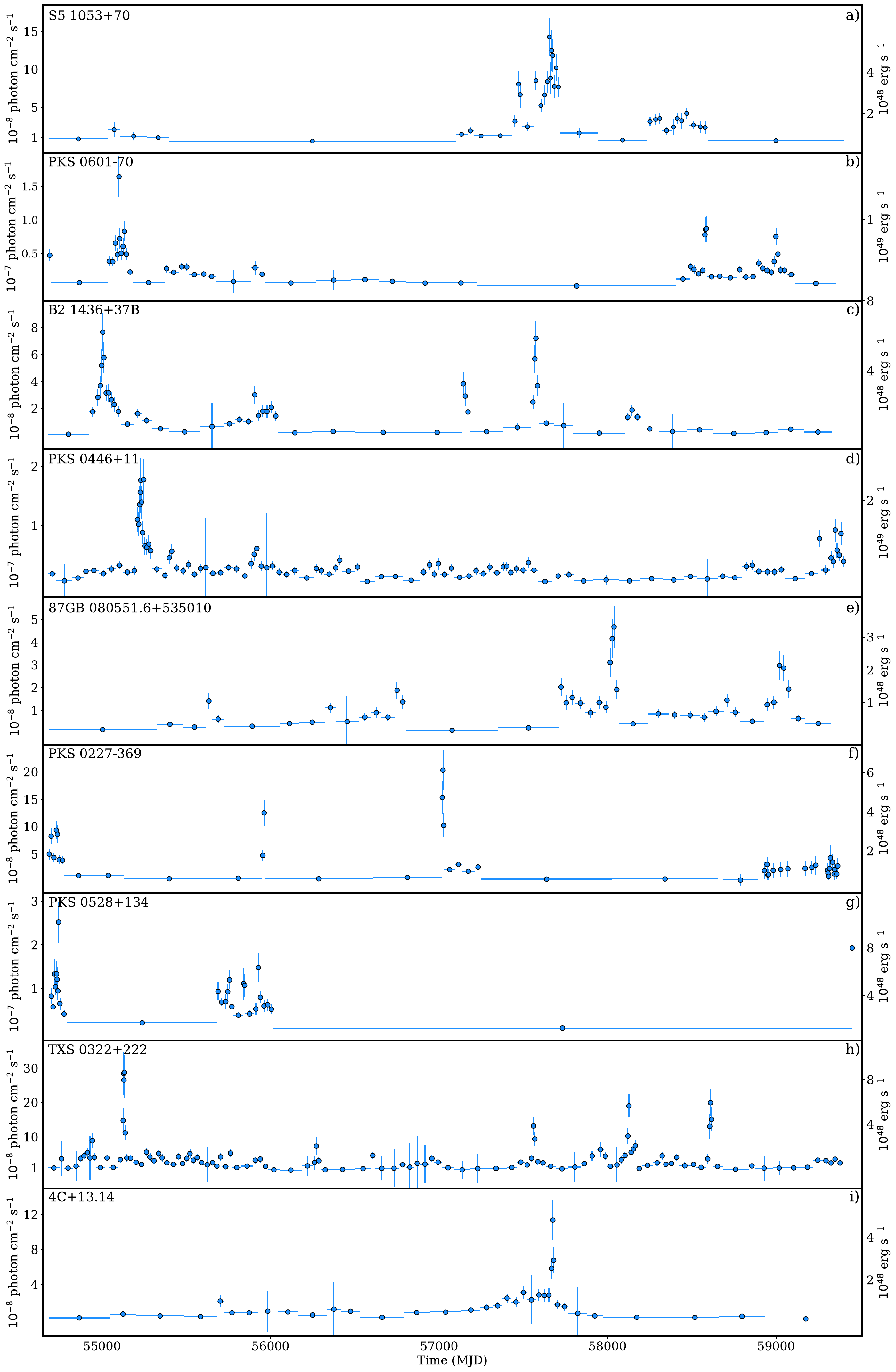}
     \caption{The \gray\ light curve of the sources analyzed in this study, which show a modest amplitude \gray\ flux increase during the flares. The left and right axes are the same as in Figure \ref{strong_gray}.}
     \label{moderate_gray}
 \end{figure*}

In Fig. \ref{moderate_gray}, the \gray\ light curves for 4C+13.14, 87GB 080551.6+535010, B2 1436+37B, PKS 0227-369, PKS 0528+134, PKS 0446+11, PKS 0601-70, S5 1053+70, and TXS 0322+222 are presented. For the majority of the time, the emission from these objects remains in a low to average state. However, during certain flaring periods, their \gray\ emission shows modest increases. These sources are typically weak, with \gray\ fluxes generally on the order of \(10^{-8}\: \text{photon\:cm}^{-2}\:\text{s}^{-1}\), but there are instances when the flux exceeds \(10^{-7}\: \text{photon\:cm}^{-2}\:\text{s}^{-1}\). For example, during an elevated \gray\ emission state, the peak flux of S5 1053+70 above \(179.20\) MeV was \((1.42 \pm 0.24) \times 10^{-7}\: \text{photon\:cm}^{-2}\:\text{s}^{-1}\) observed on MJD 57684.01. Similarly, for PKS 0601-70, the highest \gray\ flux reached \((1.64 \pm 0.30) \times 10^{-7}\: \text{photon\:cm}^{-2}\:\text{s}^{-1}\) on MJD 55136.37. For PKS 0446+11, the peak \gray\ flux was \((1.77 \pm 0.34) \times 10^{-7}\: \text{photon\:cm}^{-2}\:\text{s}^{-1}\) observed on MJD 55304.46, and so on.
 
 \subsection{X-ray variability}
 \begin{figure*}
     \centering
     \includegraphics[height=0.30\textheight,width=0.78\textwidth]{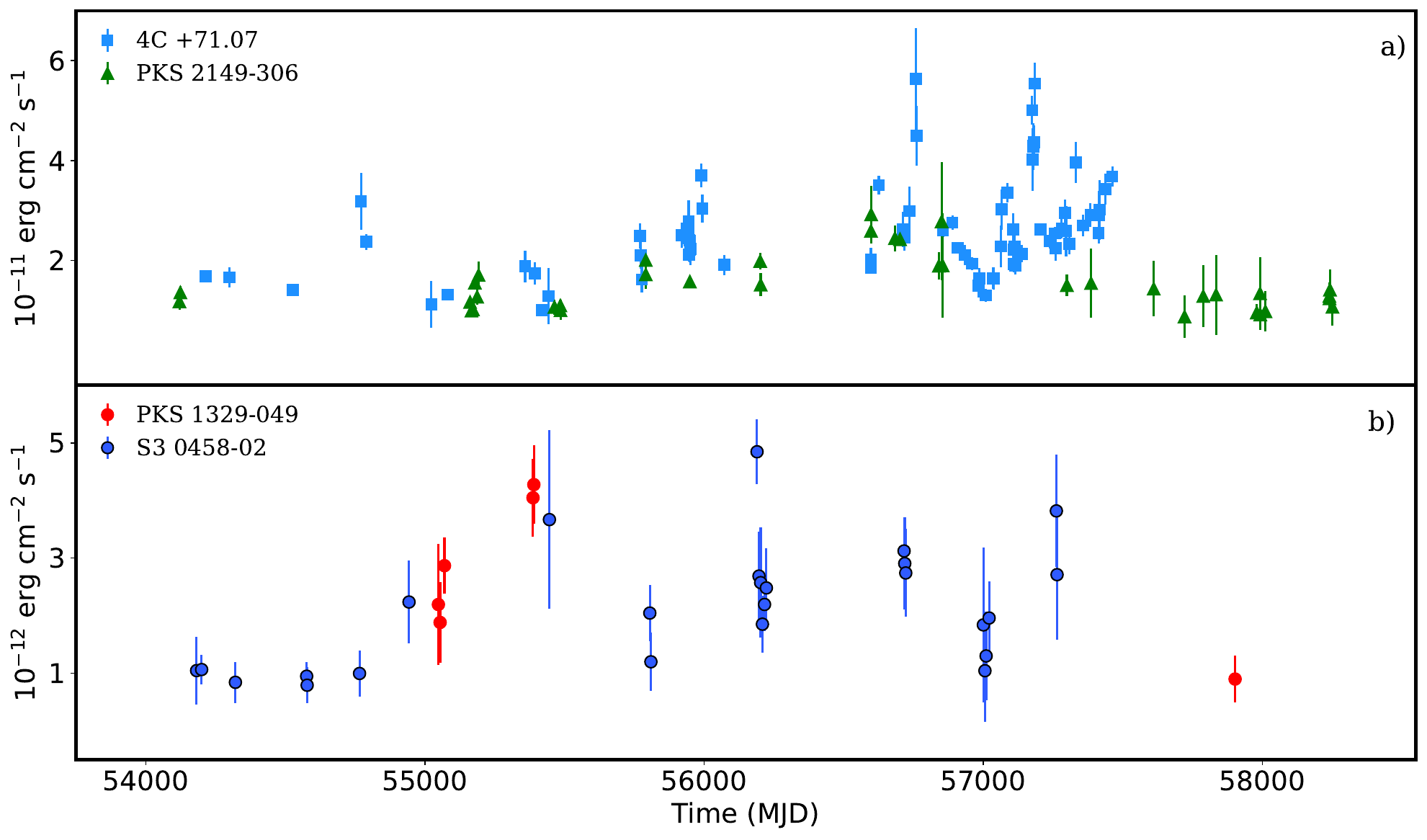}\\
     \caption{The X-ray light curve of 4C +71.07, PKS 2149-306, PKS 1329-049, and S3 0458-02, each with multiple X-ray observations, exhibits noticeable variability.}
     \label{Xvar}
 \end{figure*}
In the considered source sample, some of the sources were observed multiple times by the Swift telescope, permitting investigation of the temporal variability of X-ray flux across different years. Among the sources considered, variability in X-ray flux was observed in 4C +71.07, PKS 0226-559, PKS 1329-049, PKS 2149-306, S3 0458-02 and  S4 0917+44. For PKS 0226-559 and S4 0917+44, a limited number of observations were available but the ratio between the maximum and minimum fluxes is approximately $3-4$, indicating temporal flux variability.

The X-ray flux variations for 4C +71.07 (blue) and PKS 2149-306 (green) are shown in the upper panel of Fig. \ref{Xvar}. For 4C +71.07, the lowest observed X-ray flux was $(1.00\pm0.12)\times10^{-11}\:{\rm erg\:cm^{-2}\:s^{-1}}$ on MJD 55247.68, while the highest was $(5.63\pm1.02)\times10^{-11}\:{\rm erg\:cm^{-2}\:s^{-1}}$ on MJD 56832.97. Elevated X-ray emission states for this source were observed around MJD 56000 and MJD 57000, during which most of the X-ray observations were performed. Conversely, PKS 2149-306 exhibited lower amplitude flux changes, with its highest X-ray flux being $(2.92\pm0.58)\times10^{-11}\:{\rm erg\:cm^{-2}\:s^{-1}}$, observed on MJD 56643.02. The lower panel of Fig. \ref{Xvar} shows the X-ray flux variations for PKS 1329-048 (red) and S3 0458-02 (blue). Initially, the X-ray emission for S3 0458-02 was on the order of $\sim10^{-12}\:{\rm erg\:cm^{-2}\:s^{-1}}$, but it increased to approximately $\sim5\times10^{-12}\:{\rm erg\:cm^{-2}\:s^{-1}}$ around MJD 56200. For PKS 1329-048, the observed highest X-ray flux was $(4.28\pm0.68)\times10^{-12}\:{\rm erg\:cm^{-2}\:s^{-1}}$ on MJD 55390.19, whereas it decreased to $(8.99\pm4.03)\times10^{-13}\:{\rm erg\:cm^{-2}\:s^{-1}}$ on MJD 57902.45.

\subsection{Variability in optical/UV bands}
The observations from the Swift UVOT of selected sources allows to study the flux variability within the optical/UV bands. Investigating variability is challenging when the number of observations is limited, as changes in flux up to a factor of 2 can be observed across different filters; however, this does not provide a comprehensive understanding of the flux changes in time. Notably, clear flux variability is evident in the emissions from 4C+01.02, PKS 0226-559, and PKS 2149-306. For 4C+01.02, the initial flux measurements in the \( V \), \( B \), and \( U \) filters are approximately \( 10^{-12} \, \text{erg} \, \text{cm}^{-2} \, \text{s}^{-1} \), and approximately \( 2 \times 10^{-13} \, \text{erg} \, \text{cm}^{-2} \, \text{s}^{-1} \) in the \( W1 \), \( W2 \), and \( M2 \) filters. After MJD 56500, the source exhibits increased flux across all filters during several observations. The highest observed flux in the \( B \) filter was \( (3.91 \pm 0.33) \times 10^{-12} \, \text{erg} \, \text{cm}^{-2} \, \text{s}^{-1} \) on MJD 57363.45. For PKS 0226-559, the mean flux is \( \sim 5 \times 10^{-13} \, \text{erg} \, \text{cm}^{-2} \, \text{s}^{-1} \), but the flux in \( V \) filter increased up to \( (3.1-0.46) \times 10^{-12} \, \text{erg} \, \text{cm}^{-2} \, \text{s}^{-1} \) during the flare on MJD 58161.21. The UV emission from PKS 2149-306, in the \( W1 \), \( W2 \), and \( M2 \) filters, remains relatively stable across various observations, whereas the optical emission (\( V \), \( B \), and \( U \)) shows variability. Specifically, on MJD 53717.92, MJD 54980.86, and MJD 58586.29, the flux increased to approximately \( 3 \times 10^{-12} \, \text{erg} \, \text{cm}^{-2} \, \text{s}^{-1} \), with the lowest observed flux being around \( 10^{-13} \, \text{erg} \, \text{cm}^{-2} \, \text{s}^{-1} \).

\begin{figure*}
     \centering
     \includegraphics[width=0.45\textwidth]{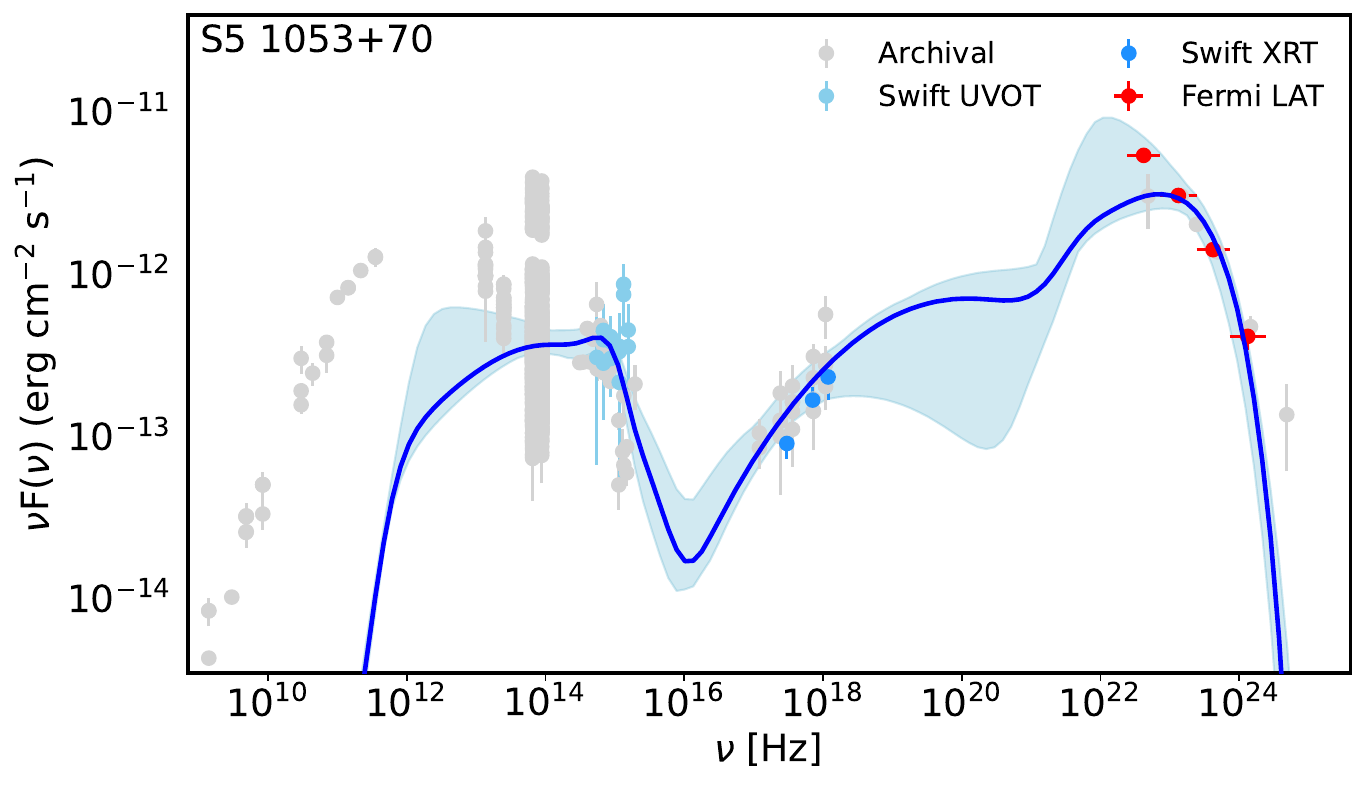}
     \includegraphics[width=0.45\textwidth]{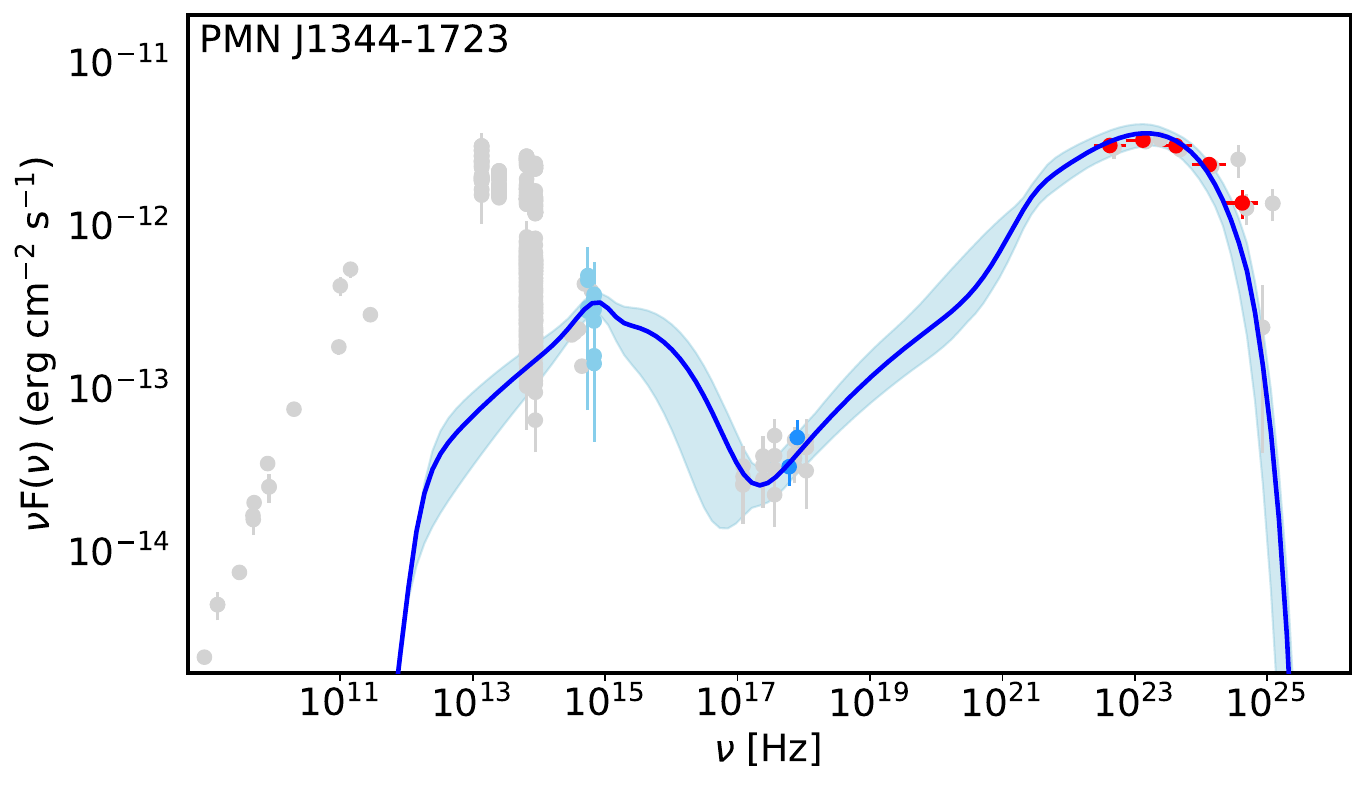}\\
     \includegraphics[width=0.45\textwidth]{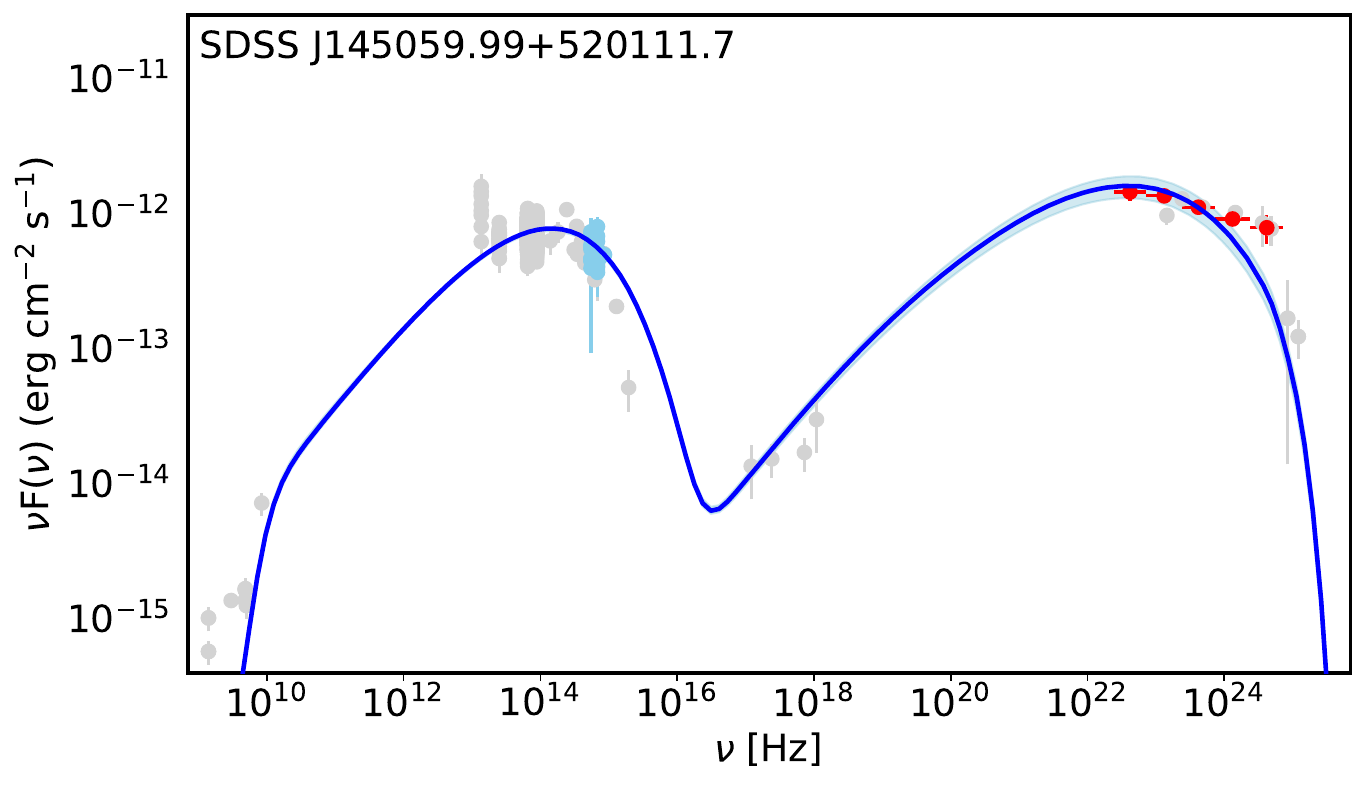}
     \includegraphics[width=0.45\textwidth]{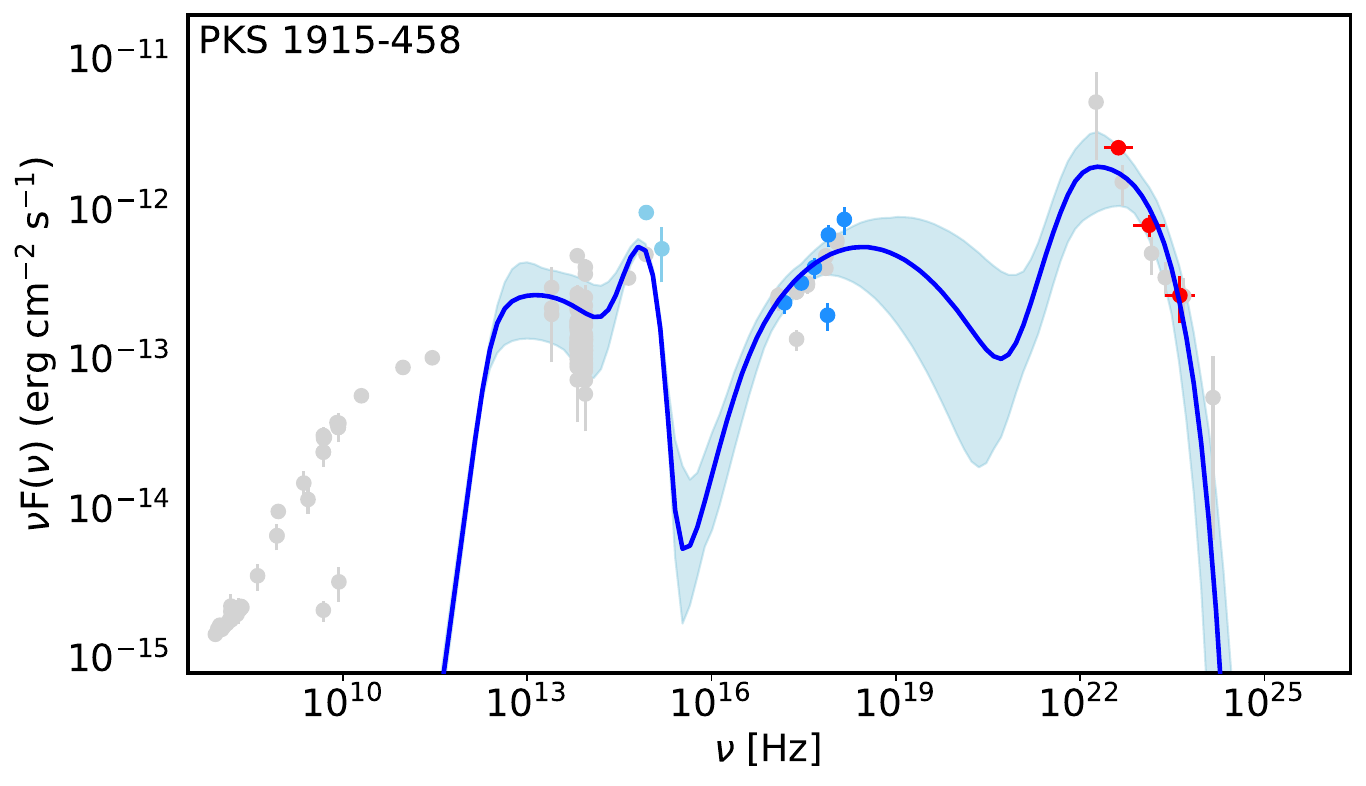}\\
     \includegraphics[width=0.45\textwidth]{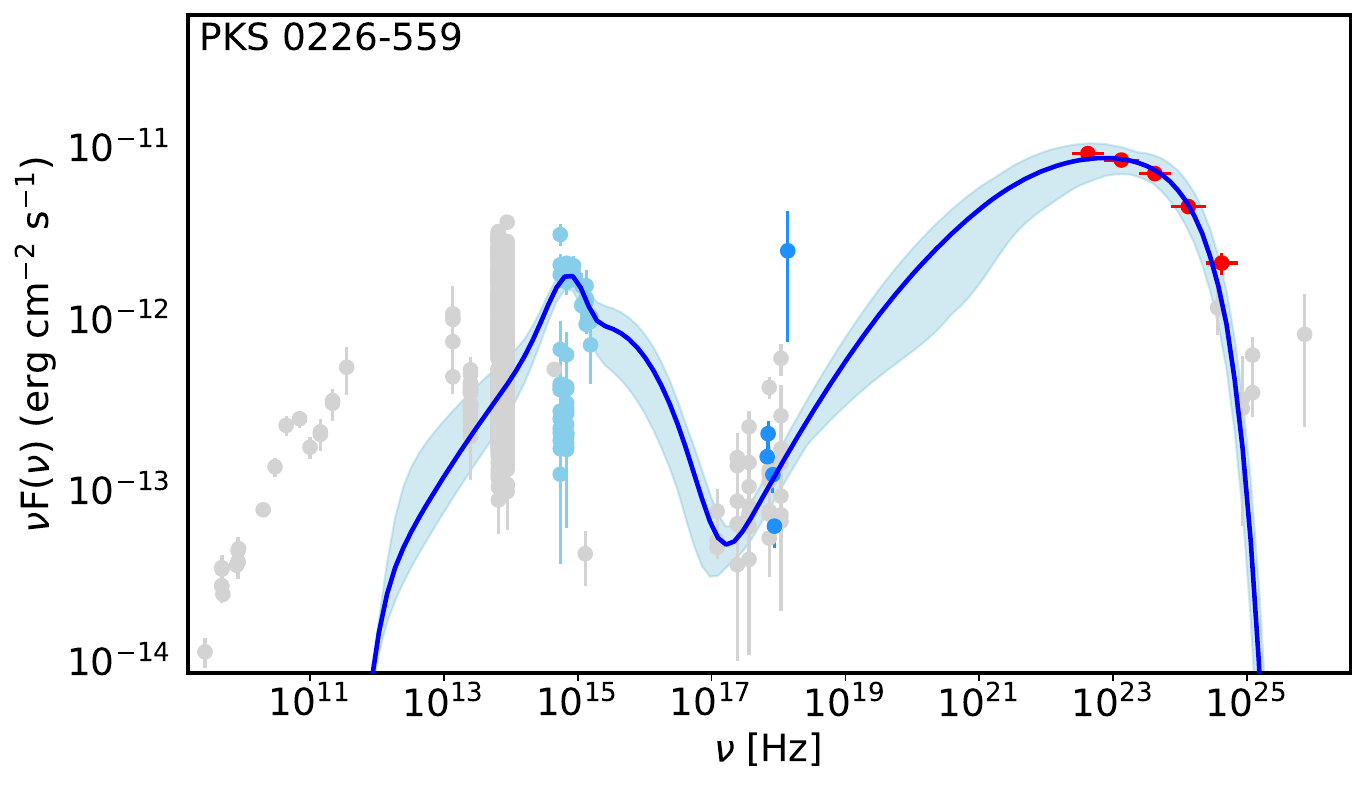}
     \includegraphics[width=0.45\textwidth]{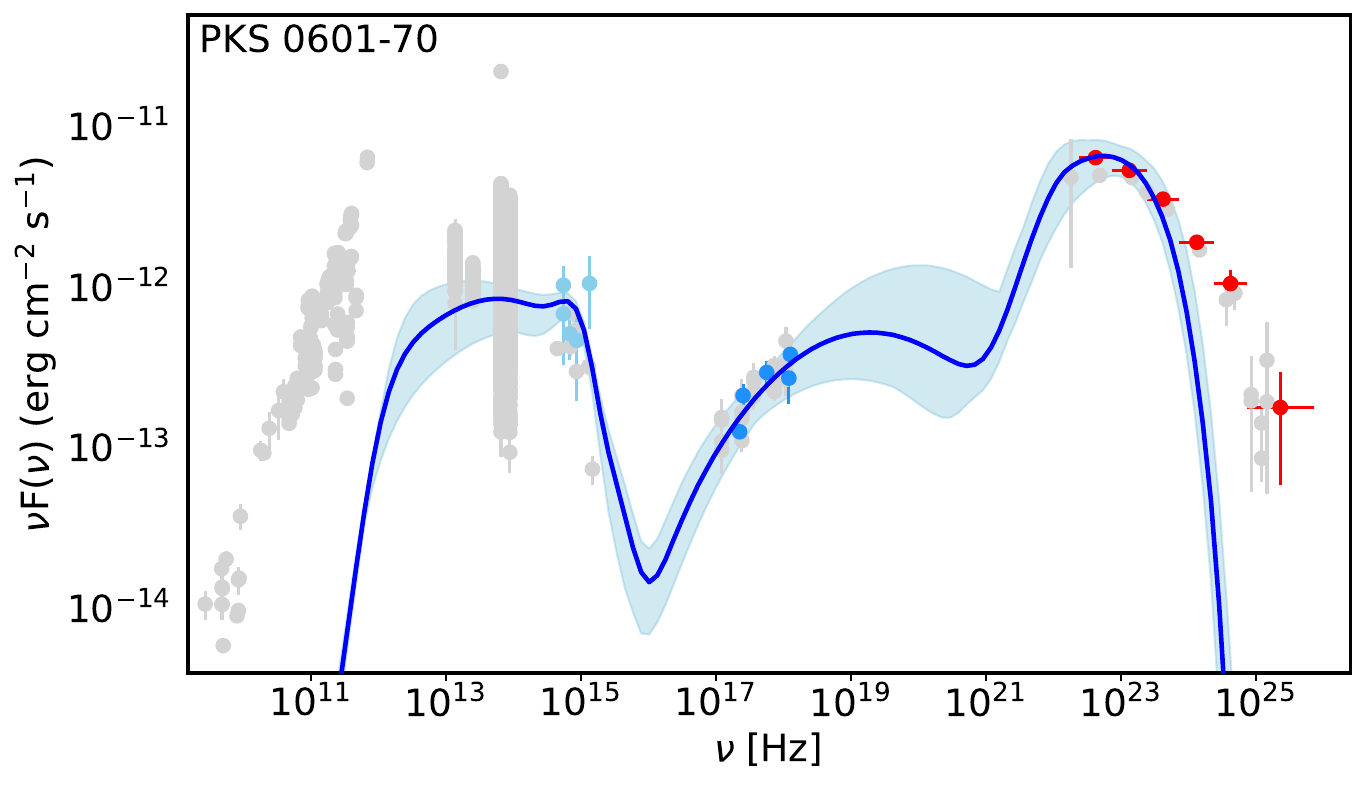}\\
     \includegraphics[width=0.45\textwidth]{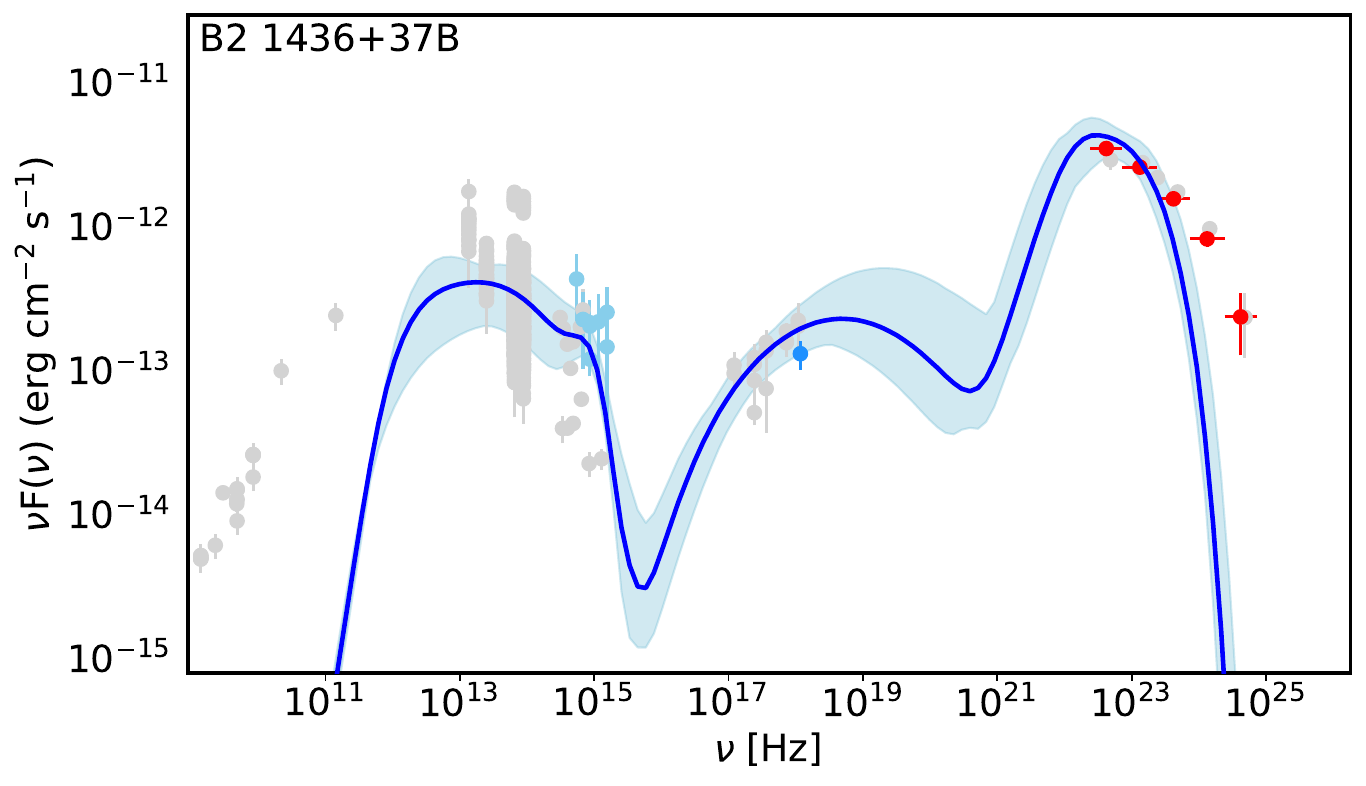}
     \includegraphics[width=0.45\textwidth]{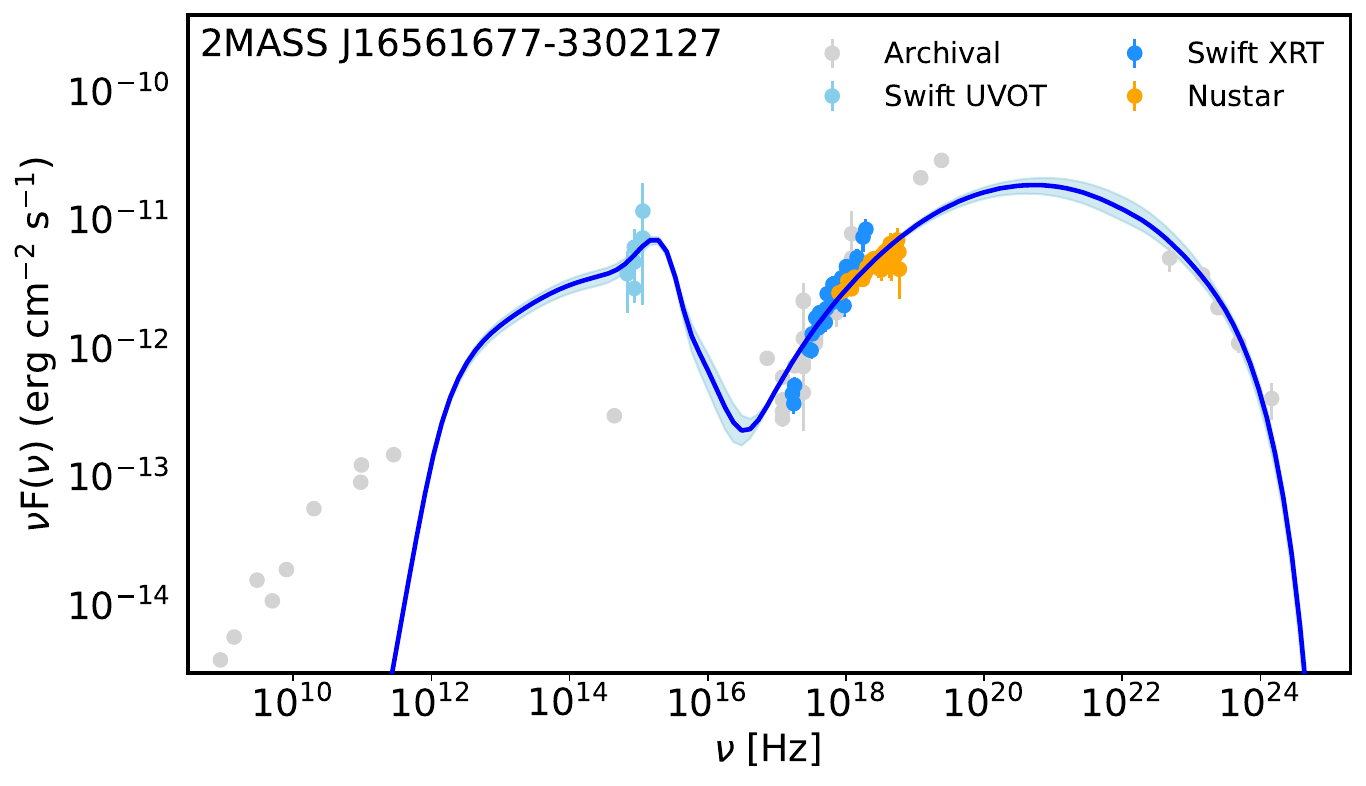}\\
     \includegraphics[width=0.45\textwidth]{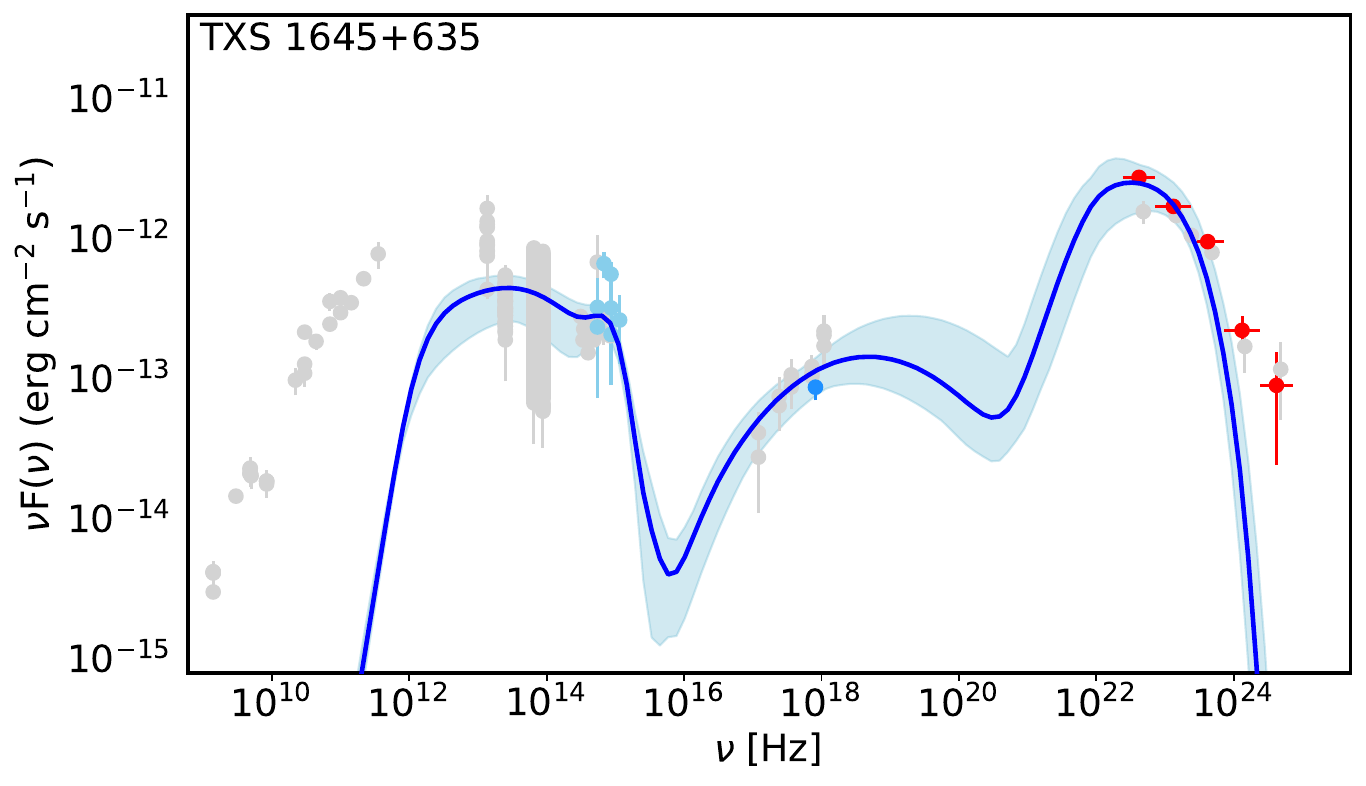}
     \includegraphics[width=0.45\textwidth]{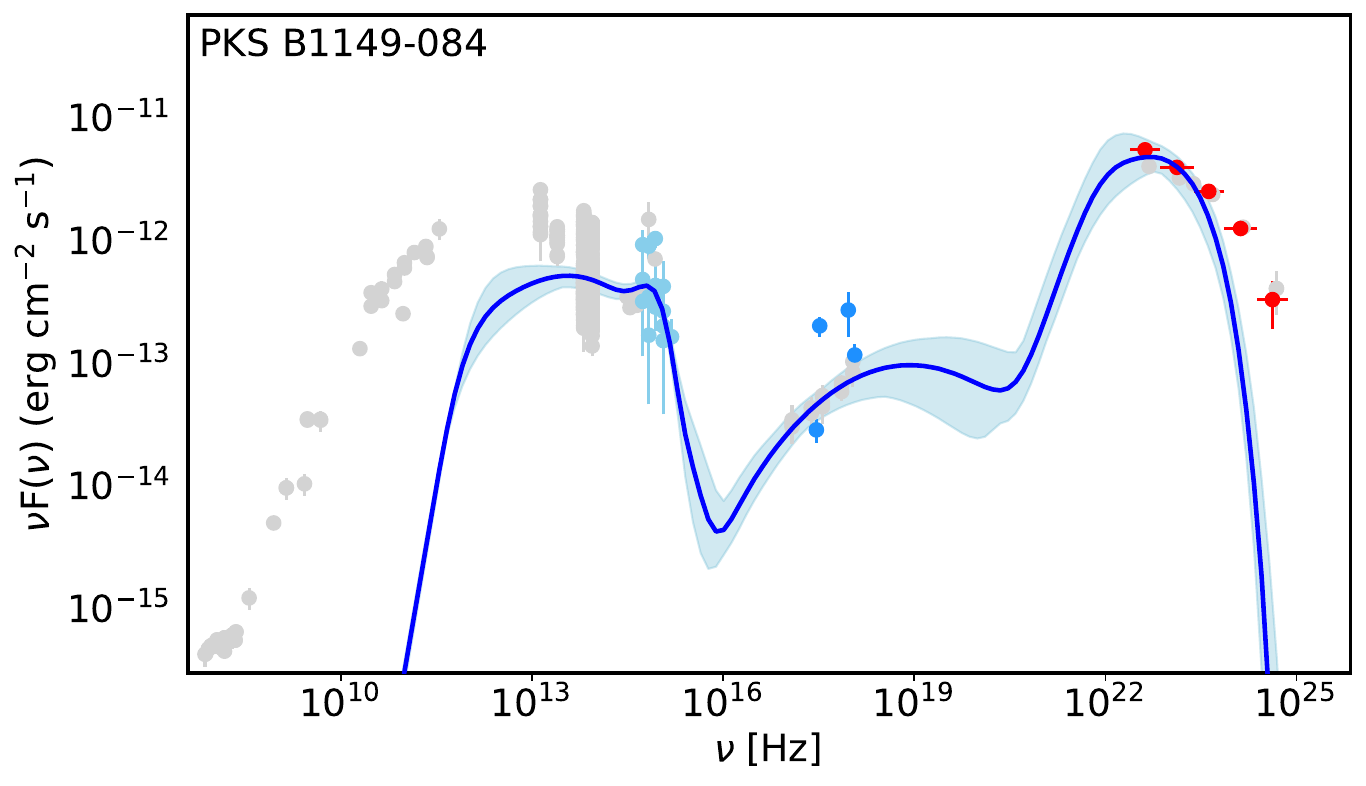}
     \caption{The broadband SED modeling results for the sources under consideration. The data analyzed from Swift UVOT, XRT, NuSTAR, and Fermi-LAT observation are depicted in cyan, blue, orange, and red, respectively, while  archival data are shown in gray. The blue line represents the combined contribution of the synchrotron, disk, SSC, and EIC-BLR components. The blue shaded area indicates the region of uncertainty associated with the model.}
     \label{sed}
\end{figure*}
\begin{figure*}
     \centering
     \ContinuedFloat
     \includegraphics[width=0.45\textwidth]{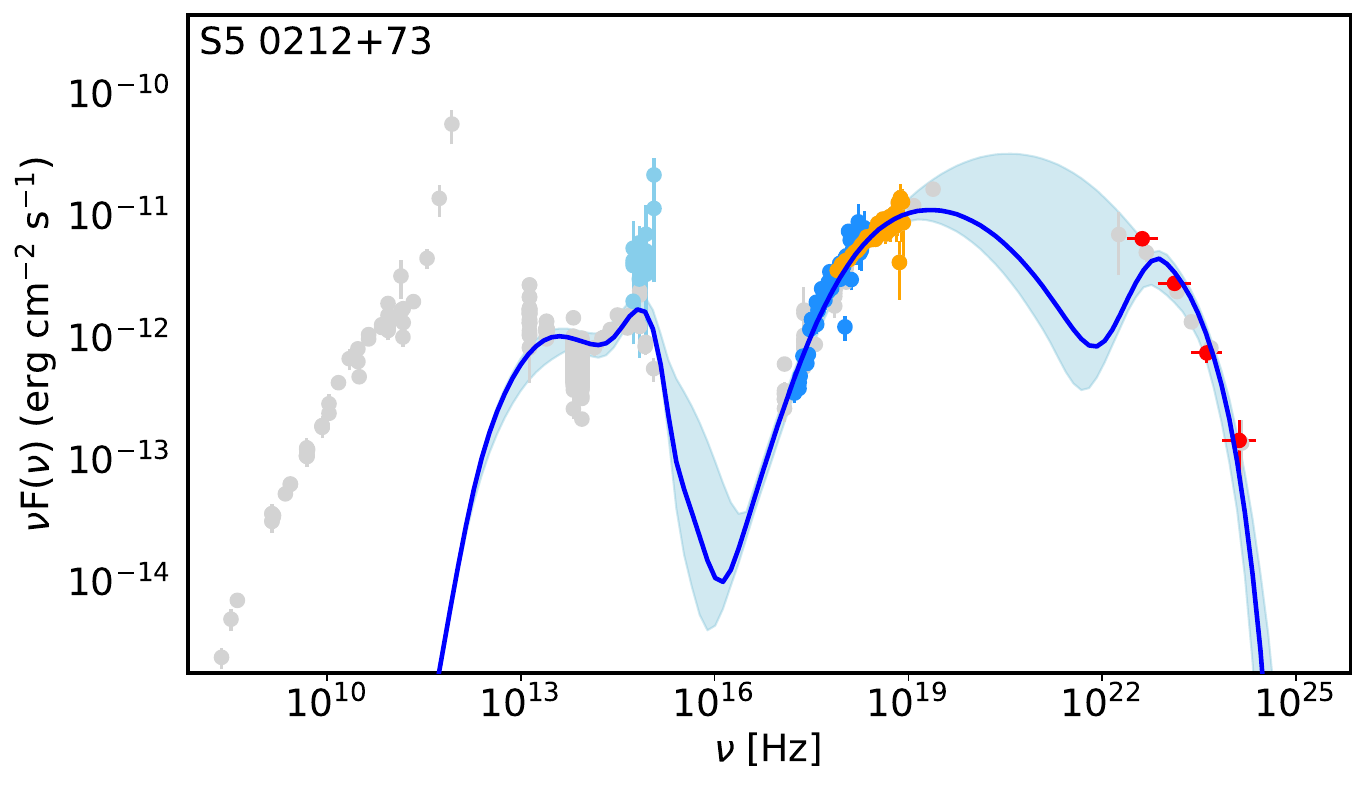}
     \includegraphics[width=0.45\textwidth]{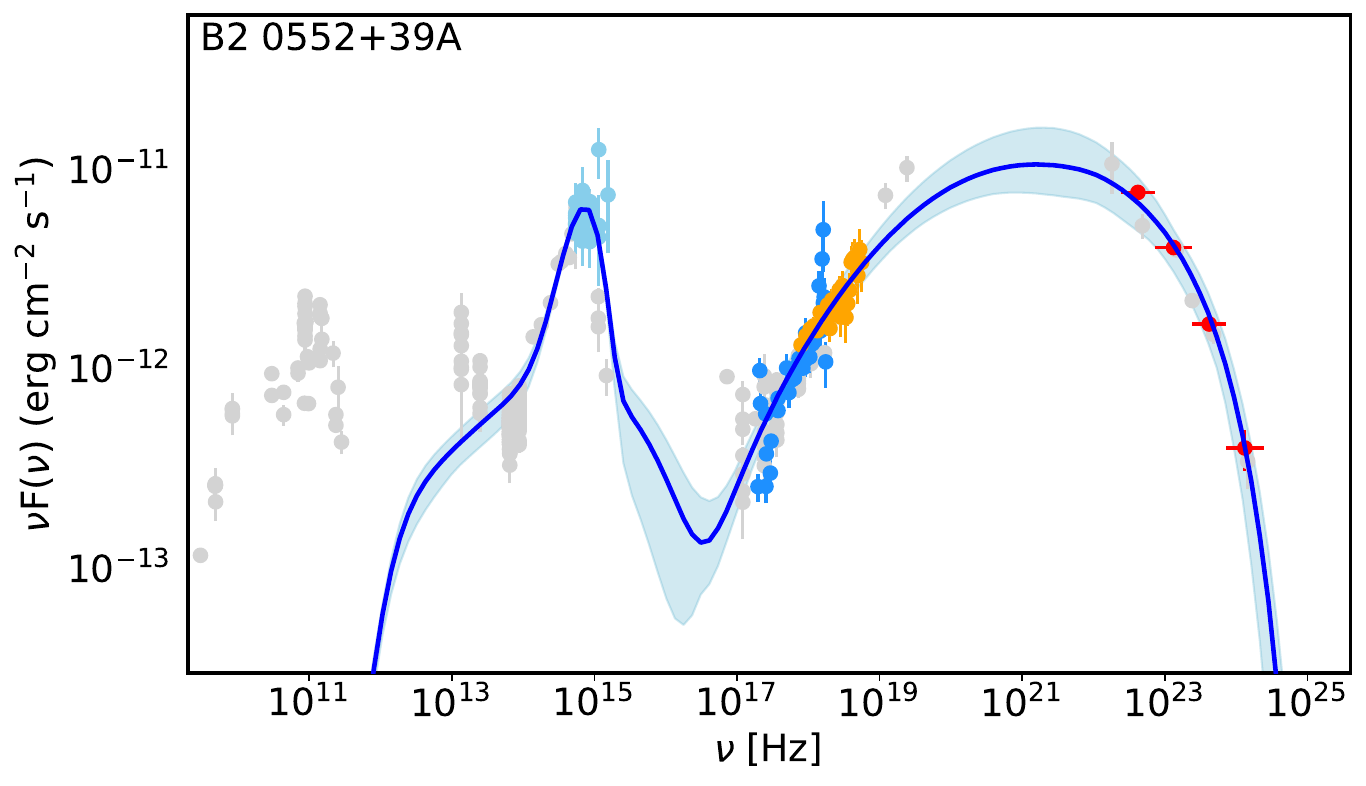}\\
     \includegraphics[width=0.45\textwidth]{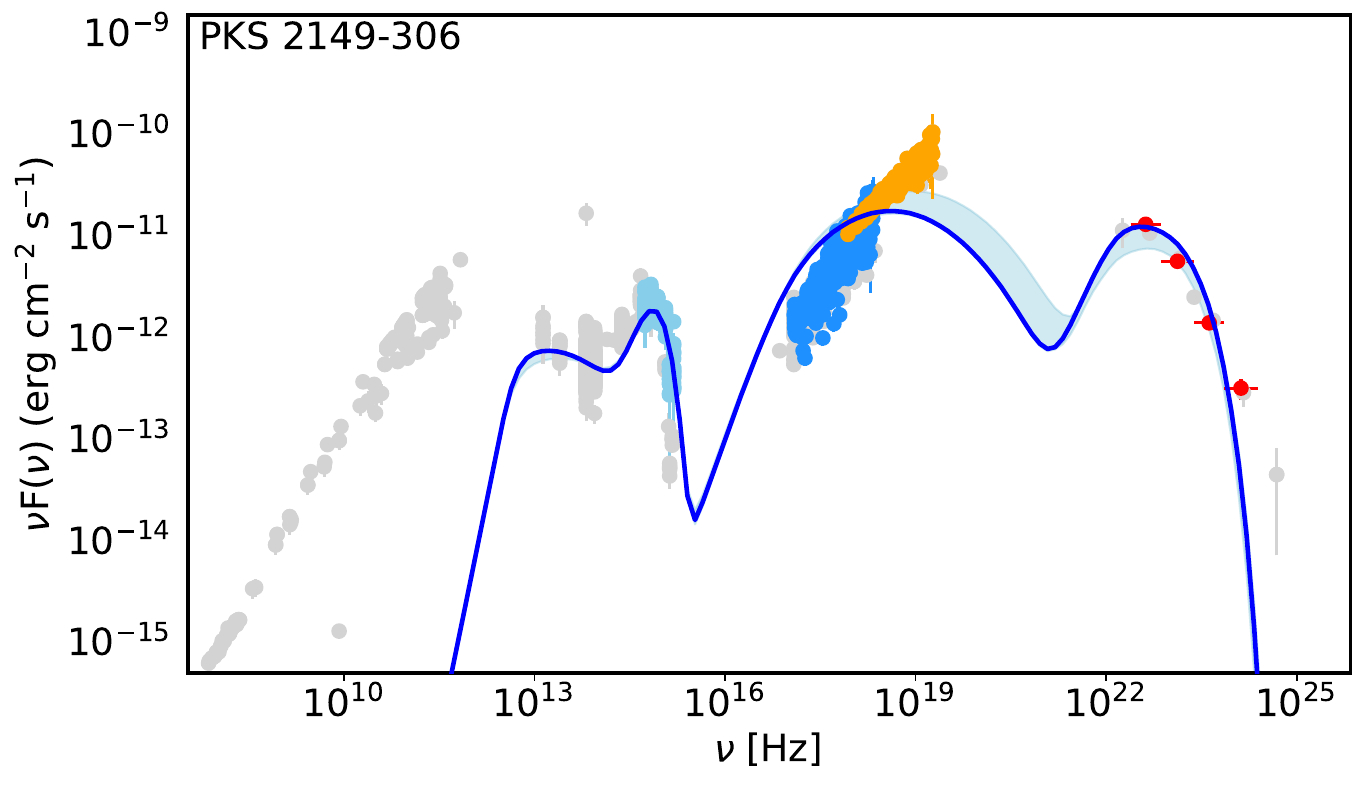}
     \includegraphics[width=0.45\textwidth]{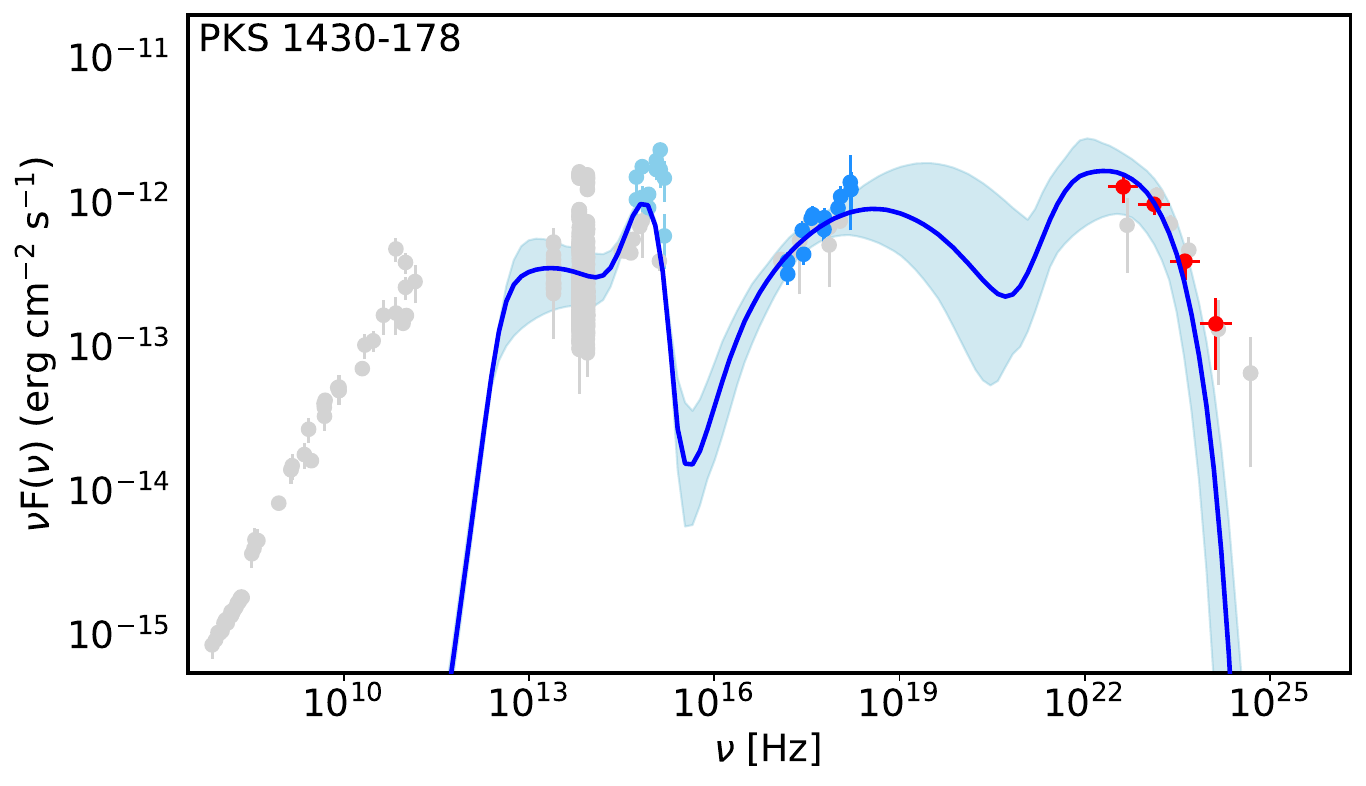}\\
     \includegraphics[width=0.45\textwidth]{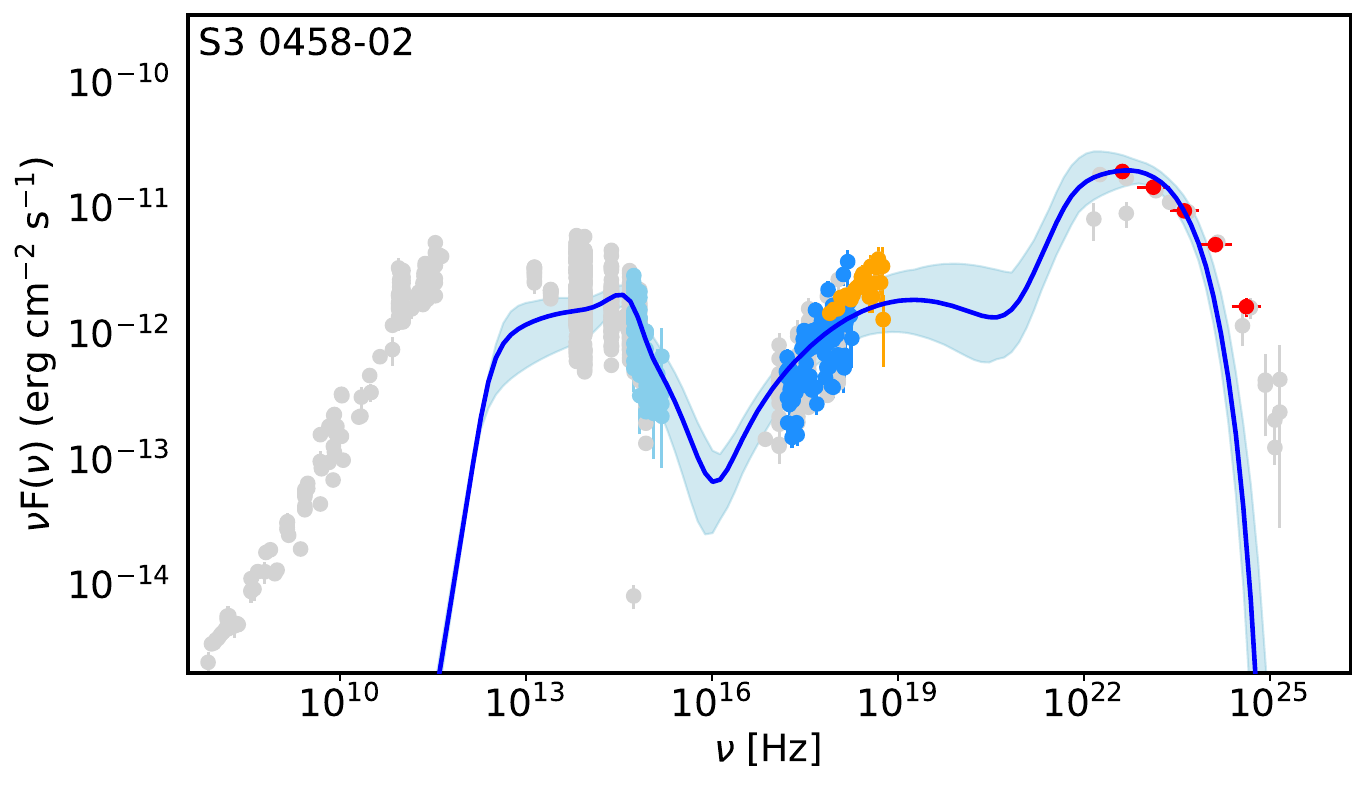}
     \includegraphics[width=0.45\textwidth]{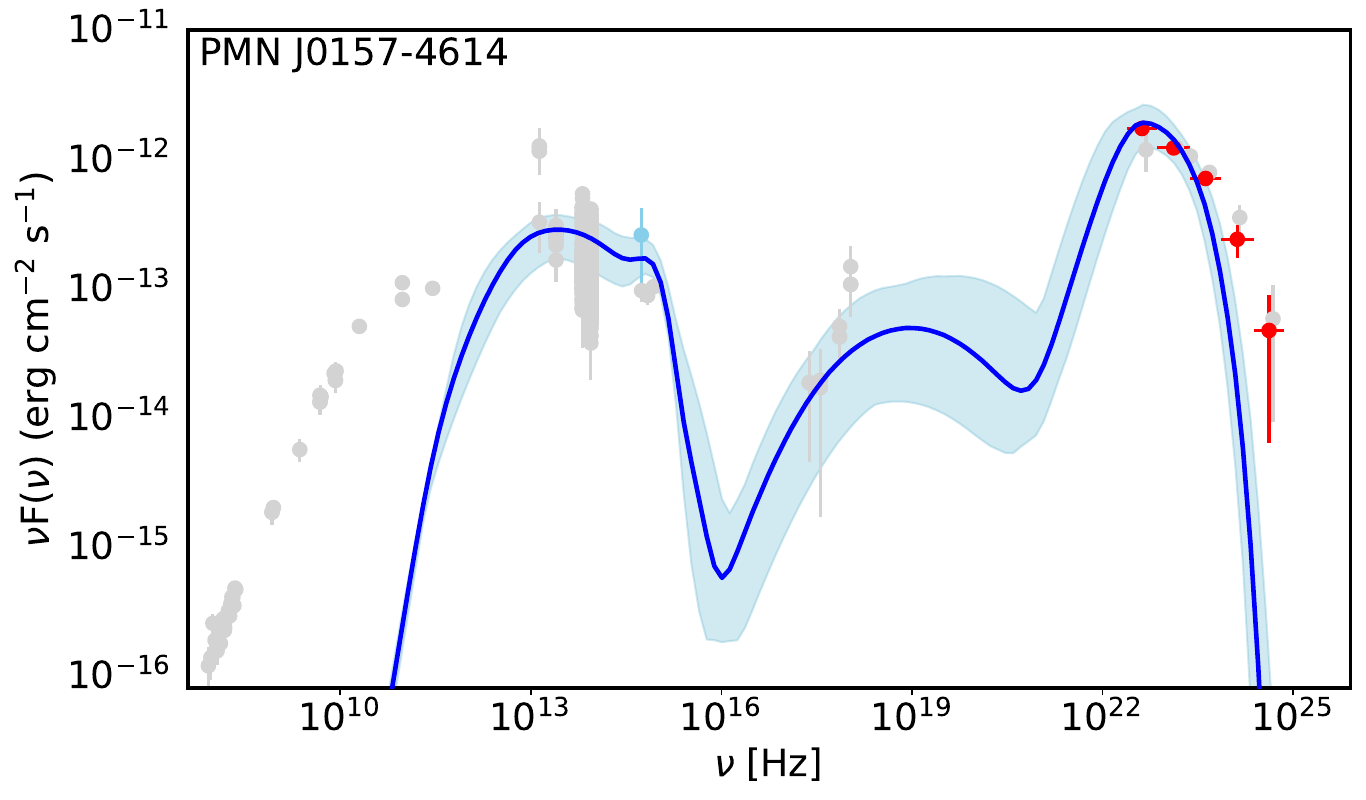}\\
     \includegraphics[width=0.45\textwidth]{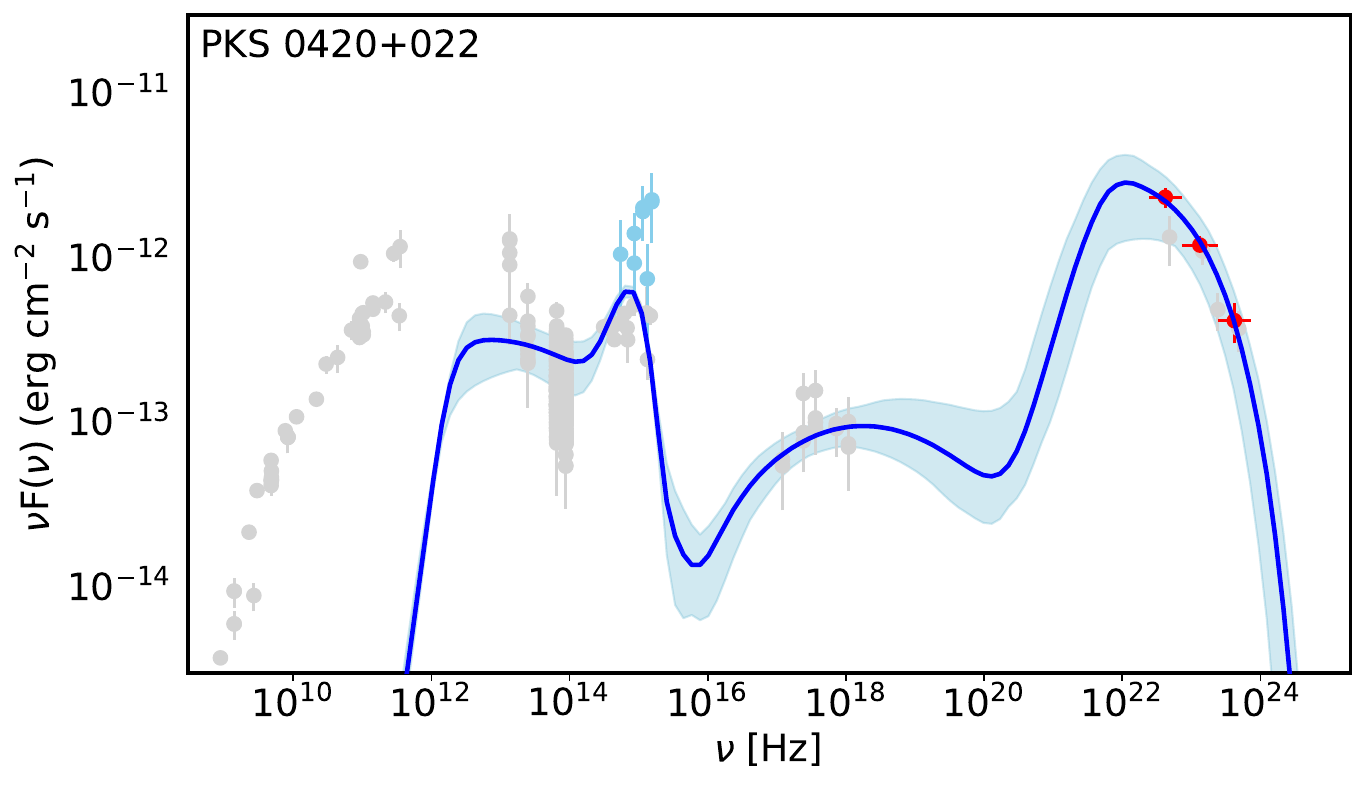}
     \includegraphics[width=0.45\textwidth]{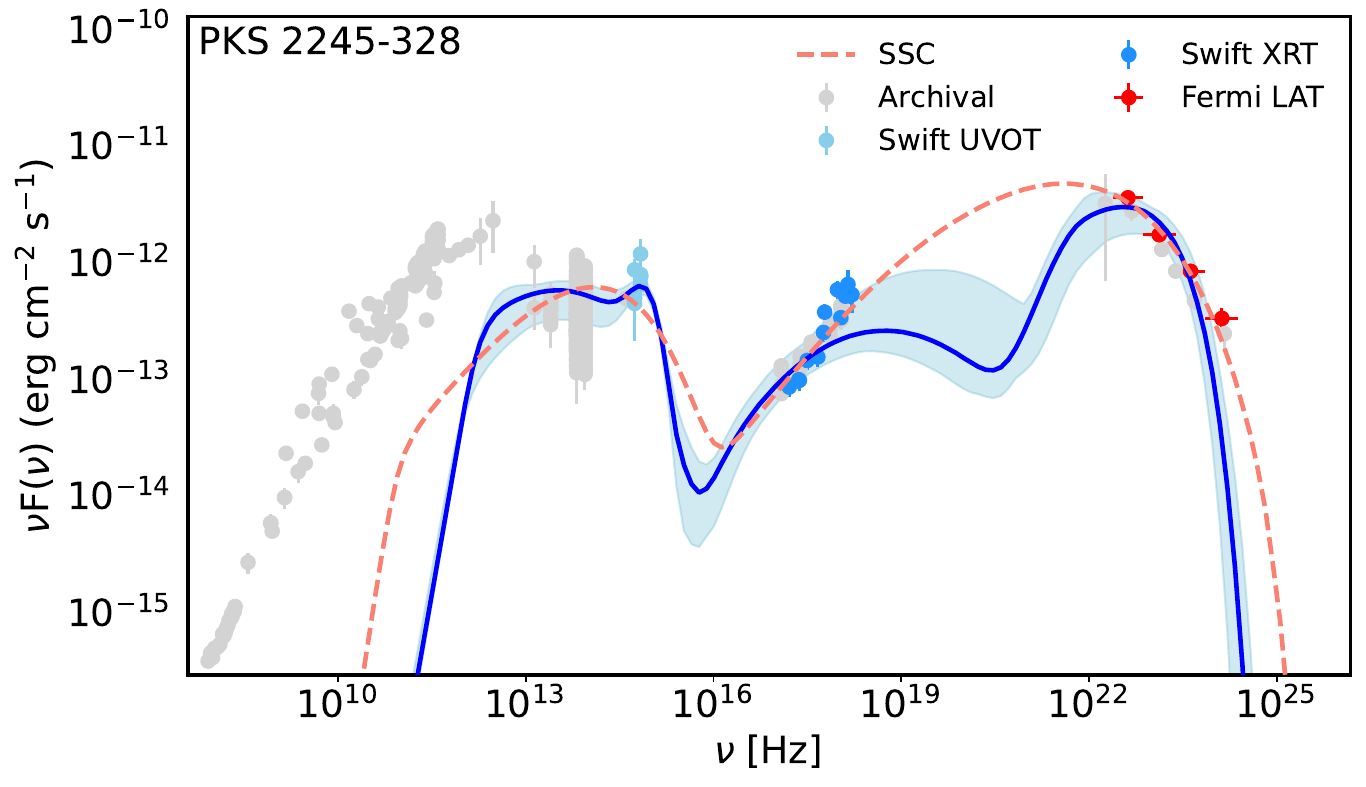}\\
     \includegraphics[width=0.45\textwidth]{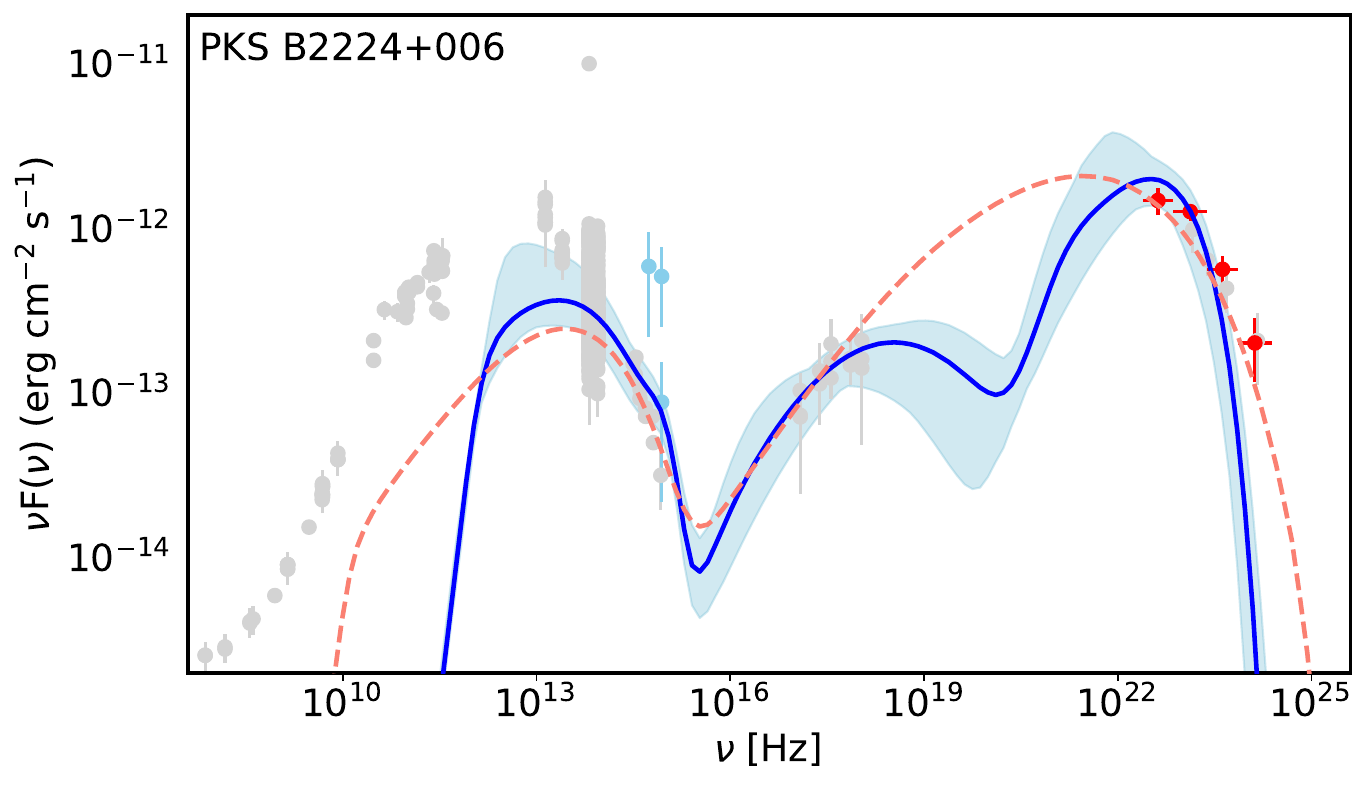}
     \includegraphics[width=0.45\textwidth]{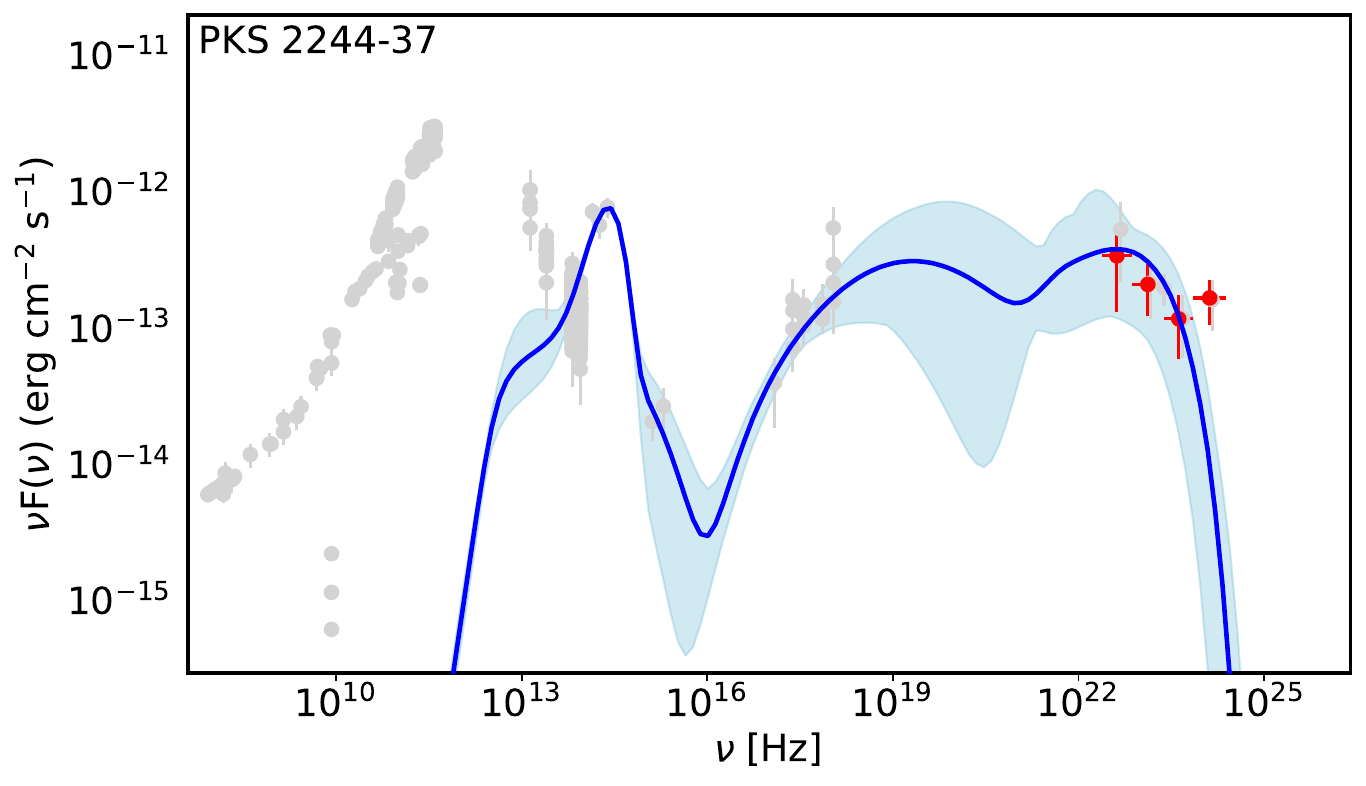}\\
     \caption{(continued)}
\end{figure*}
\begin{figure*}
     \centering
        \ContinuedFloat
     \includegraphics[width=0.45\textwidth]{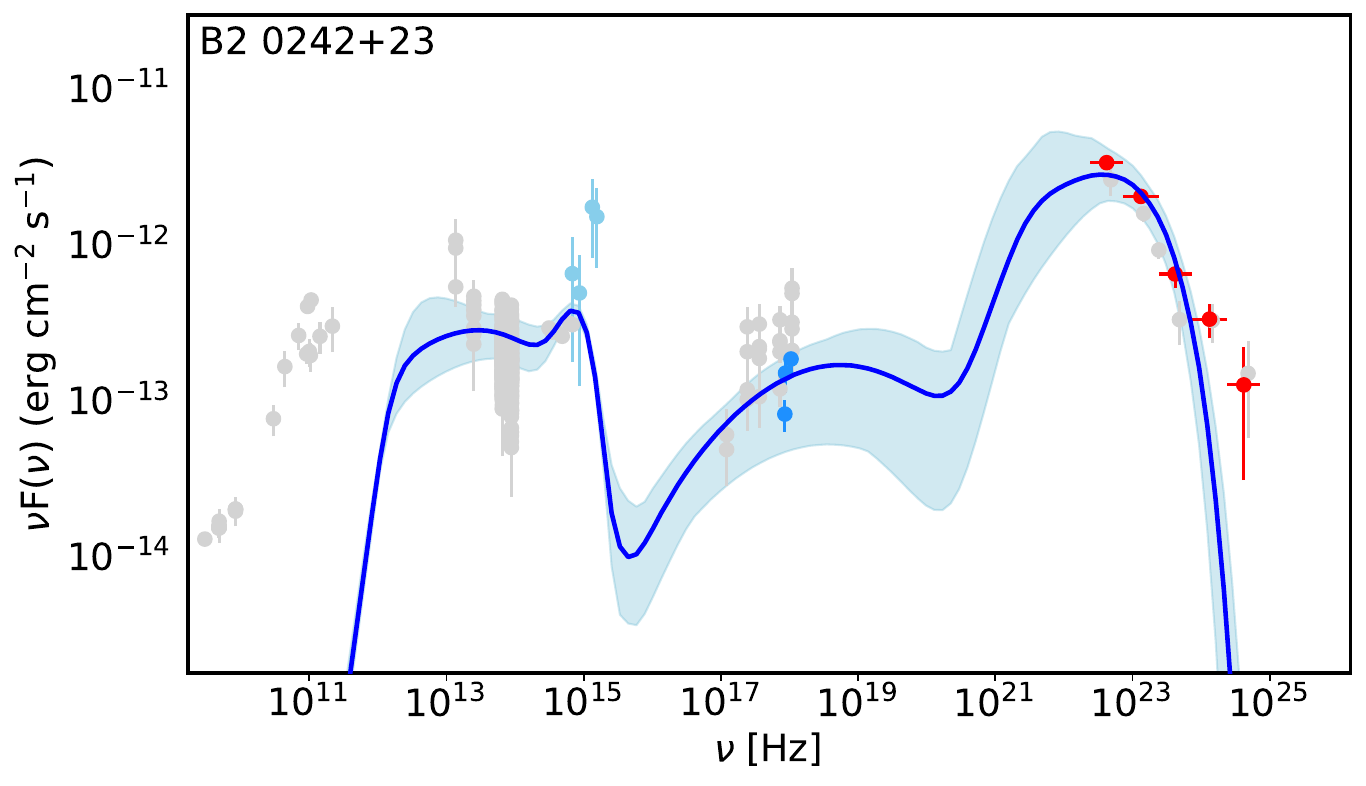}
     \includegraphics[width=0.45\textwidth]{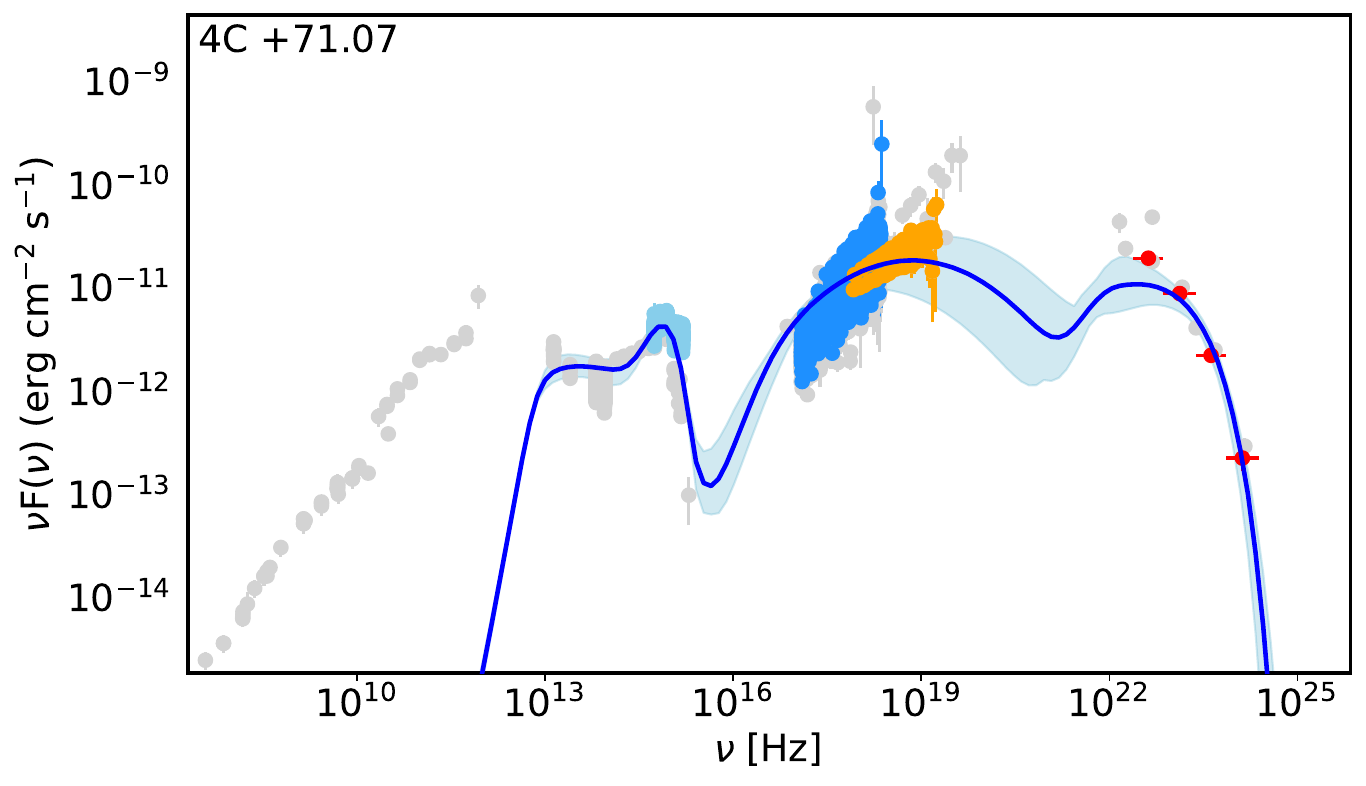}\\
     \includegraphics[width=0.45\textwidth]{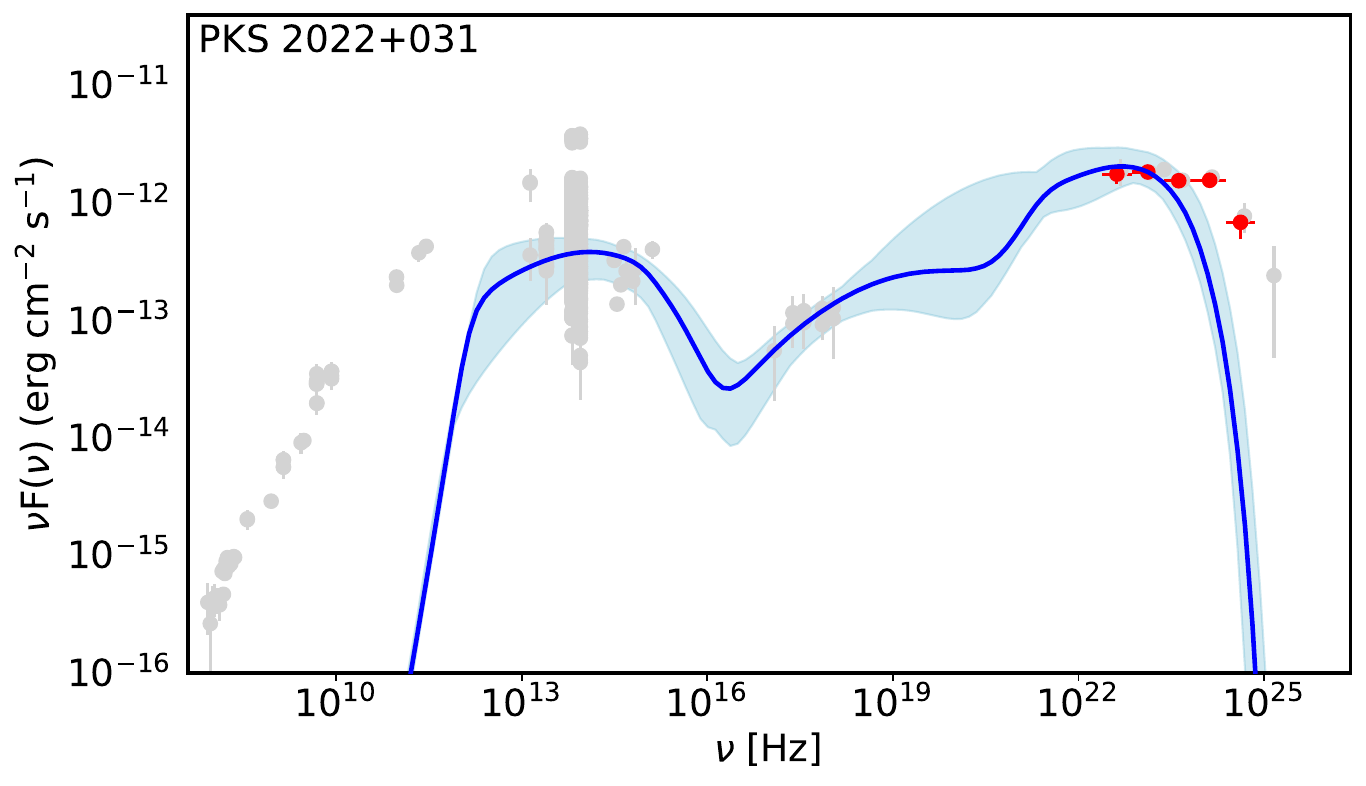}
     \includegraphics[width=0.45\textwidth]{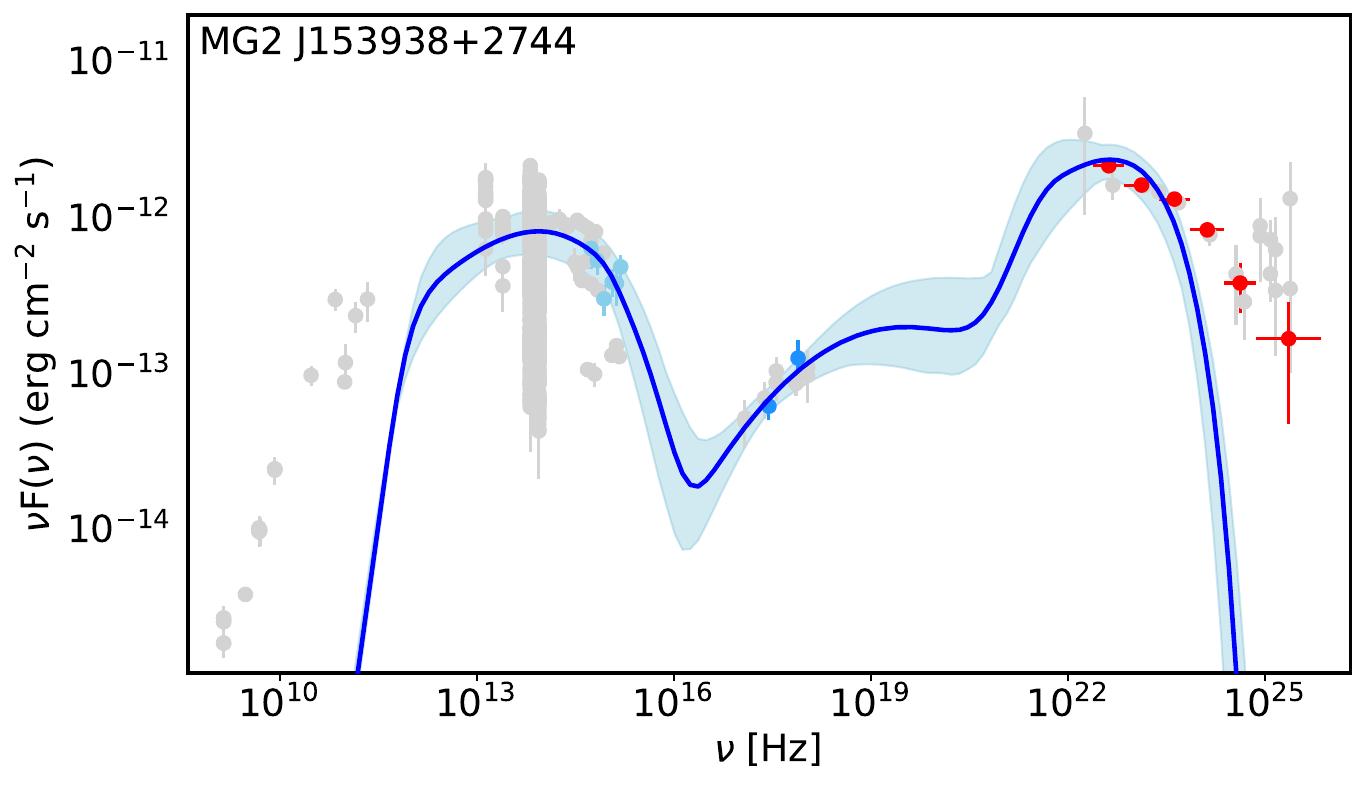}\\
     \includegraphics[width=0.45\textwidth]{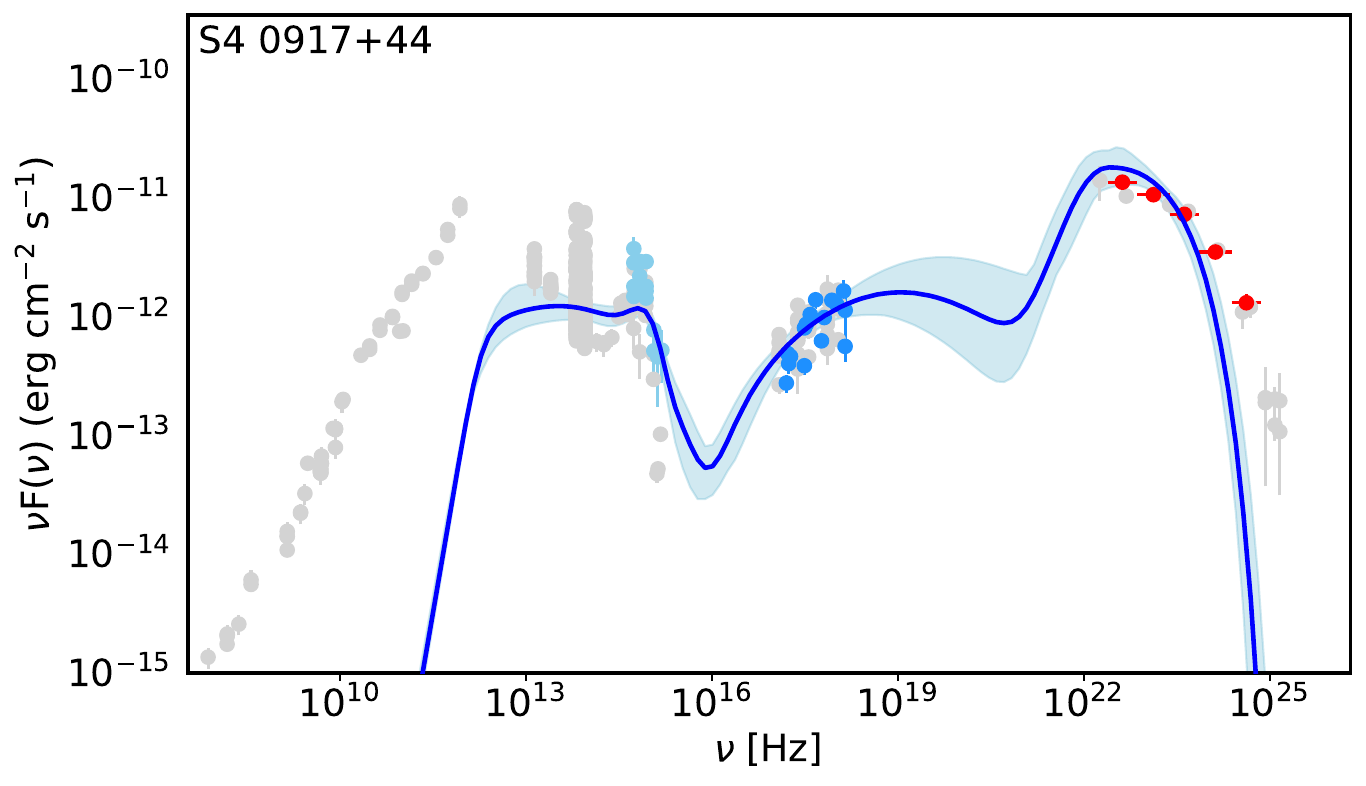}
     \includegraphics[width=0.45\textwidth]{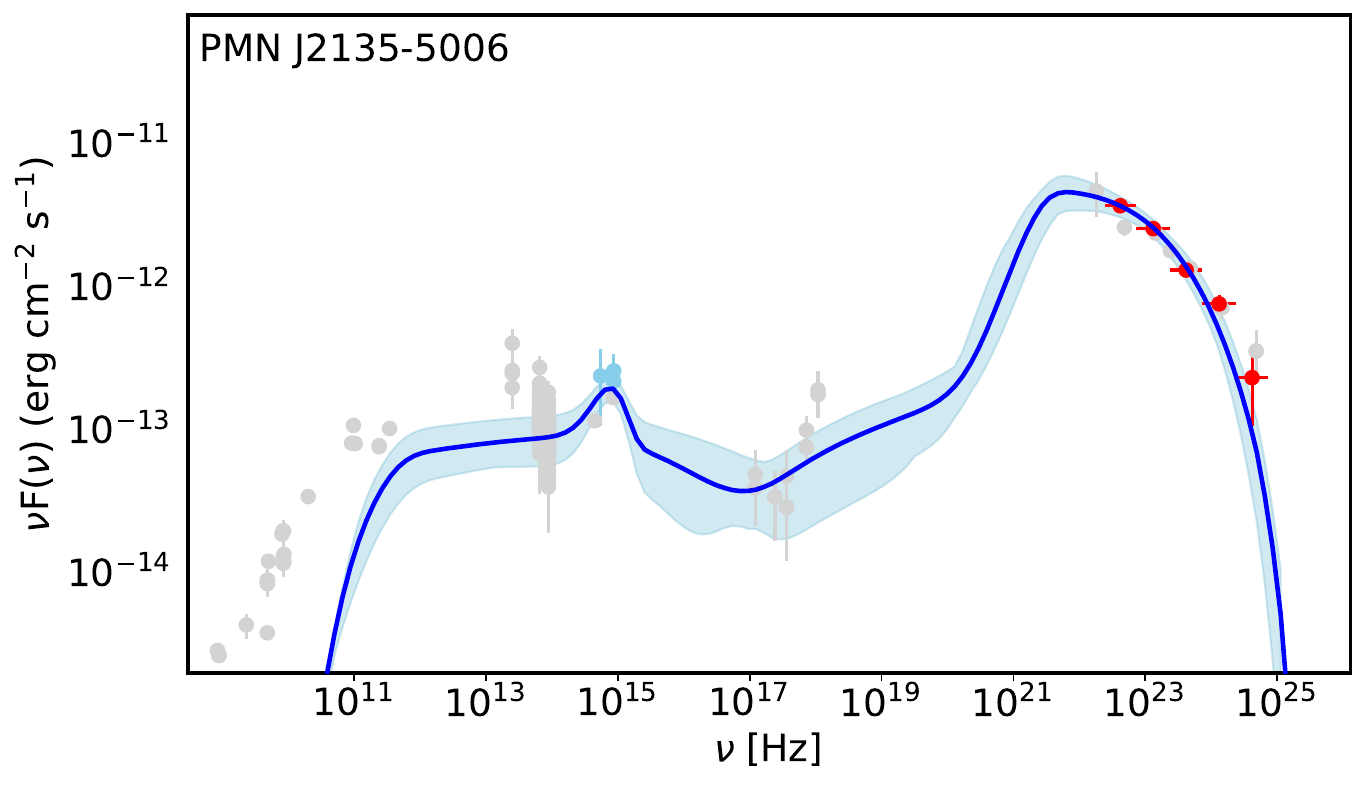}\\
     \includegraphics[width=0.45\textwidth]{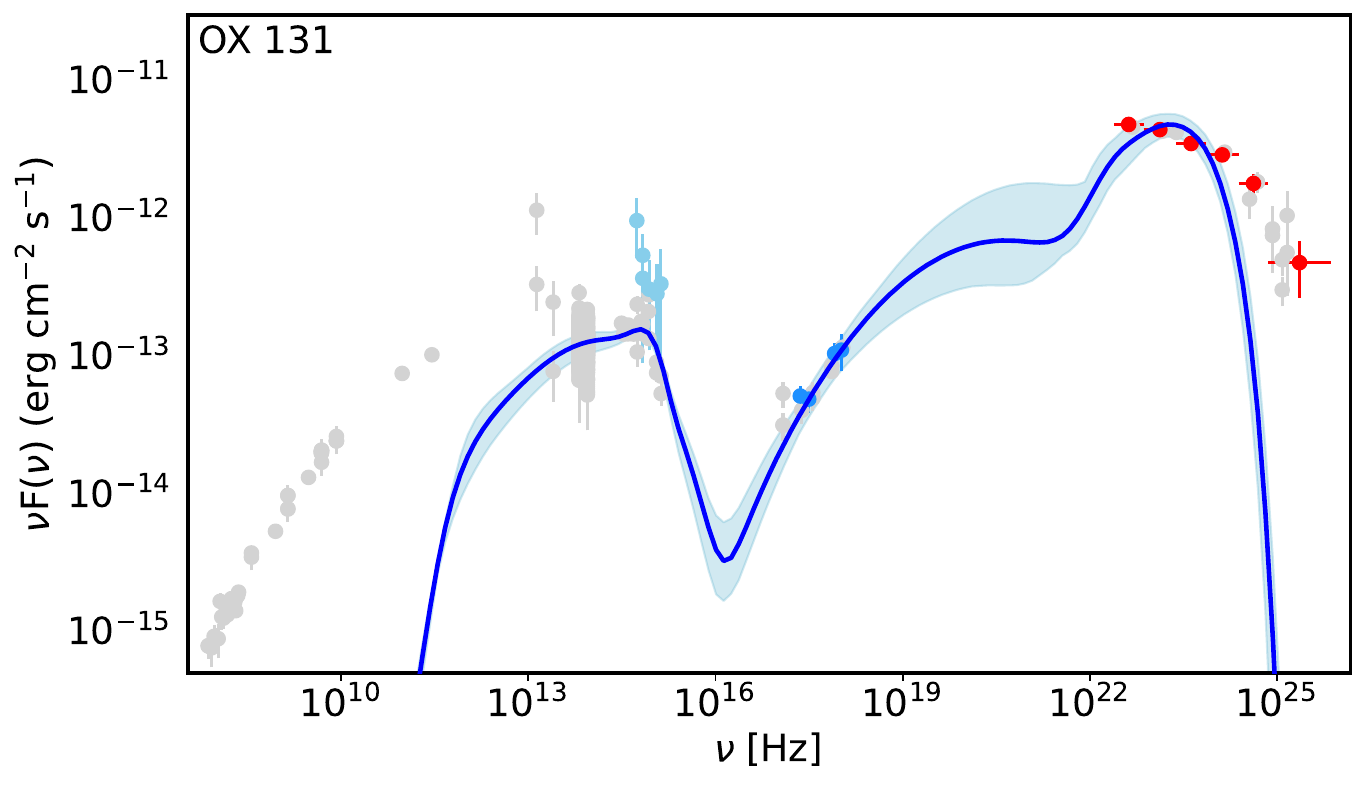}
     \includegraphics[width=0.45\textwidth]{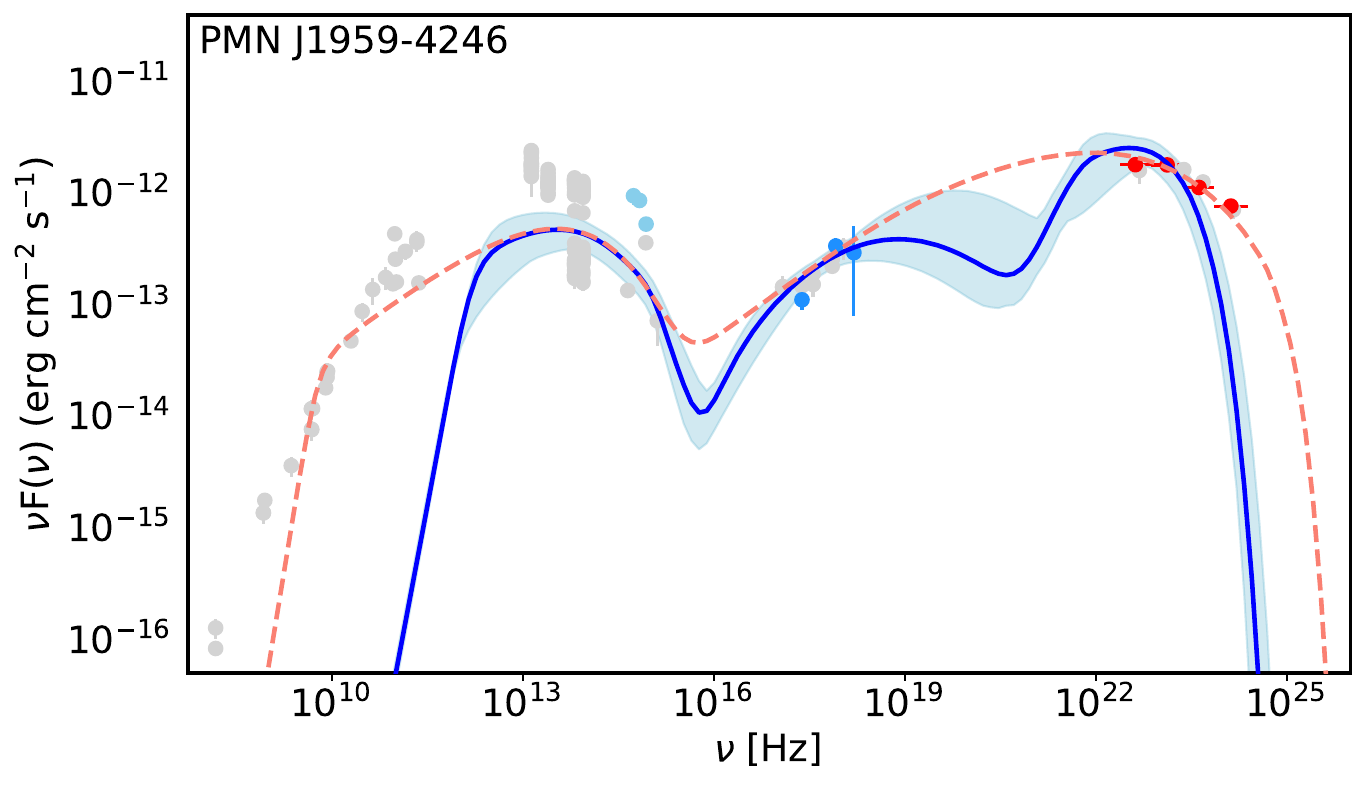}\\
     \includegraphics[width=0.45\textwidth]{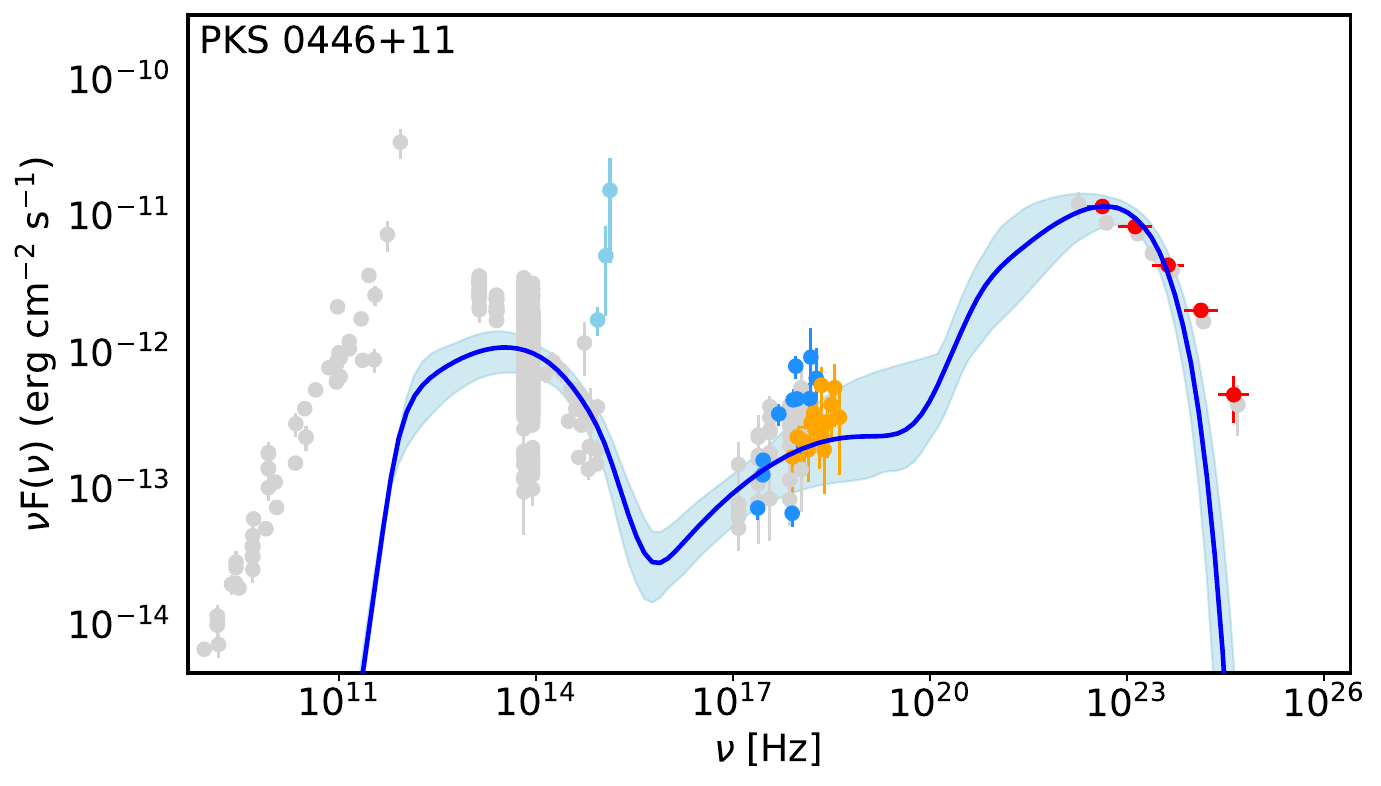}
     \includegraphics[width=0.45\textwidth]{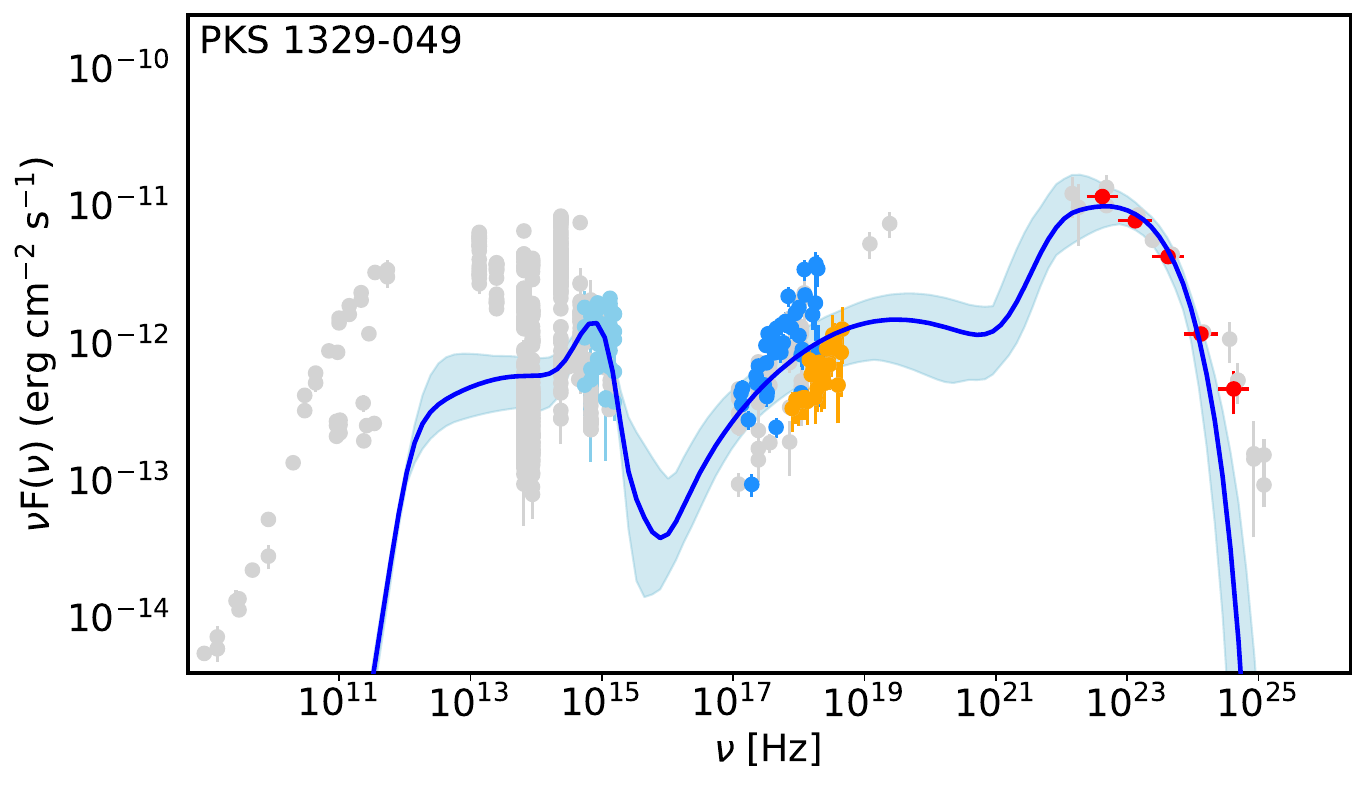}
     \caption{(continued)}
\end{figure*}
\begin{figure*}
     \centering
        \ContinuedFloat
     \includegraphics[width=0.45\textwidth]{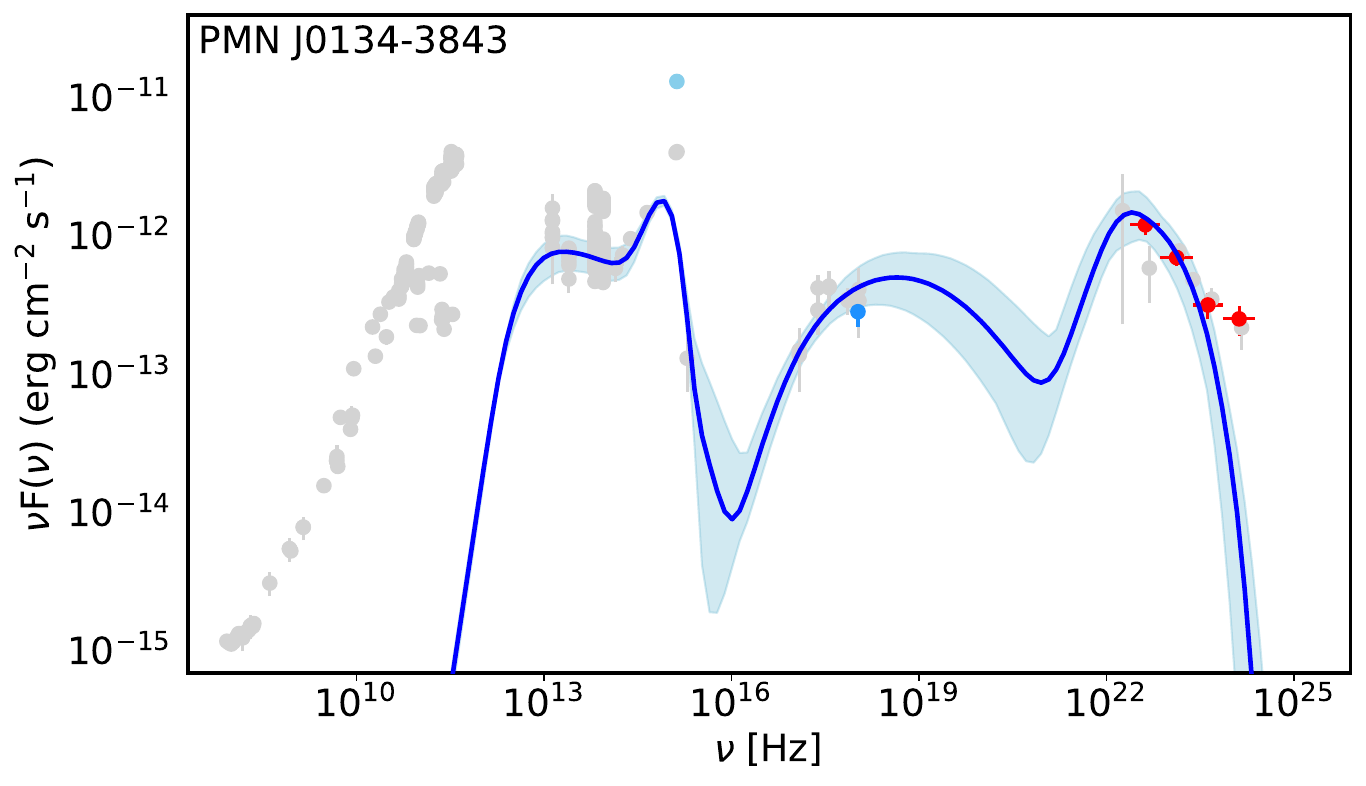}
     \includegraphics[width=0.45\textwidth]{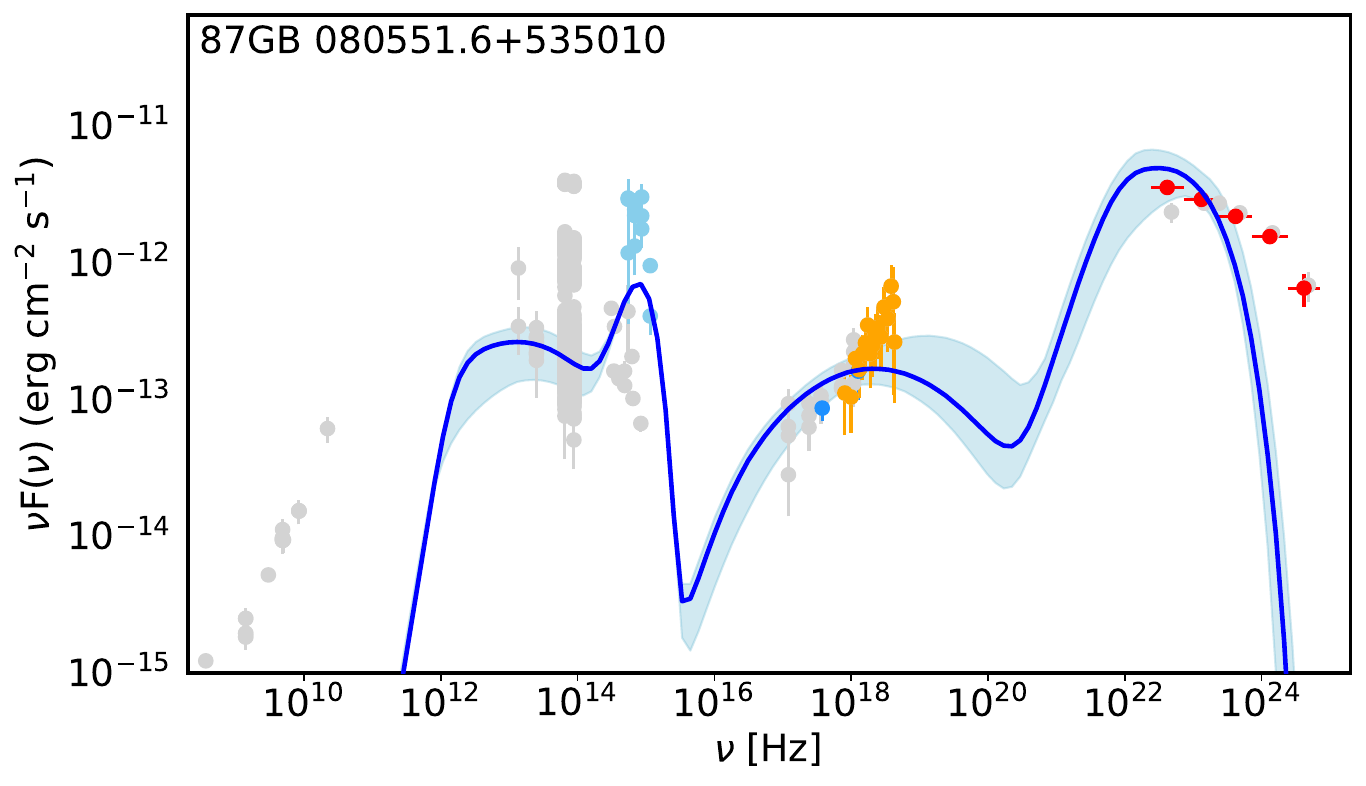}\\
     \includegraphics[width=0.45\textwidth]{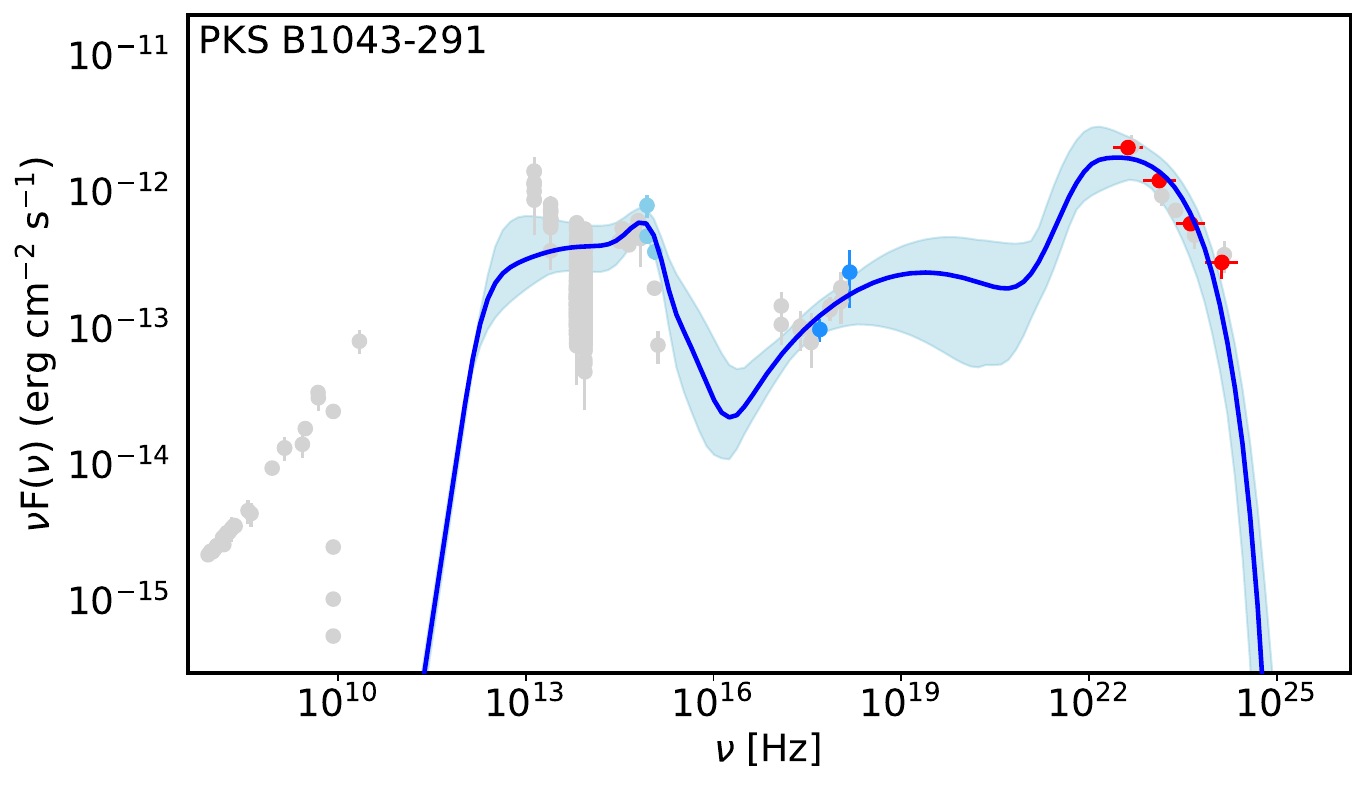}
     \includegraphics[width=0.45\textwidth]{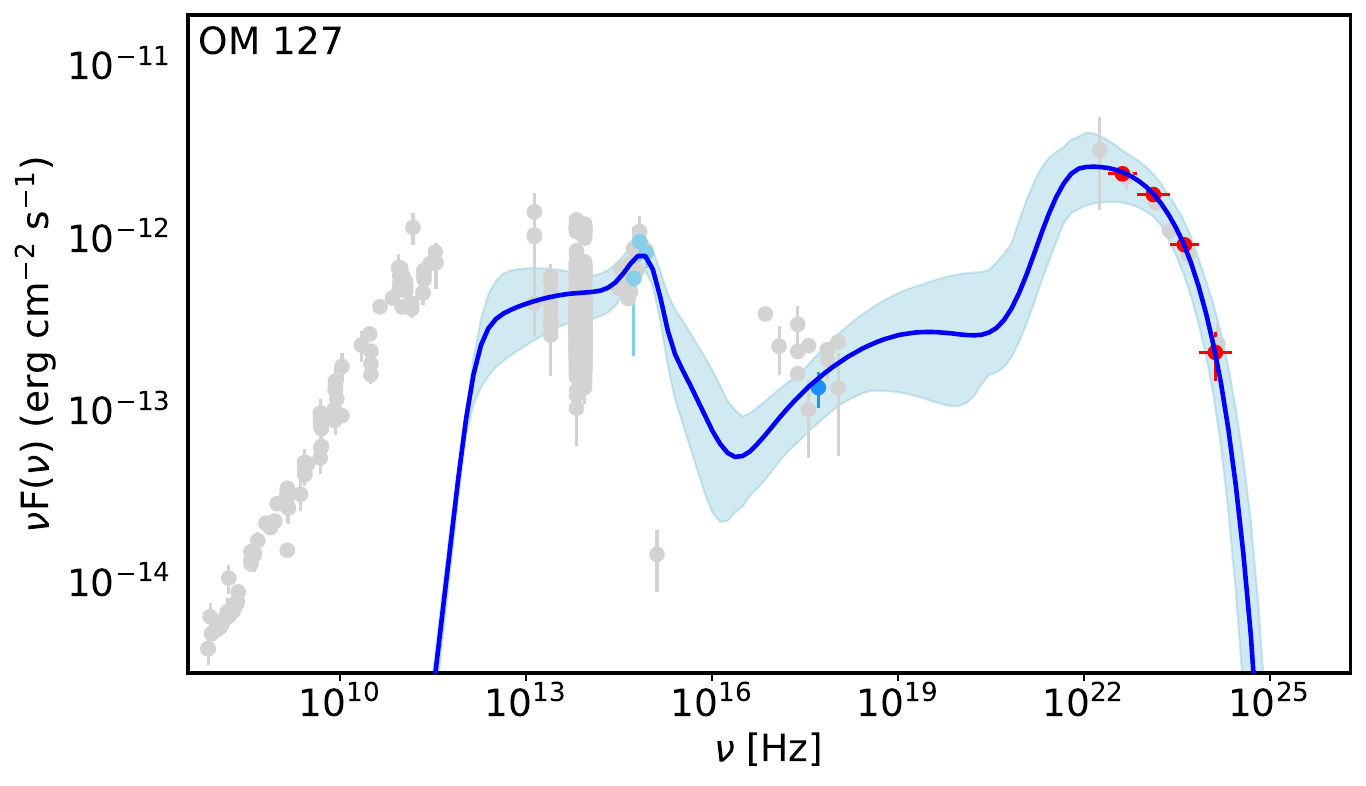}\\
     \includegraphics[width=0.45\textwidth]{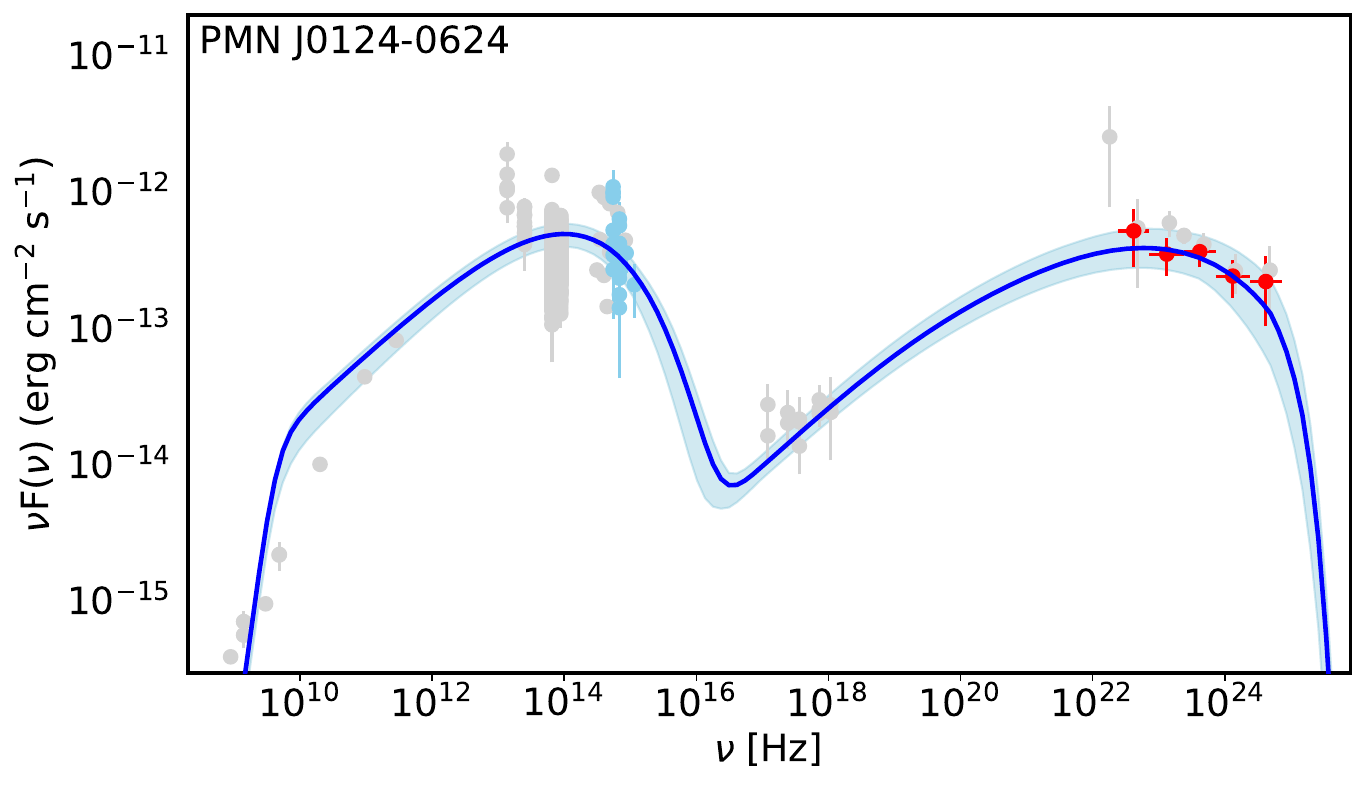}
     \includegraphics[width=0.45\textwidth]{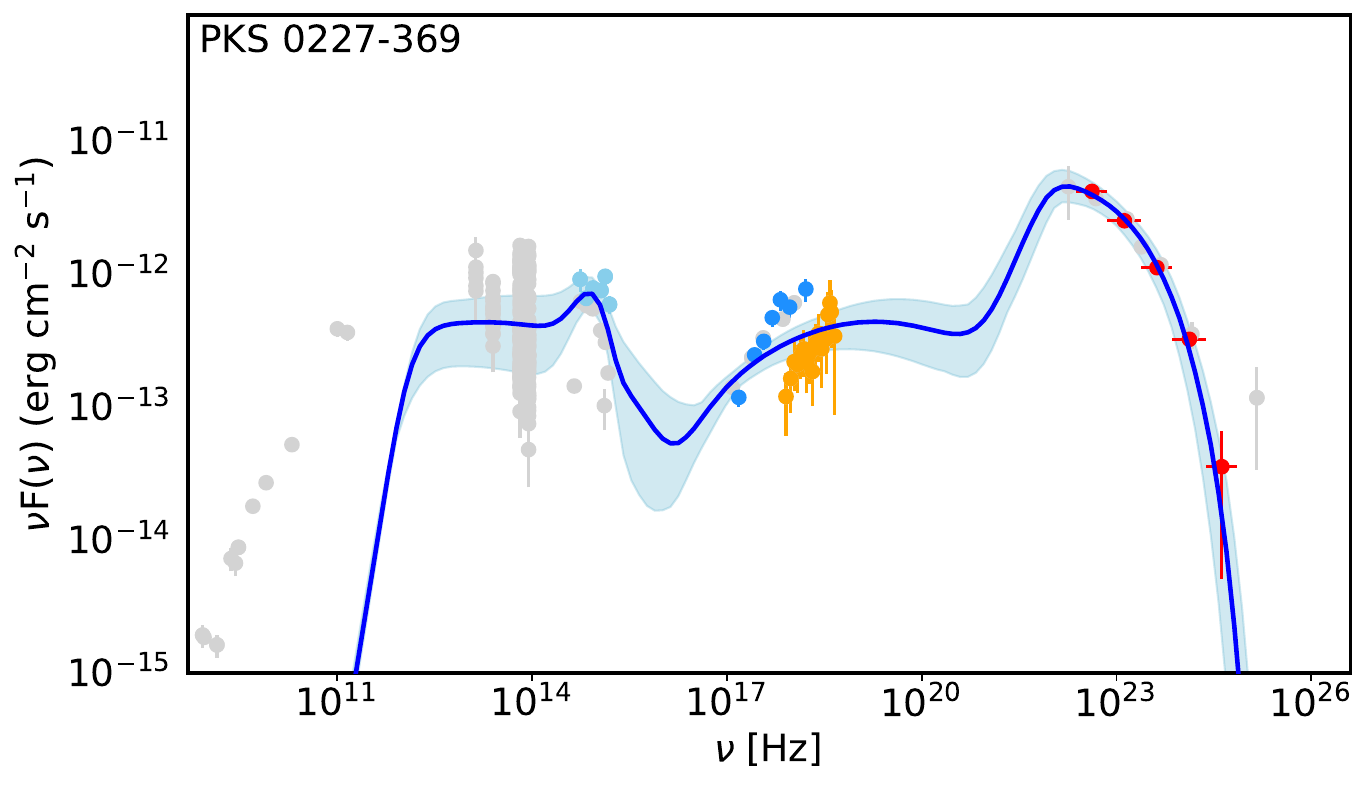}\\
     \includegraphics[width=0.45\textwidth]{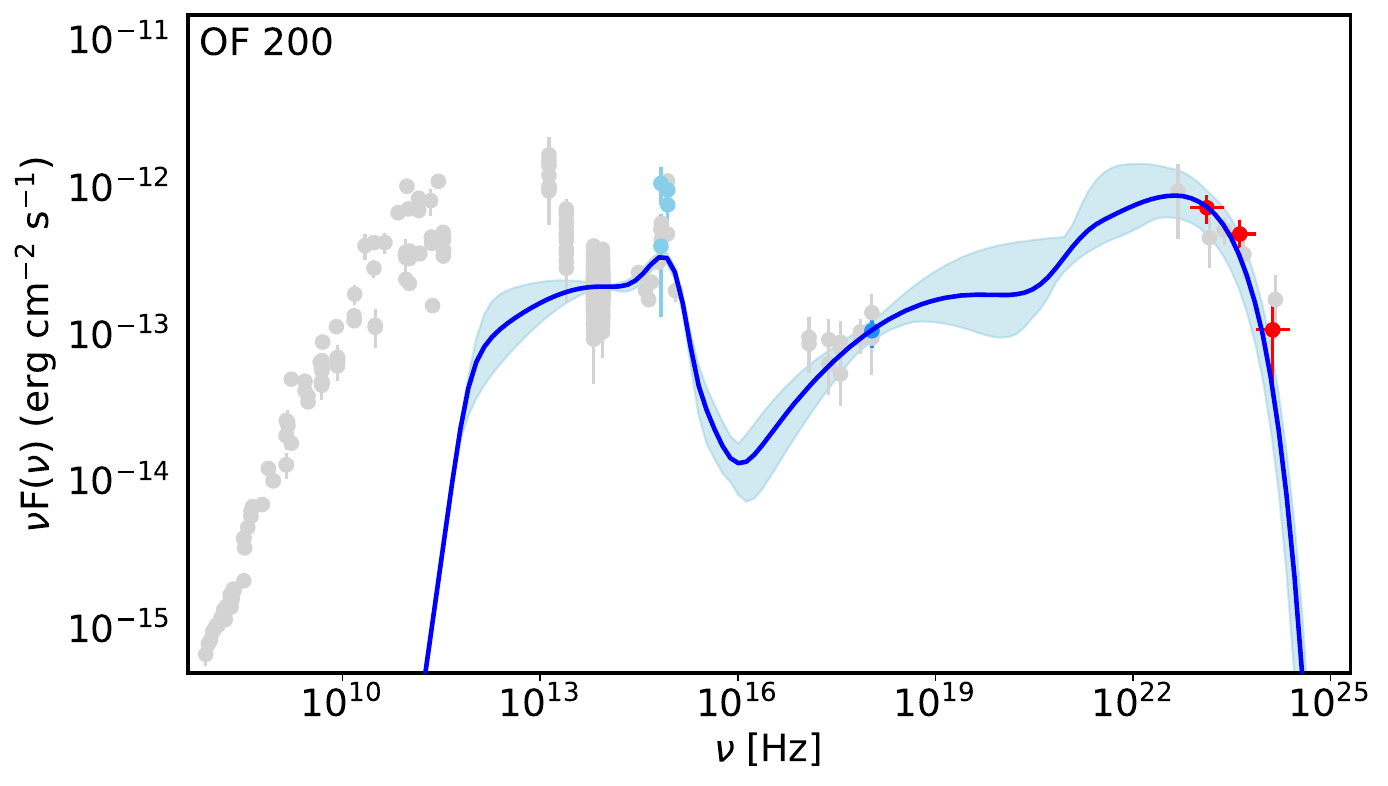}
     \includegraphics[width=0.45\textwidth]{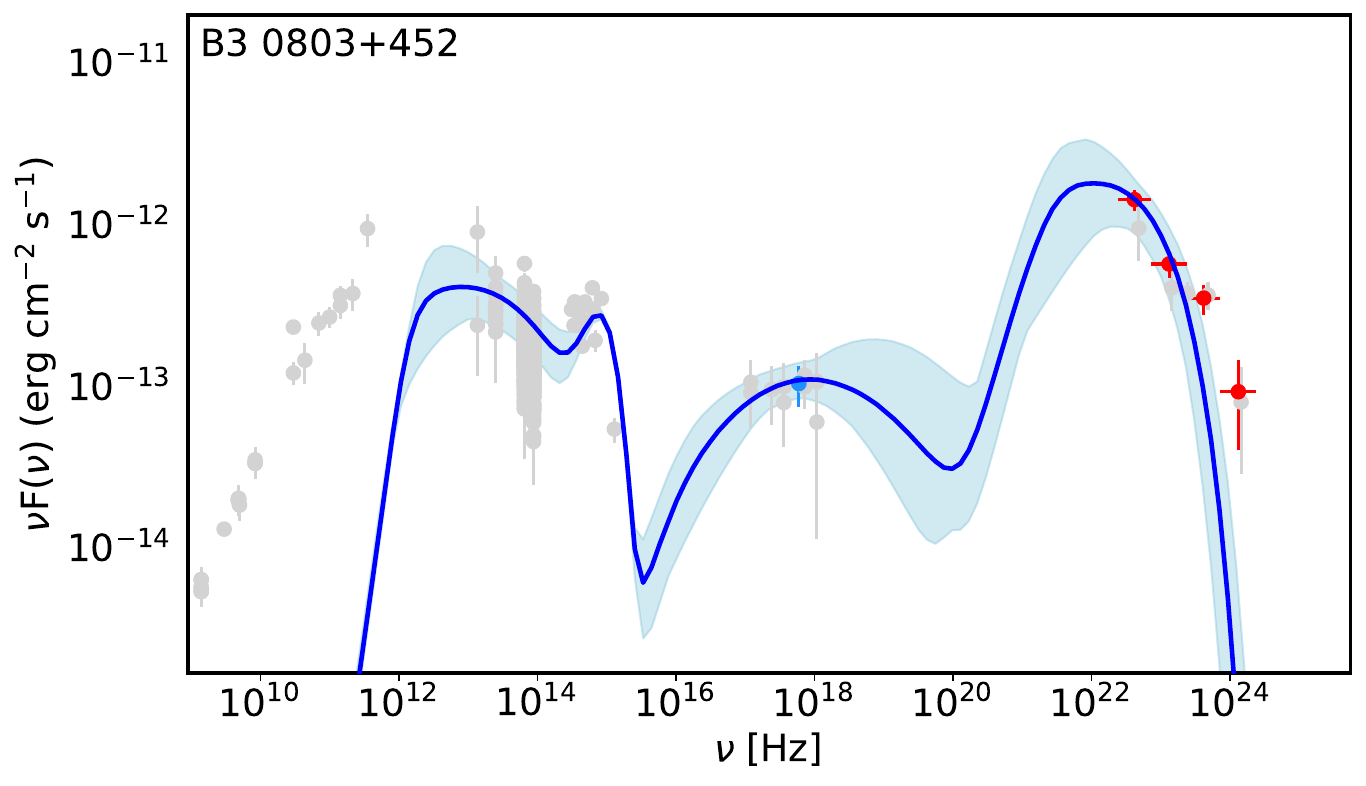}\\
     \includegraphics[width=0.45\textwidth]{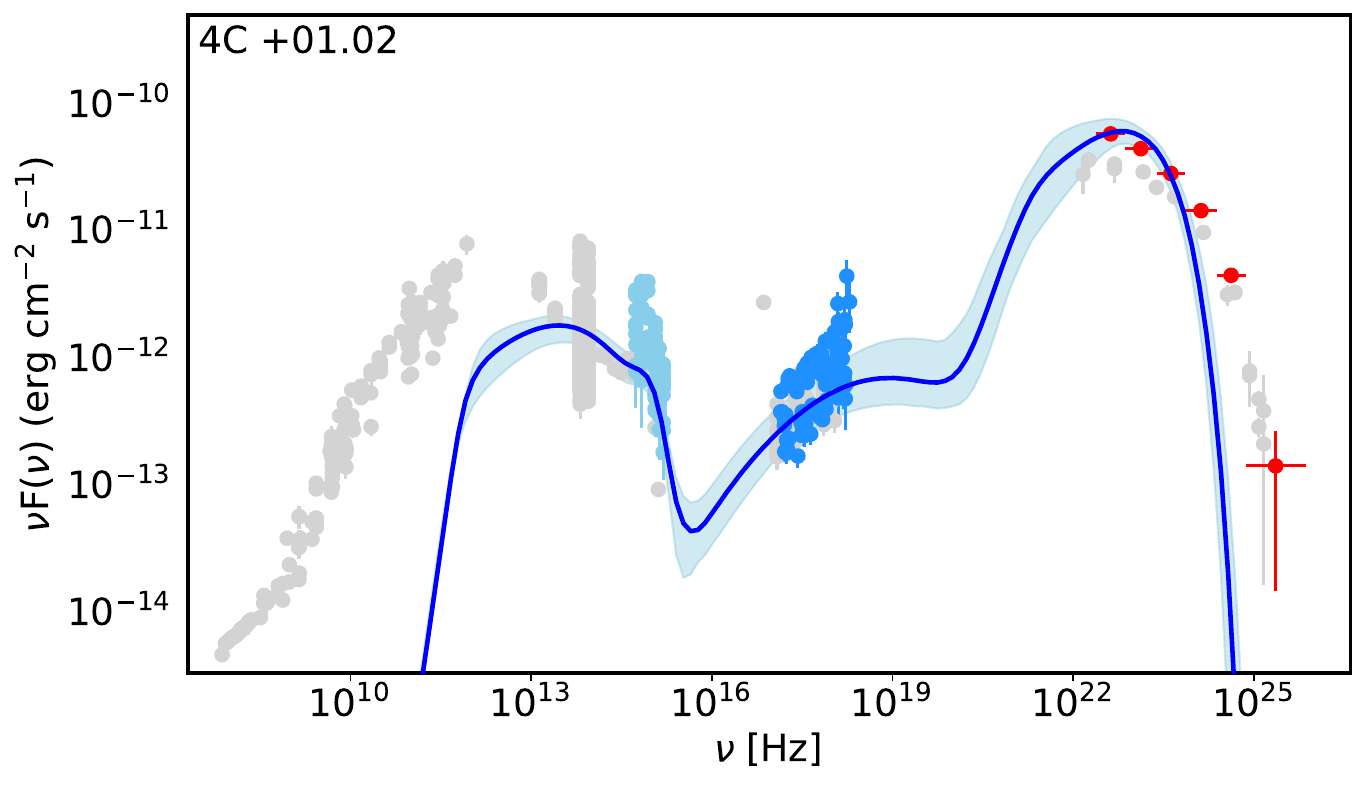}
     \includegraphics[width=0.45\textwidth]{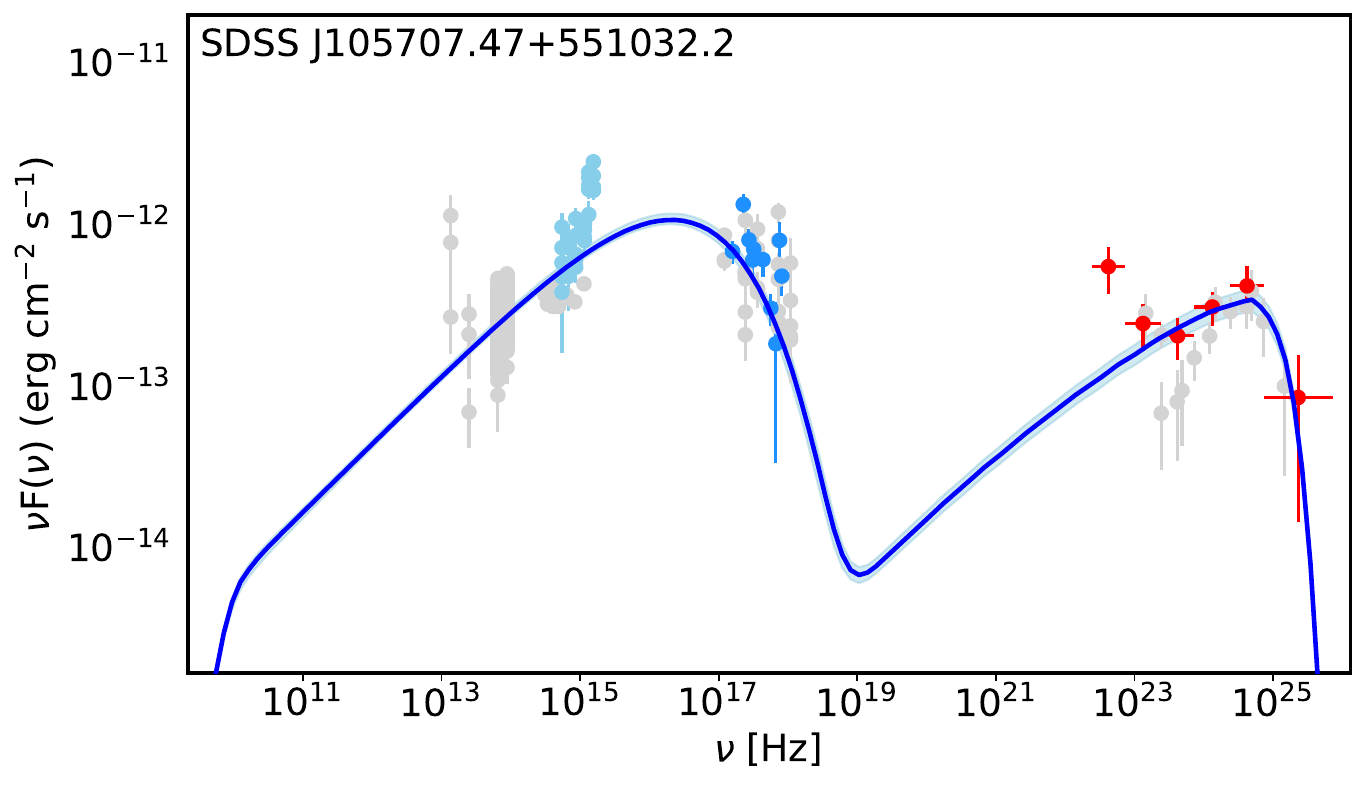}
     \caption{(continued)}
\end{figure*}
\begin{figure*}
     \centering
        \ContinuedFloat
     \includegraphics[width=0.45\textwidth]{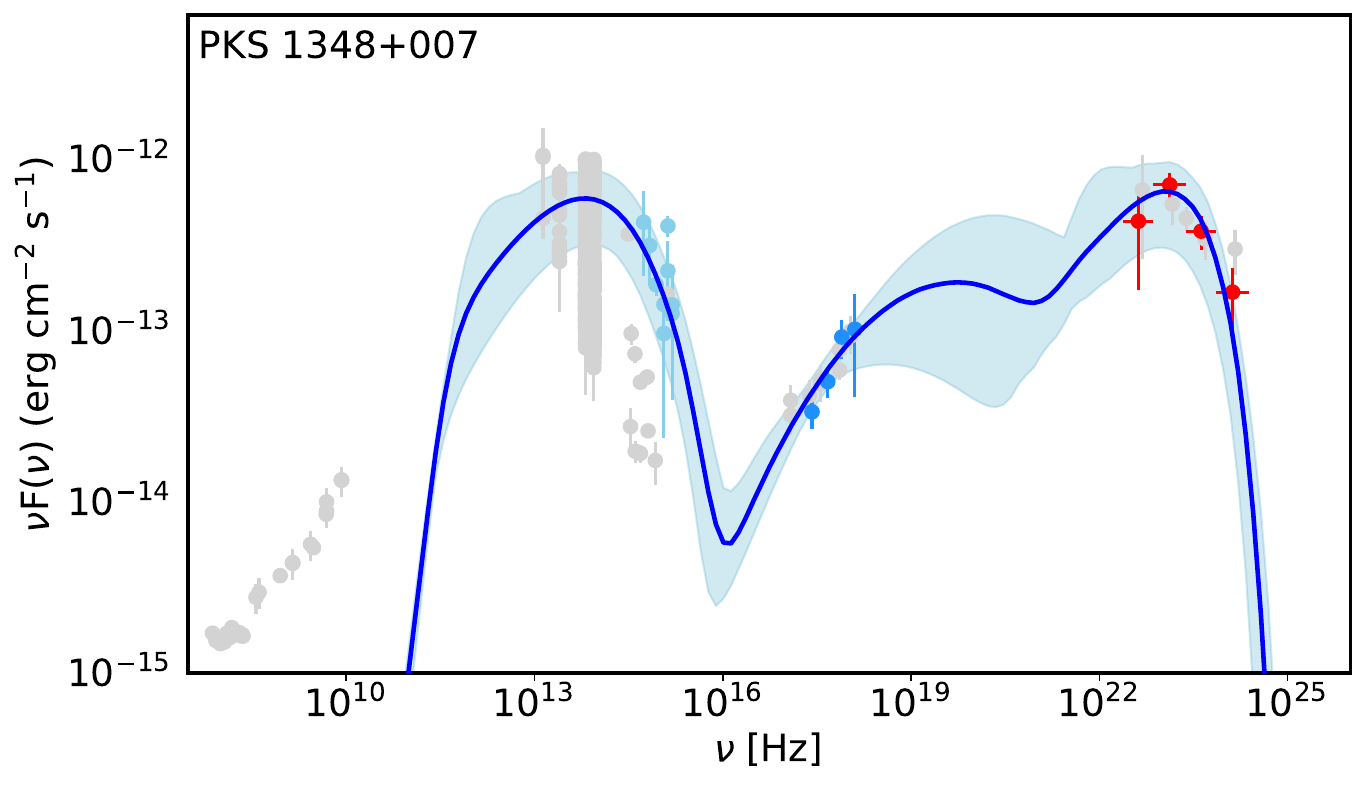}
     \includegraphics[width=0.45\textwidth]{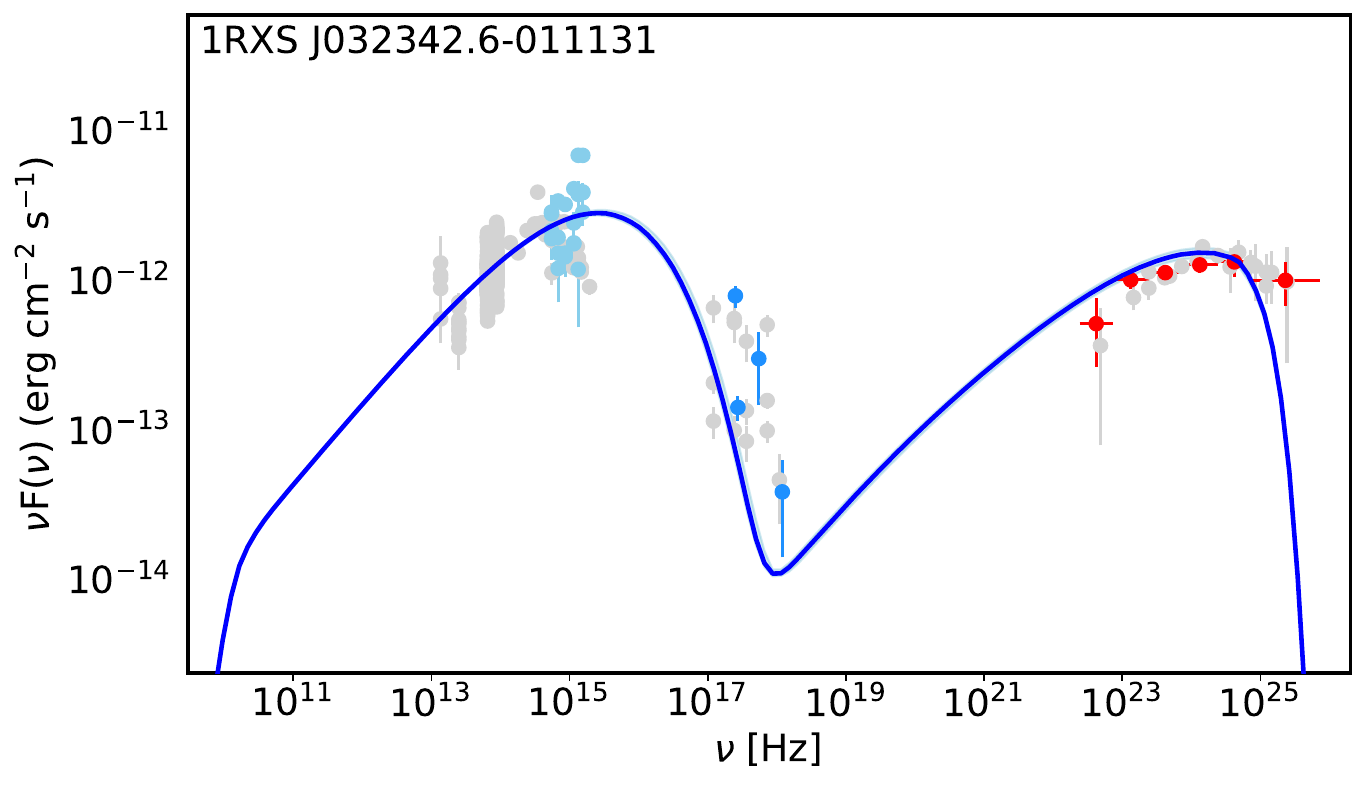}\\
     \includegraphics[width=0.45\textwidth]{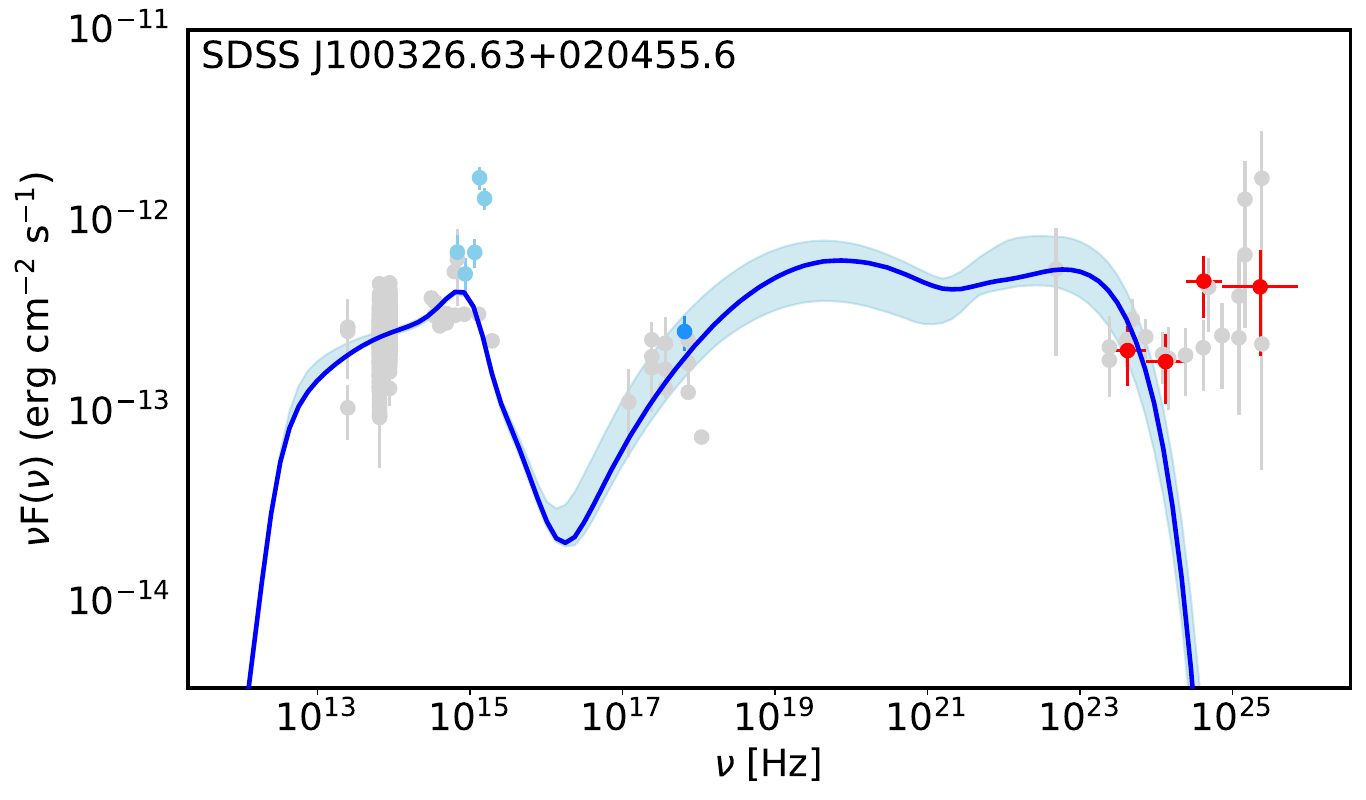}
     \includegraphics[width=0.45\textwidth]{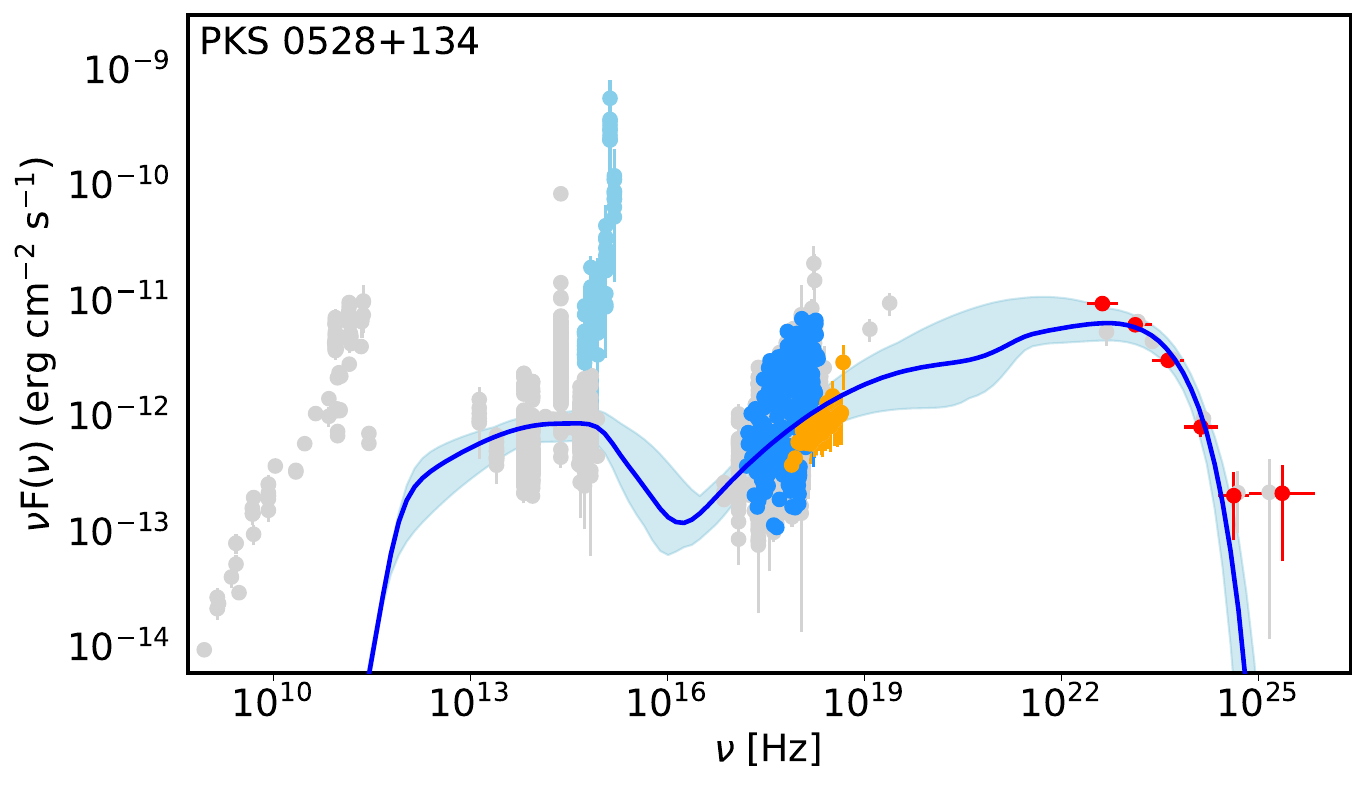}\\
     \includegraphics[width=0.45\textwidth]{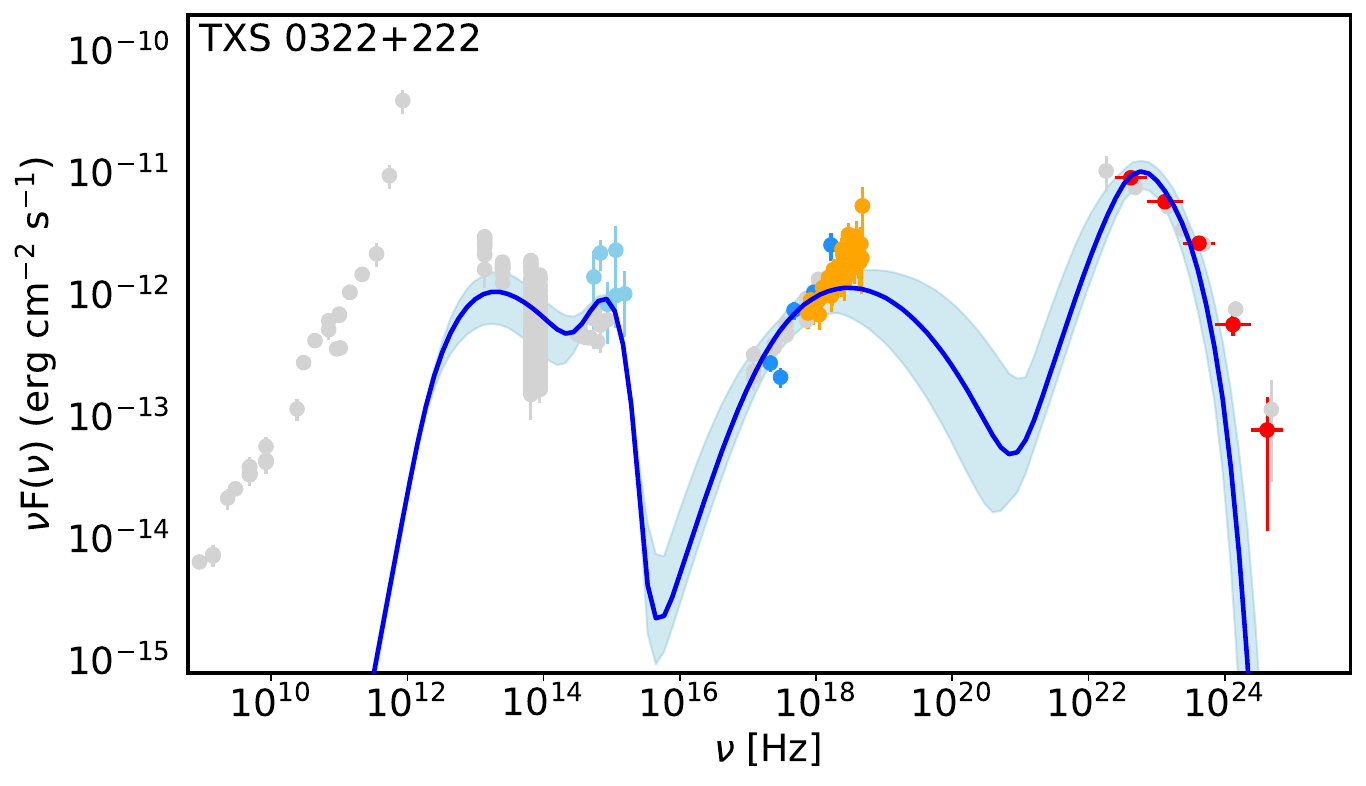}
     \includegraphics[width=0.45\textwidth]{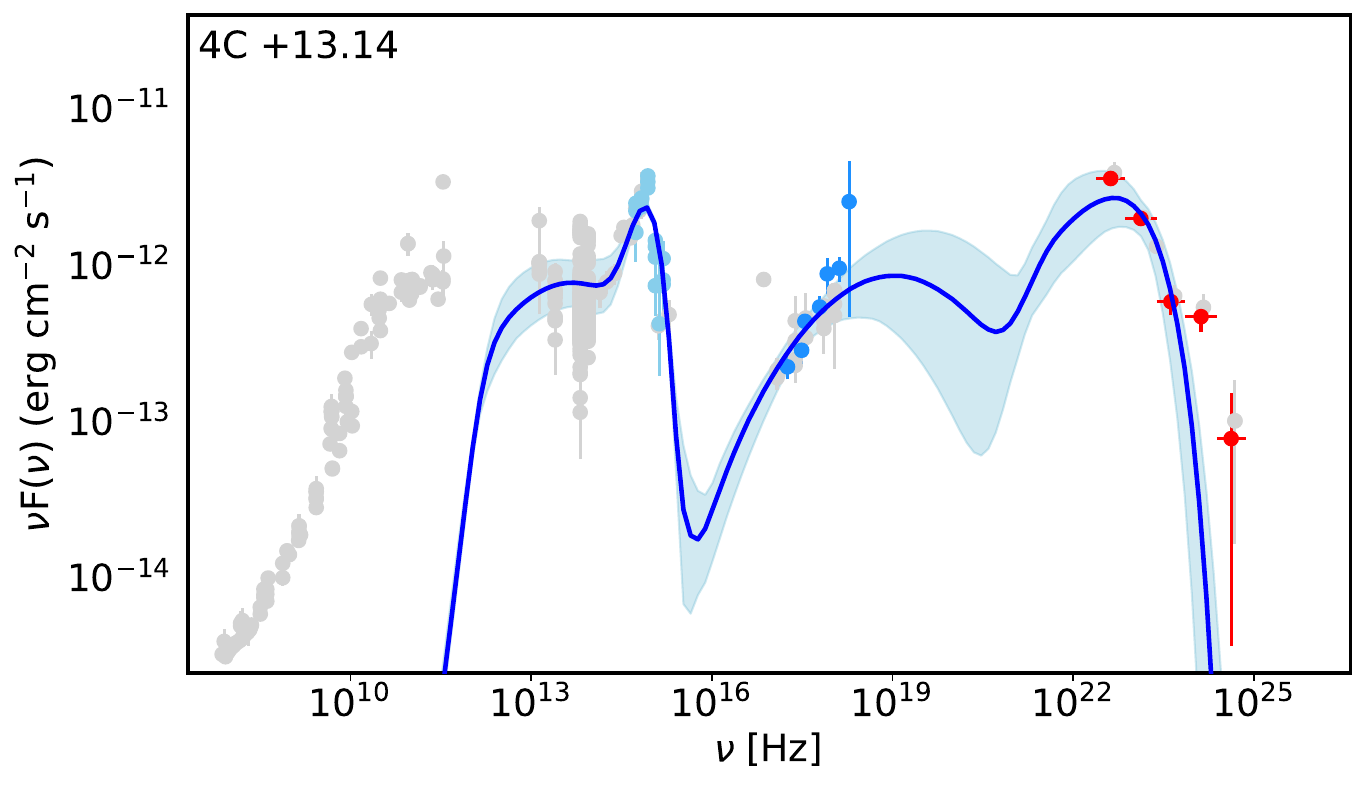}\\
     \includegraphics[width=0.45\textwidth]{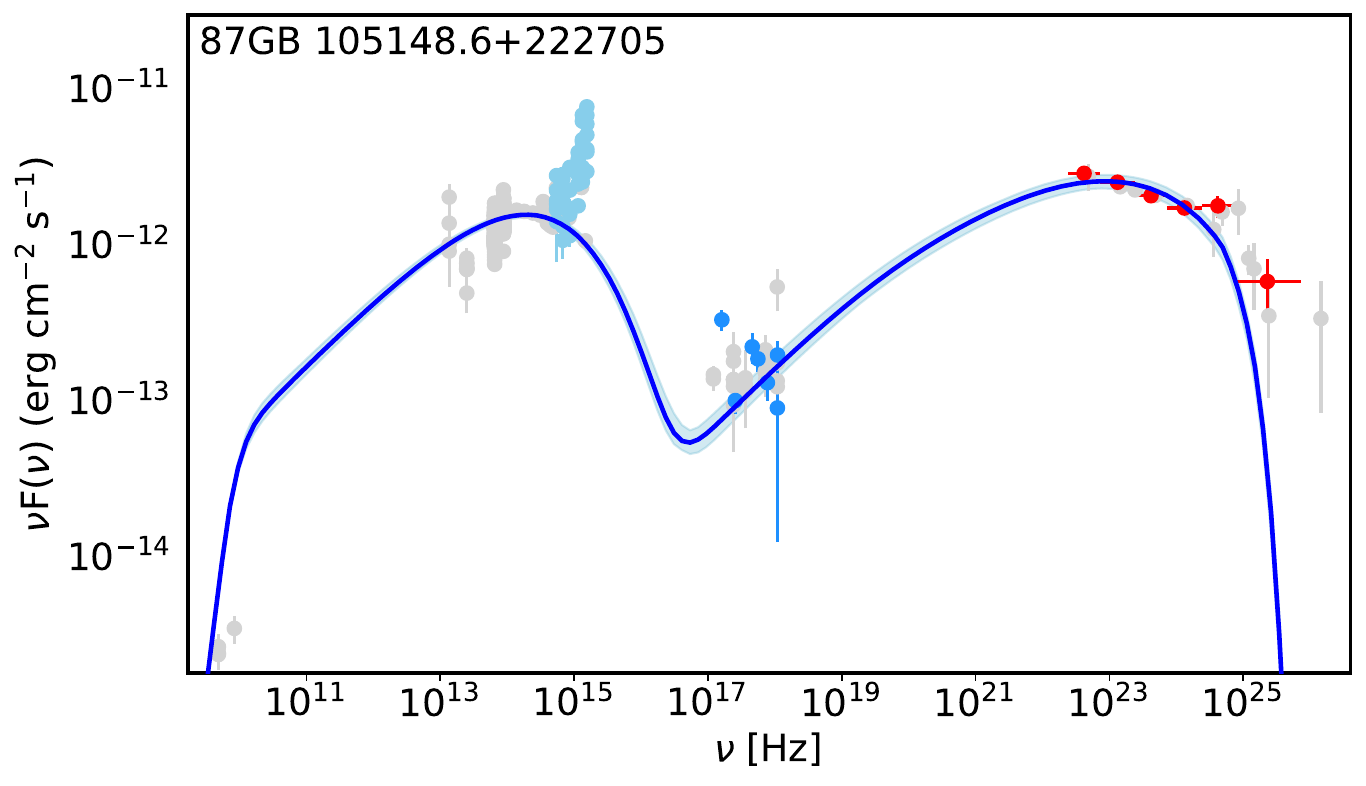}
     \includegraphics[width=0.45\textwidth]{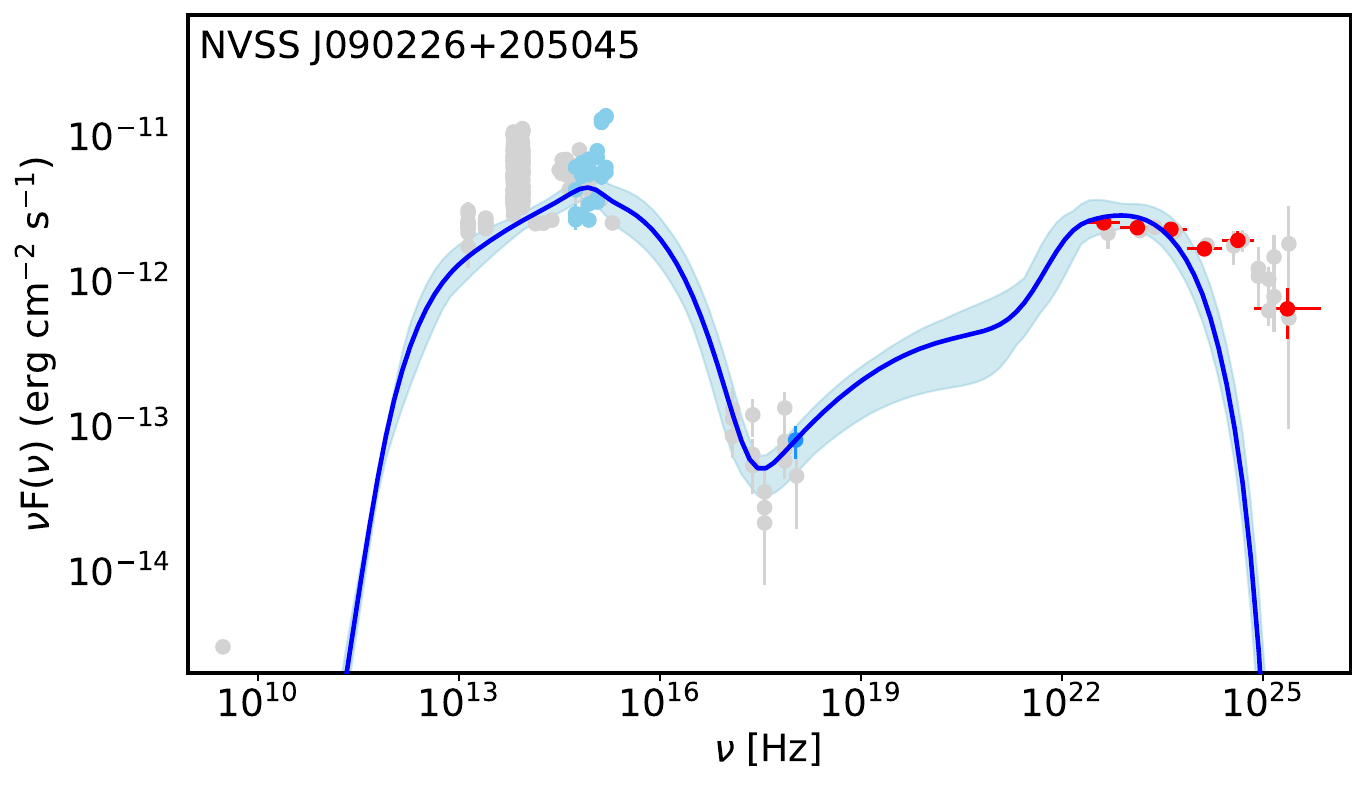}\\
     \includegraphics[width=0.45\textwidth]{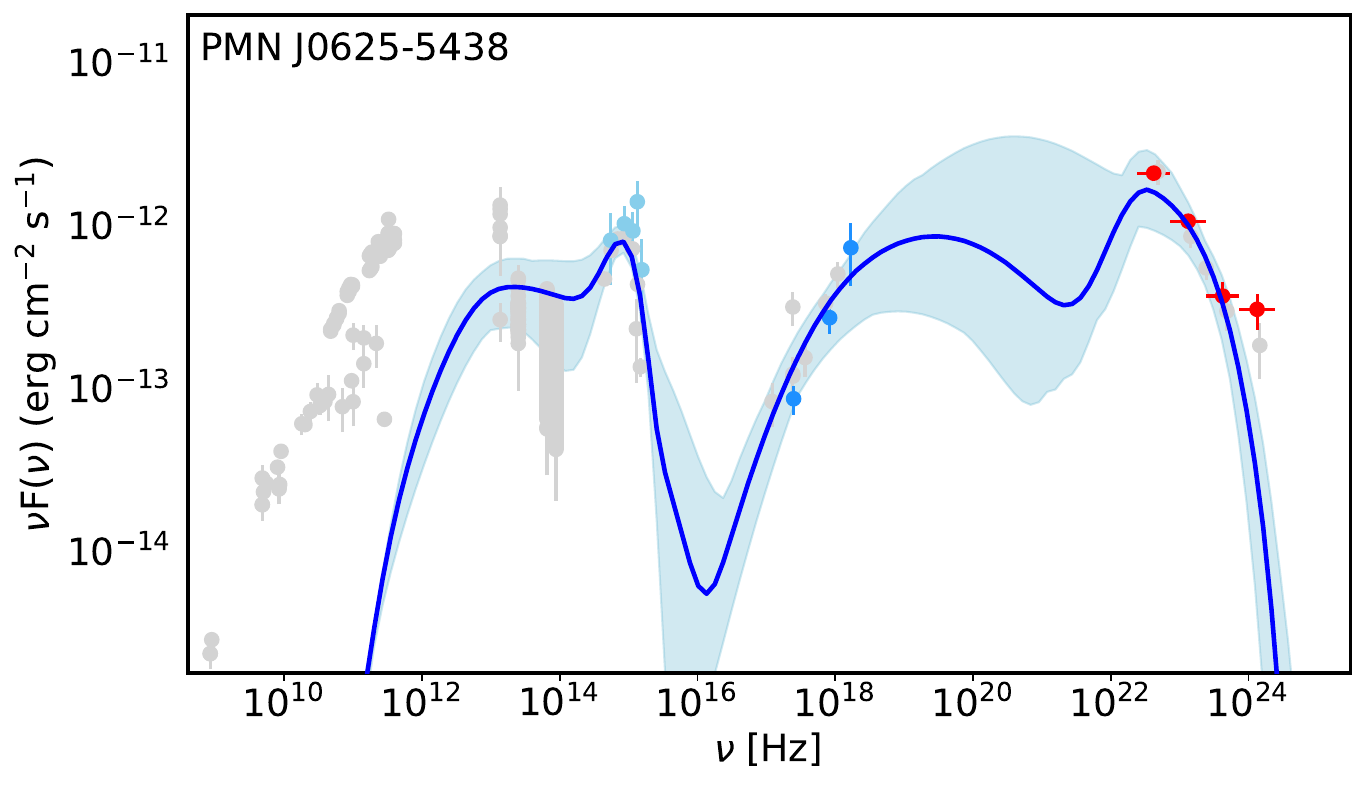}
     \includegraphics[width=0.45\textwidth]{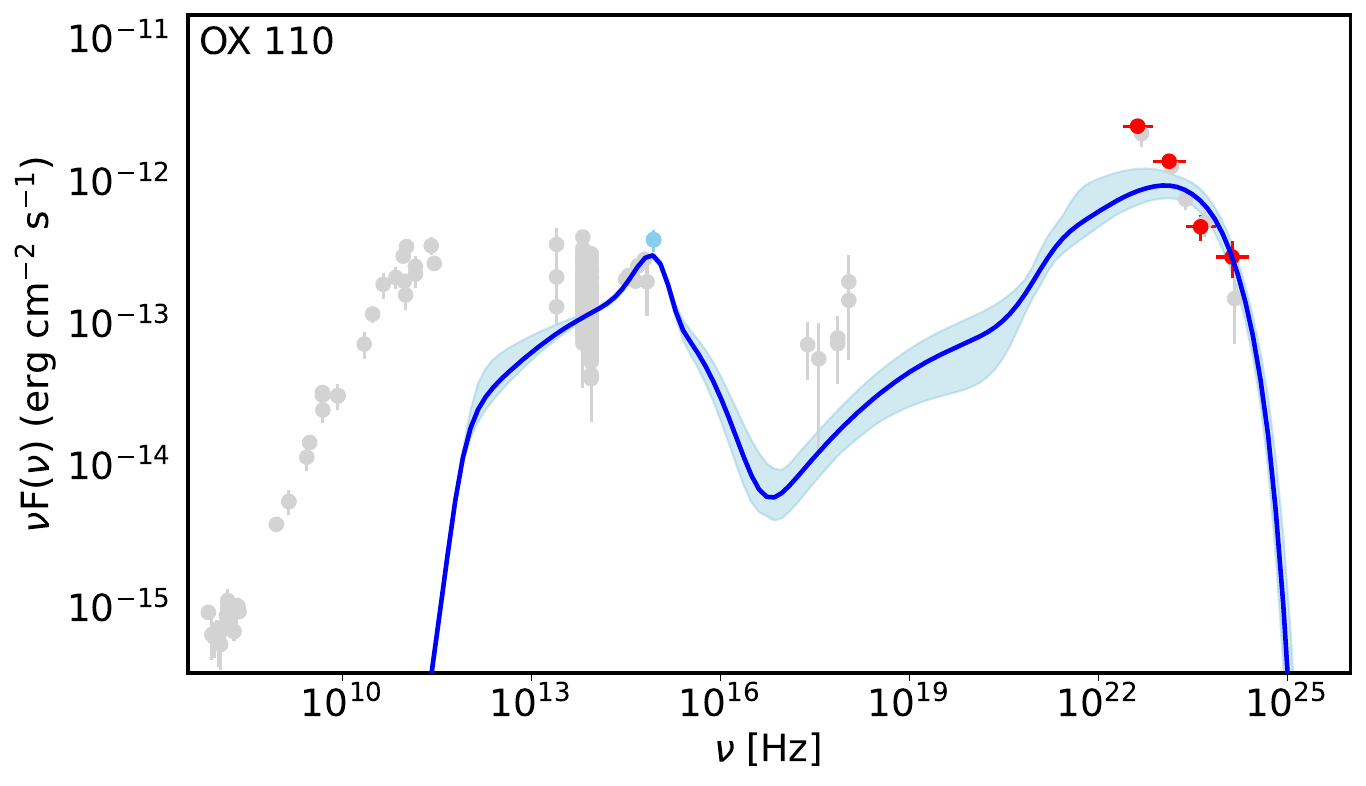}
     \caption{(continued)}
\end{figure*}
\begin{figure*}
     \centering
        \ContinuedFloat
     \includegraphics[width=0.45\textwidth]{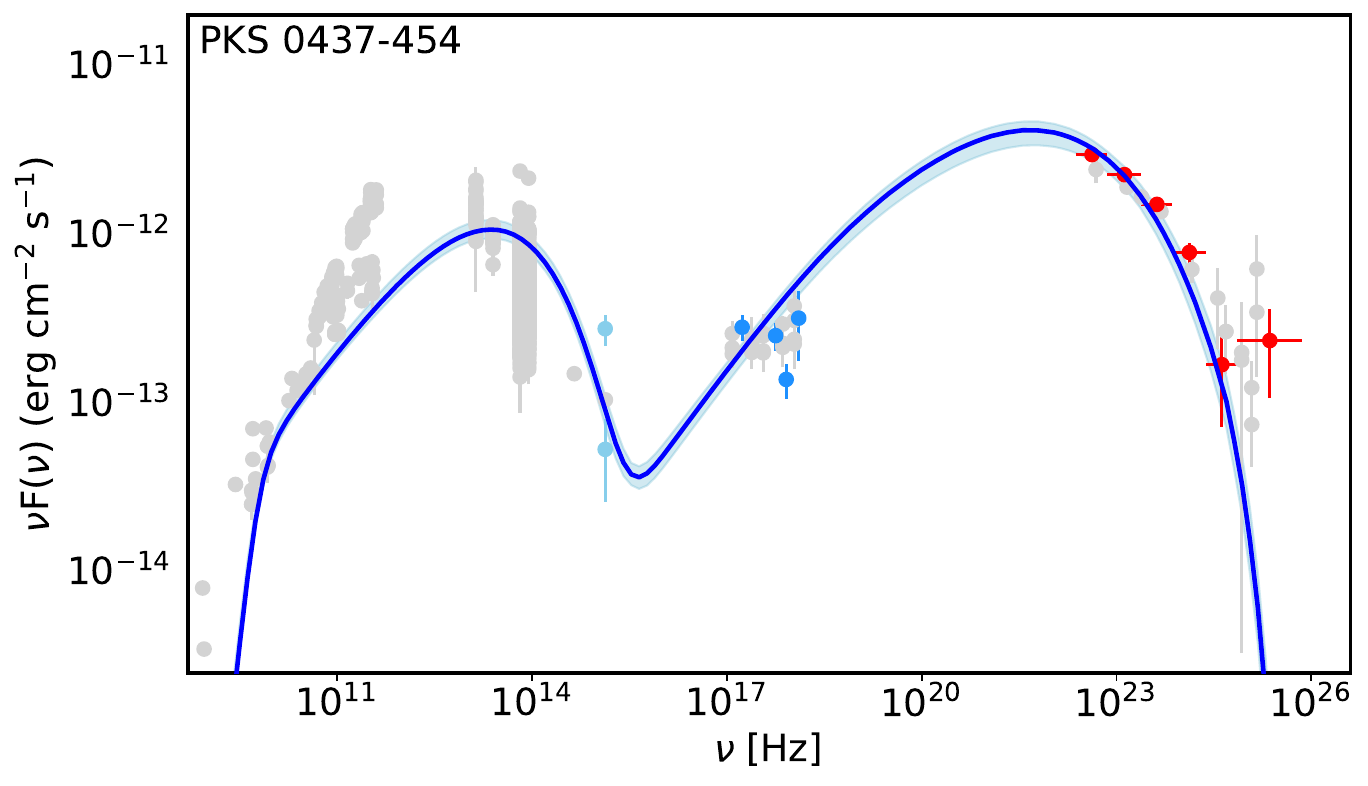}
     \includegraphics[width=0.45\textwidth]{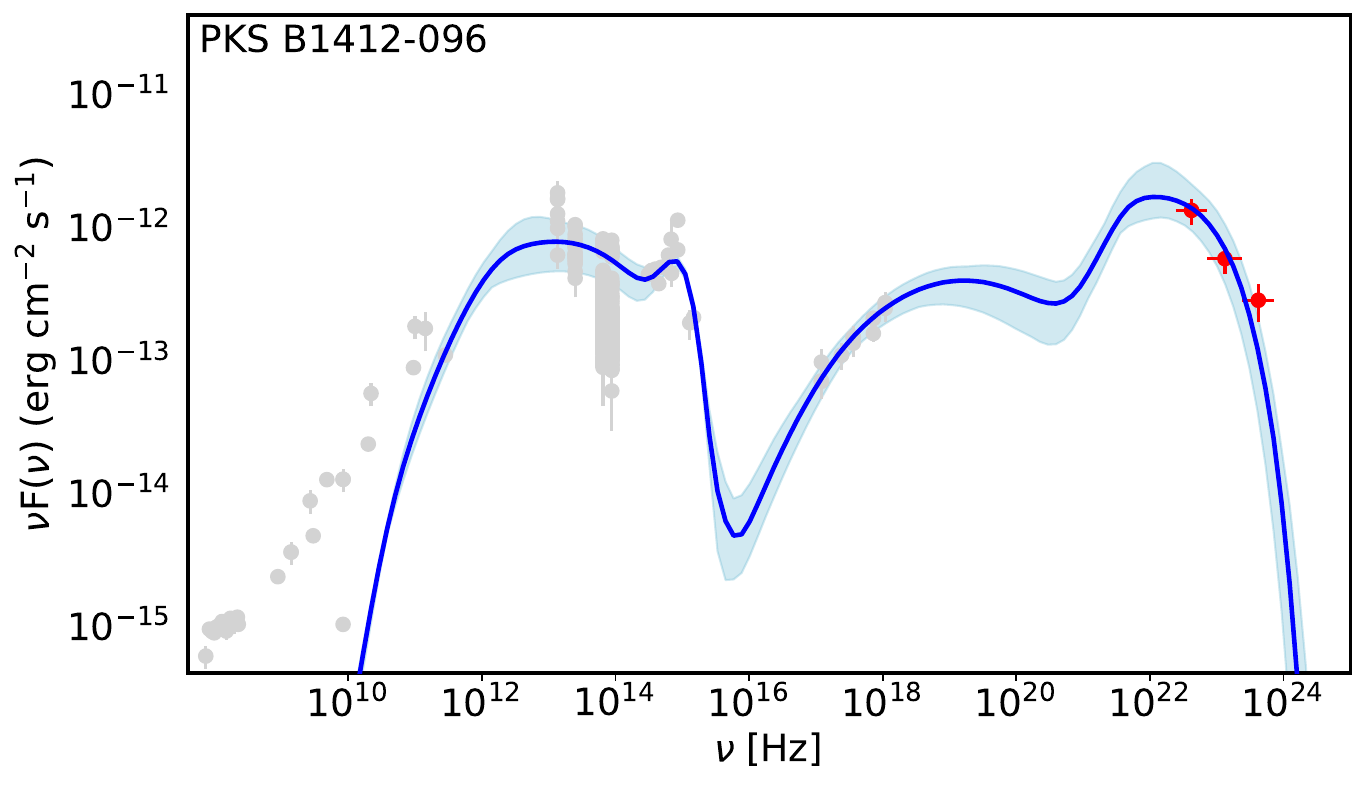}\\
     \includegraphics[width=0.45\textwidth]{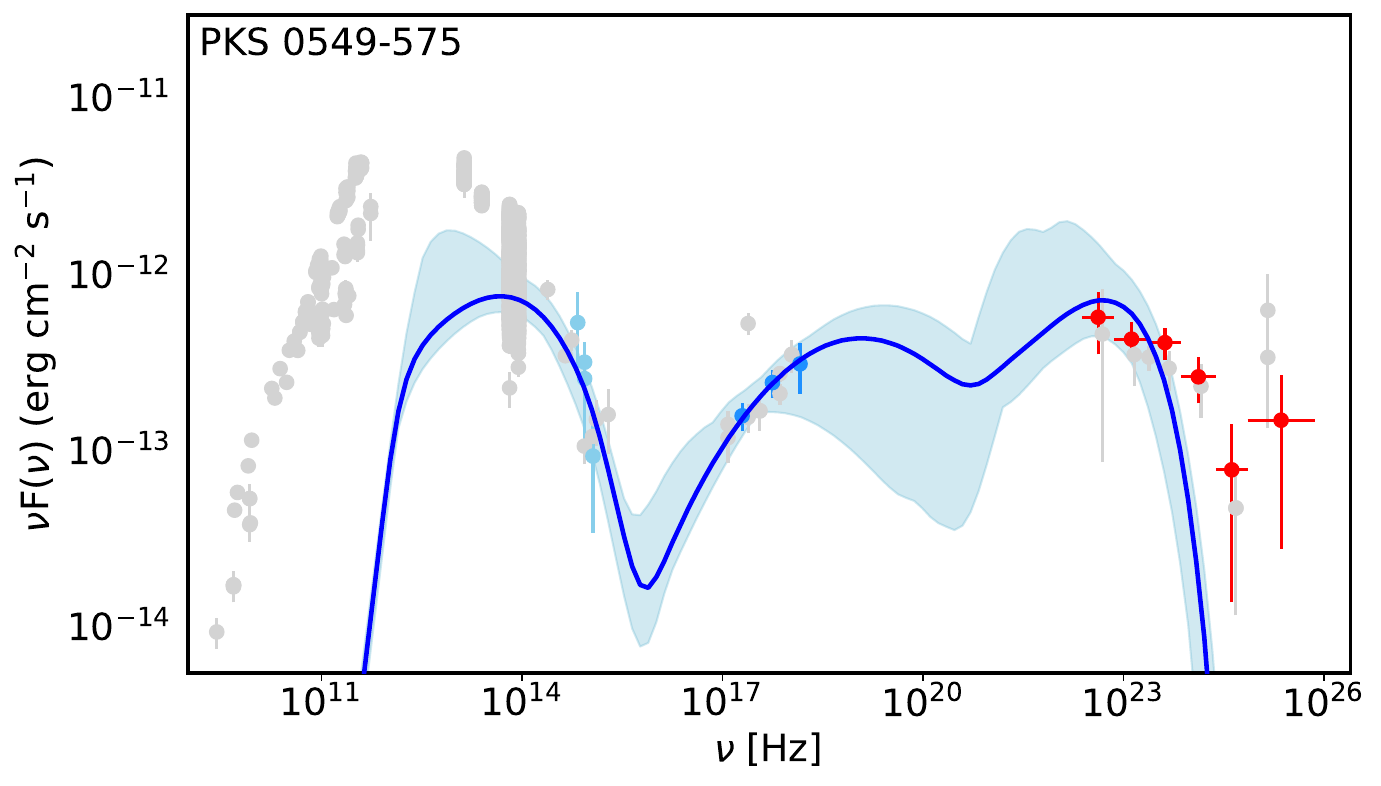}
     \caption{(continued)}
\end{figure*}

\section{Modeling of multiwavelength SEDs}\label{theormodel}
The data analyzed in this study, combined with those collected from the archives, enabled to construct the SEDs for the selected sources from the radio to the HE \gray\ bands. The multiwavelength SEDs are presented in Fig. \ref{sed} where the data analyzed in this work are highlighted in different colors (see the legend), while archival data retrieved using the VOU-Blazar tool are depicted in gray. All the SEDs exhibit the traditional double-humped structure with a clear Compton dominance (i.e., the Compton peak luminosity is significantly higher than the synchrotron peak luminosity).

The selected SEDs were modeled using a leptonic one-zone synchrotron and inverse Compton model. In this scenario, the emission region is assumed to be located at a distance \( R_{\text{dis}} \) from the central black hole and is assumed to have a spherical geometry with a radius \( R \). This region moves along the jet with a bulk Lorentz factor \( \Gamma \) and is observed at a small viewing angle, resulting in the amplification of radiation by a factor of \( \delta \approx \Gamma \). The emitting region is populated with nonthermal particles (electrons and positrons), whose energy distribution follows a power law with an exponential cutoff described as:
\begin{equation}
    N(\gamma) = N_0 \gamma^{-p} e^{-\gamma / \gamma_{\text{cut}}} \quad \text{for} \quad \gamma > \gamma_{\text{min}}
\end{equation}
where \( p \) is the spectral index of the power-law distribution for the emitting electrons, \( \gamma_{\text{cut}} \) is the cutoff Lorentz factor, and \( \gamma_{\text{min}} \) represents the minimum energy. Here, \( N_0 \) denotes the normalization constant of the electron distribution, which defines the total energy content of the electrons as given by \( U_{\text{e}} = m c^2 \int \gamma N(\gamma) d\gamma \).

Within the emitting region (blob), electrons lose energy through synchrotron emission under the magnetic field \( B \), resulting in the first bump in the multiwavelength SED. The second bump is attributed to inverse Compton scattering of low-energy photons, which may originate either internally or externally to the jet. Synchrotron photons can be inverse Compton scattered to higher energies via SSC radiation \citep{1985A&A...146..204G, 1992ApJ...397L...5M, 1996ApJ...461..657B}, and the inverse Compton scattering of external photons may also contribute to the formation of the second component. Depending on the emitting region's proximity to the central black hole, different low-energy photon fields can be inverse Compton up-scattered: photons directly from the accretion disk \citep{1992A&A...256L..27D, 1994ApJS...90..945D}, photons reprocessed by the BLR \citep{1994ApJ...421..153S}, or photons from the dusty torus \citep{2000ApJ...545..107B}. In the current study, we assume that the emitting region is within the BLR (\( R_{\rm dis}<R \)), and that accretion disk photons reprocessed by the BLR (EIC-BLR) alongside synchrotron photons are the primary targets for inverse Compton scattering. However, we note that alternative scenarios, wherein the emitting region lies significantly closer to the central black hole or outside the BLR, cannot be excluded. Constraining the location of the emitting region requires high-quality and high-resolution data at low energy bands, which is unavailable for distant blazars and falls outside the scope of this study.

During the modeling, the accretion disk luminosity (\( L_{\rm disk} \)) and the temperature (energy) of the photons were estimated by fitting the UV band's excess emission with a blackbody component. If the thermal component could not be distinguished, an upper limit was established by ensuring that the disk's emission does not outshine the observed nonthermal emission from the jet. With \( L_{\rm disk} \), the radius of the BLR is calculated using the relation \( R_{\rm BLR}=10^{17} L_{\rm 45, disk} \) \citep{2015MNRAS.448.1060G}. Subsequently, the BLR is represented as a spherical shell, with an inner boundary of \( R_{\rm in, BLR}=0.9 \times R_{\rm BLR} \) and an outer boundary of \( R_{\rm out, BLR}=1.2 \times R_{\rm BLR} \).\\ 
\indent The SEDs were fitted using the publicly available JetSet code \citep{2006A&A...448..861M, 2009A&A...501..879T, 2011ApJ...739...66T,2020ascl.soft09001T}, which facilitates the comparison of numerical models with observed data. For the FSRQs, the synchrotron/EIC-BLR model was used, while for the BL Lacs, the synchrotron/SSC model. Only one BCU has sufficient data for the modeling, SDSS J100326.63+020455 which was also modeling using synchrotron/SSC-BLR modeling considering the shape of the SED. The model consists of a combined contribution from synchrotron, disk, SSC, and EIC-BLR components, which are optimized during the fitting process.  The free parameters in the model are the spectral index (\(p\)), the cutoff Lorentz factor (\(\gamma_{\rm cut}\)), the minimum Lorentz factor (\(\gamma_{\rm min}\)), the magnetic field (\(B\)), the Doppler factor (\(\delta\)), the size of the emitting region (\(R\)), and the energy density of the electrons (\(U_{\rm e}\)). Optimization is performed in two stages: initial fitting is done with the Minuit optimizer, followed by refinement using the Markov Chain Monte Carlo method. We note that \(R\) can also be constrained by variability considerations. However, due to insufficient data for all sources, variability analysis could not be conducted. Therefore, during the fitting process, \(R\) was treated as a parameter dependent on the variability timescale (\(t_{\rm var}\)) and \(\delta\), via the relation \(R=\delta c t_{\rm var}/(1+z)\), with \(t_{\rm var}\) assumed to be within a 0.1 to 7-day range. However, we refrain from commenting on \(t_{\rm var}\) (and consequently \(R\)) as it can only be accurately determined with high-quality data. The pair-production absorption effect due to interaction with EBL photons was incorporated using the model of \citet{2008A&A...487..837F}.
 
\section{Results}\label{res}
The modeling results are shown in Fig. \ref{sed} and the results are given in Table \ref{table:sed_params}. For this modeling, we selected only those sources with sufficient data, particularly those having the X-ray data, as the HE component cannot be constrained without them. There is significant amplitude variability in some sources across optical, IR, and X-ray bands (e.g., PKS 1329-049, PKS 0446+11, 4C +01.02, etc.), which our current modeling does not adequately interpret. To accurately model these flaring periods, a careful selection of (quasi) contemporaneous data is required. However, this is beyond the scope of the current paper, whose primary aim is to estimate the time-averaged properties of the selected sources.

The EIC-BLR model was used to model all the SED of all FSRQs (46 sources). In Fig. \ref{sed}, the sum of all components – the combined contribution of the synchrotron, disk, SSC, and EIC-BLR components – is depicted in blue. The emission at lower energy bands (up to X-rays) is dominated by the synchrotron and disk components, while at HEs, the SSC and EIC-BLR components are dominant. The modeling facilitates the estimation of parameters describing the jet (\(B\), \(\delta\)) as well as those characterizing the emitting electrons (\(p\), \(\gamma_{\rm min}\), and \(\gamma_{\rm cut}\)). From the modeling of the SEDs of the FSRQs, the power-law index of the electrons varies within a range from \(p=1.56\) to \(p=2.85\), predominantly constrained by the SSC fit of the X-ray data. The hardest spectral indices are observed in sources lacking high-quality X-ray data. For instance, \(p=1.56\pm0.064\) for PKS 0226-559, \(p=1.60\pm0.12\) for OX 131, and \(p=1.70\pm0.21\) for PKS 1348+007, etc. The power-law index of the electrons, inferred from the SSC fitting of BL Lacs (7 sources), is harder than that for FSRQs, ranging between $1.71$ and $2.25$. Relatively hard indices of $1.71\pm0.01$ and $1.79\pm0.01$ were estimated for SDSS J145059.99+520111.7 and NVSS J090226+205045, respectively.

For the FSRQs, the cutoff Lorentz factor (\( \gamma_{\text{cut}} \)) for the considered sources varies within a range from $237.6$ to $3.53\times10^3$, determined by the optical/UV and/or X-ray data. A high value of $(3.55\pm0.49)\times10^4$ was estimated for PMN J2135-5006, which has sparse X-ray data and is therefore not considered in the discussions. The lowest value of $237.6\pm21.8$ was estimated for TXS 0322+222, as the shape of the low-energy data (below optical) and the optical/UV data impose a strong constraint on this parameter. A comparably high cutoff Lorentz factor, \( \gamma_{\text{cut}} = (3.53 \pm 0.31) \times 10^3 \), was estimated for PMN J1344-1723. In the modeling of BL Lacs, the cutoff Lorentz factor (\( \gamma_{\text{cut}} \)) is $(1.53-25.41)\times 10^4$ which is larger than those typically estimated for EIC-BLR model, as the average energy of the synchrotron photons is lower than that of BLR photons. The highest cutoff Lorentz factor, \( \gamma_{\text{cut}} = (2.54 \pm 0.05) \times 10^5 \), was estimated for SDSS J105707.47+551032.2, which is due to the fact that the synchrotron emission of this source extends up to the X-ray band. The minimum Lorentz factor ranges from $1.08$ to $207.0$ and is given in Table \ref{table:sed_params}, constrained by the requirement that the emission at radio bands does not overshoot the observed data.

For the FSRQs the estimated magnetic field ranges between $1.30$ and $15.17$ G. The highest value, $15.17\pm0.90$ G, was estimated for PMN J0134-3843 which is attributed to the fact that the synchrotron component of this source exhibits a comparable flux compared to its inverse Compton component. Consequently, when the external photons are boosted into the jet frame (with a \(\delta=24.71\) for this object), their density increases. This requires the presence of lower density electrons, thereby requiring a higher magnetic field to account for the observed synchrotron emission. In contrast to the modeling of FSRQs, when considering BL Lacs, the magnetic field is found to be in the range of $0.001$ to $0.04$ G. For the FSRQs, except for the bordering cases of PKS B1412-096, PMN J2135-5006, and PMN J1344-1723, where \(\delta\) of $5.24$, $5.45$, and $47.18$ were estimated respectively, \(\delta\) varies within a narrow range of $12.52$ to $33.82$. This range is typical for blazars. For PMN J1344-1723, a \(\delta\) of $47.18$ was estimated due to the pronounced Compton dominance observed in this object. In contrast, a lower \(\delta\) of $5.24$ for PKS B1412-096 was estimated because its Compton dominance is nearly 1. Meanwhile, \(\delta=5.45\) for PMN J2135-5006 was estimated due to its relatively high cutoff Lorentz factor (\( \gamma_{\text{cut}} \)), which is unconstrained by the data. For the BL Lacs, larger \(\delta\) in the range of $28.94$ to $49.84$ were estimated. This is because the synchrotron component of these objects peaks around $10^{13}-10^{14}$ Hz, so a higher \(\delta\) is required to explain the second peak, which is in the MeV band.

\begin{figure}

     \centering
     
     \includegraphics[width=0.45\textwidth]{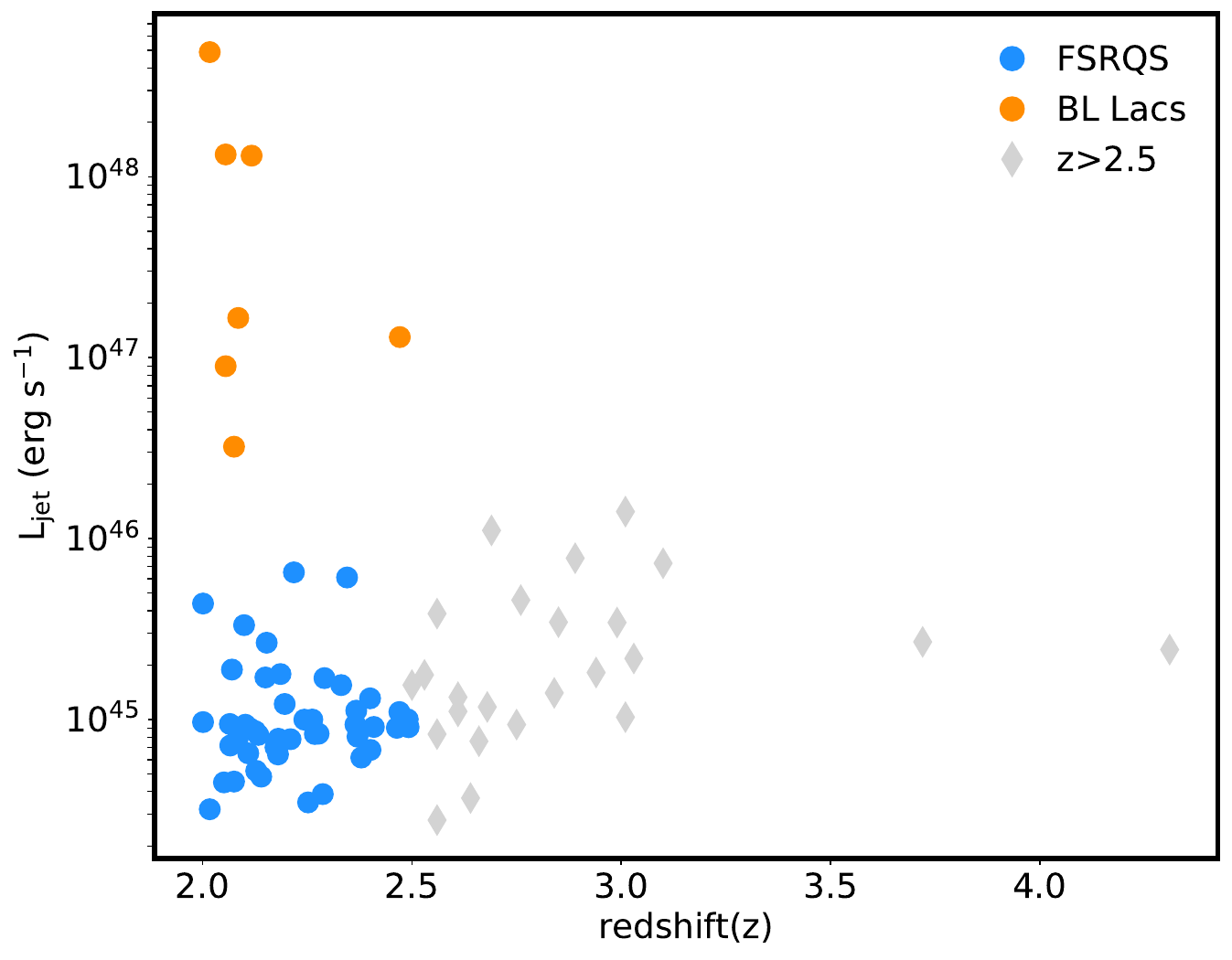}\\
     \includegraphics[width=0.45\textwidth]{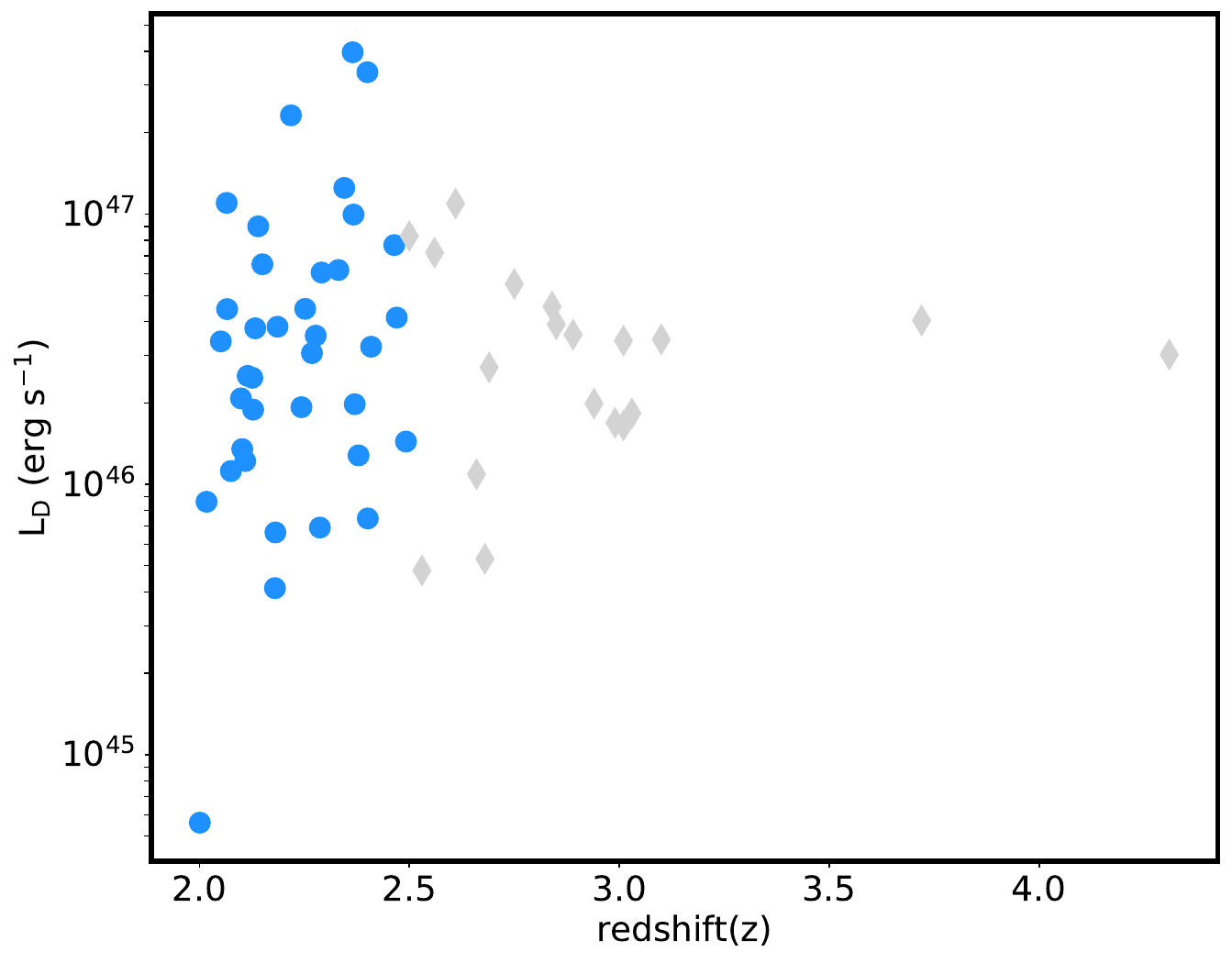}\\
     \includegraphics[width=0.45\textwidth]{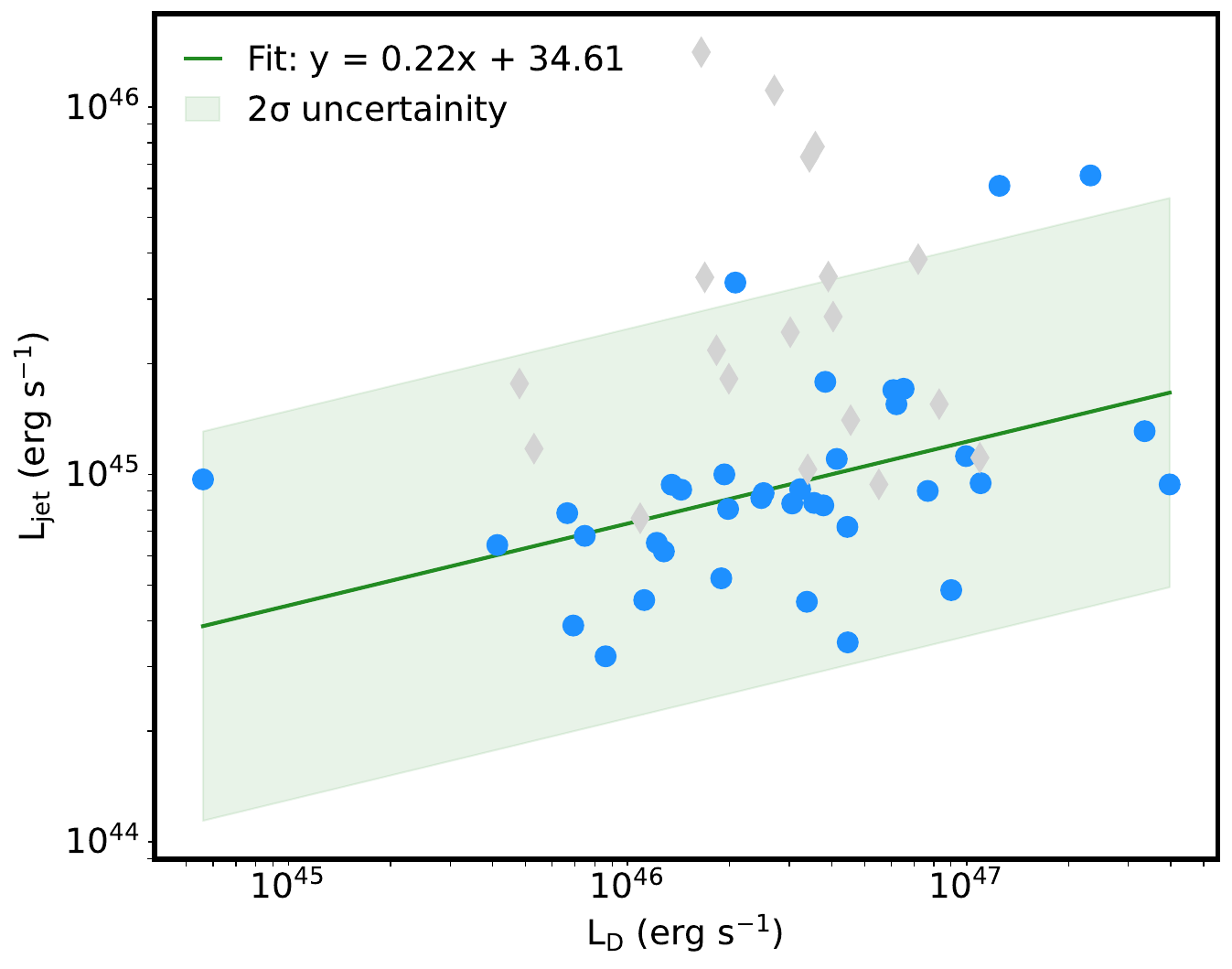}
     \caption{{\it Upper panel:} The dependence of the total jet luminosity (\(L_{\rm jet}\)) on redshift. {\it Middle panel:} The dependence of the disc luminosity (\(L_{\rm D}\)) on redshift. {\it Lower panel:} The correlation between jet luminosity and disc luminosity. The sources from \citet{2020MNRAS.498.2594S} are shown in gray.}
     \label{lumplot}
\end{figure}

Another key parameter to assess following the modeling is the jet luminosity. The jet power, in the form of electrons and magnetic field, is computed as \( L_{e}=\pi c R_b^2 \Gamma^2 U_{e} \) and \( L_{B}=\pi c R_b^2 \Gamma^2 U_{B} \), respectively. Assuming one proton per relativistic emitting electron and that protons are 'cold' in the comoving frame, the total jet luminosity is calculated as \( L_{\rm jet}=L_{e}+L_{B}+L_{p} \). The corresponding \( L_{e} \) and \( L_{B} \) luminosities are presented in Table \ref{table:sed_params}. For the FSRQs, \( L_{e} \) varies within the range from \( 7.24\times10^{43}\:{\rm erg\:s^{-1}} \) to \( 2.94\times10^{45}\:{\rm erg\:s^{-1}} \), while \( L_{B} \) ranges from \( 2.72\times10^{42}\:{\rm erg\:s^{-1}} \) to \( 4.06\times10^{45}\:{\rm erg\:s^{-1}} \). Among the considered sources, 33 exhibit an \( L_e/L_B \) ratio in the range of 0.1 to 10, suggesting that the emitting region is close to equipartition. For PKS B1412-096 the ratio \( L_e/L_B = 0.05 \) is estimated which is because of the synchrotron component is at the same level as the inverse Compton component. Conversely, for OX 131, 4C +71.07, and PKS 2149-306, a significantly higher ratio with \( L_e/L_B > 100 \) is estimated, due to the dominance of the SSC component over the synchrotron component. In the modeling of BL Lacs, \( L_{e} \) is in the range of \( (0.04-3.85)\times10^{47}\:{\rm erg\:s^{-1}} \), which is higher than the \( L_{B} \), ranging from \( (0.10-3.92)\times10^{43}\:{\rm erg\:s^{-1}} \). The relationship between the total jet luminosity (\(L_{\rm jet}\)) and redshift is illustrated in Fig. \ref{lumplot} upper panel, where our current source sample (FSRQs in blue and BL Lacs in orange) is compared with sources from \citet{2020MNRAS.498.2594S} (gray), all of which are FSRQs. Clearly, BL Lacs exhibit different properties and tend to have higher luminosities. However, discussions about the distribution of BL Lacs remain inconclusive. This is because BL Lacs are rarely identified at distances \( z>2.0 \) but a significant fraction of BL Lacs lack robust redshift determination, suggesting that BL Lacs might indeed be present at larger redshifts. 
In contrast, FSRQs show a similar luminosity range to those at redshifts \(z>2.5\). For the FSRQs the highest luminosity of \( 6.51\times10^{45}\:{\rm erg\:s^{-1}} \) was observed for 4C +71.07 at a redshift of \(z=2.218\). Upon comparing both samples, it is evident that the jet luminosity tends to increase with source distance, with two notable exceptions: PKS 1351-018 and GB 1508+5714, which have the highest redshifts of \(z=3.72\) and \(z=4.31\), respectively.

The fit presented in Fig. \ref{sed} also enables the estimation of disk luminosity for cases where a thermal blue-bump component is observable. This excess is evident in a total of 37 sources. The disk luminosity, \( L_{D} \), varies from \( 5.60\times10^{44} \) erg s\(^{-1}\) to \( 3.97\times10^{47} \) erg s\(^{-1}\). The relationship between disk luminosity and redshift for \(\gamma\)-ray blazars with \( z>2.0 \) is depicted in the middle panel of Fig. \ref{lumplot}. High disk luminosities of \( 1.25\times10^{47} \) erg s\(^{-1}\), \( 2.32\times10^{47} \) erg s\(^{-1}\), \( 3.35\times10^{47} \) erg s\(^{-1}\), and \( 3.97\times10^{47} \) erg s\(^{-1}\) were estimated for PKS 2149-306, 4C +71.07, 2MASS J16561677-3302127, and B2 0552+39A, respectively. Furthermore, the relationship between disk luminosity and jet luminosity is examined in the lower panel of Fig. \ref{lumplot}. This relationship can be described by the equation \( \log_{10}(L_{\rm jet}) = 0.22 \times \log_{10}(L_{\rm D}) + 34.60 \), represented by the green line in Fig. \ref{lumplot}, with the \( 2\sigma \) uncertainty region shown in light green. Notably, sources such as PKS 2149-306 and 4C +71.07 exhibit disk luminosities exceeding \( 10^{47} \) erg s\(^{-1}\) and jet luminosities exceeding \( 10^{45} \) erg s\(^{-1}\), highlighting their significant energetic output.
\begin{table*}
    \centering
\caption{Broadband SED Modeling of the Considered Sources. [1] Source Name. [2] Doppler Factor. [3] Power-Law index of emitting electrons. [4] Minimum Lorentz factor of the electrons and [5] cutoff Lorentz factor in units of 100. [6] Magnetic field, expressed in units of $G$. [7] Accretion Disk Luminosity, in units of \(10^{46}\) erg s\(^{-1}\). [8] and [9] Power of the jet in the form of relativistic electrons (\(L_{e}\)) and Magnetic field (\(L_{B}\)), both in units of \(10^{44}\) erg s\(^{-1}\). [10] Luminosity of the jet in the form of protons assuming one proton per relativistic emitting electron in units of \(10^{45}\) erg s\(^{-1}\).}
\label{table:sed_params}
\resizebox{\textwidth}{!}{%
\begin{tabular}{lccccccccc}
\toprule
Source & $\delta$ & $p$ & $\gamma_{\text{min}}$ & $\gamma_{\text{cut}}$ & $B$ & $L_{D}$ & $L_{e}$ & $L_{B}$ & $L_{p}$ \\
$[1]$ & [2] & [3] & [4] & [5] & [6] & [7] & [8] & [9] & [10]\\
\midrule
S5 1053+70 &$18.42 \pm 1.42$ & $2.08 \pm 0.21$ & $69.49 \pm 7.30$ & $12.50 \pm 1.22$ & $3.96 \pm 0.42$ & $1.44$ & $4.40$ & $0.76$ & $0.39$\\
PMN J1344-1723 &$47.18 \pm 1.66$ &$2.10 \pm 0.06$ &$18.56 \pm 1.83$ &$35.32 \pm 3.13$ &$4.21 \pm 0.27$ &$0.84$ &$2.93$ &$0.08$ &$0.70$\\
PKS 1915-458 &$24.53 \pm 1.14$ &$2.29 \pm 0.22$ &$65.94 \pm 6.50$ &$4.09 \pm 0.69$ &$7.62 \pm 0.80$ &$4.14$ &$4.55$ &$0.24$ &$0.62$\\
PKS 0226-559 &$33.13 \pm 2.03$ &$1.56 \pm 0.06$ &$17.32 \pm 2.23$ &$27.07 \pm 2.54$ &$4.25 \pm 0.33$ &$7.68$ &$4.37$ &$0.12$ &$0.45$\\
PKS 0601-70 &$24.91 \pm 1.20$ &$1.92 \pm 0.17$ &$71.10 \pm 6.49$ &$6.39 \pm 0.59$ &$5.81 \pm 0.62$ &$3.23$ &$3.28$ &$2.43$ &$0.34$\\
B2 1436+37B &$21.06 \pm 0.96$ &$2.00 \pm 0.21$ &$95.07 \pm 10.48$ &$4.69 \pm 0.42$ &$5.07 \pm 0.57$ &$0.75$ &$2.64$ &$1.71$ &$0.25$\\
2MASS J16561677-3302127 &$17.15 \pm 0.22$ &$1.93 \pm 0.02$ &$97.22 \pm 1.59$ &$14.50 \pm 0.29$ &$9.91 \pm 0.24$ &$33.50$ &$6.39$ &$2.56$ &$0.42$\\
TXS 1645+635 &$24.58 \pm 0.94$ &$1.92 \pm 0.18$ &$67.97 \pm 8.10$ &$3.87 \pm 0.37$ &$7.46 \pm 0.70$ &$1.28$ &$1.70$ &$2.40$ &$0.21$\\
PKS B1149-084 &$25.62 \pm 0.93$ &$1.92 \pm 0.19$ &$64.32 \pm 6.41$ &$4.86 \pm 0.59$ &$5.98 \pm 0.39$ &$1.98$ &$1.64$ &$4.50$ &$0.19$\\
S5 0212+73 &$23.24 \pm 0.65$ &$2.66 \pm 0.14$ &$207.00 \pm 9.11$ &$7.63 \pm 0.73$ &$8.48 \pm 0.38$ &$9.96$ &$7.23$ &$0.13$ &$0.39$\\
B2 0552+39A &$13.93 \pm 0.82$ &$2.12 \pm 0.07$ &$105.80 \pm 7.80$ &$27.18 \pm 2.45$ &$6.91 \pm 0.77$ &$39.70$ &$5.60$ &$0.43$ &$0.34$\\
PKS 2149-306 &$25.83 \pm 0.83$ &$1.82 \pm 0.04$ &$83.20 \pm 1.31$ &$3.02 \pm 0.12$ &$5.14 \pm 0.06$ &$12.50$ &$29.40$ &$0.04$ &$3.16$\\
PKS 1430-178 &$26.31 \pm 1.22$ &$2.25 \pm 0.17$ &$43.81 \pm 5.82$ &$4.17 \pm 0.62$ &$9.80 \pm 0.83$ &$6.21$ &$5.32$ &$0.20$ &$1.00$\\
S3 0458-02 &$33.82 \pm 1.37$ &$2.21 \pm 0.11$ &$55.51 \pm 7.02$ &$7.16 \pm 0.75$ &$7.21 \pm 0.59$ &$6.08$ &$6.60$ &$1.09$ &$0.93$\\
PMN J0157-4614 &$22.59 \pm 1.43$ &$1.94 \pm 0.21$ &$133.10 \pm 21.74$ &$4.26 \pm 0.68$ &$6.59 \pm 0.63$ &$0.69$ &$0.72$ &$2.63$ &$0.05$\\
PKS 0420+022 &$26.55 \pm 1.18$ &$2.78 \pm 0.14$ &$48.92 \pm 4.10$ &$8.32 \pm 0.95$ &$8.01 \pm 0.55$ &$3.55$ &$2.31$ &$1.54$ &$0.45$\\
PKS 2245-328 &$24.98 \pm 1.46$ &$1.94 \pm 0.23$ &$47.52 \pm 6.02$ &$4.15 \pm 0.51$ &$8.18 \pm 0.74$ &$3.06$ &$2.47$ &$1.86$ &$0.40$\\
PKS B2224+006 &$21.38 \pm 1.01$ &$1.79 \pm 0.20$ &$29.27 \pm 4.49$ &$3.13 \pm 0.38$ &$8.02 \pm 0.87$ &$0.25$ &$2.78$ &$1.33$ &$0.59$\\
PKS 2244-37 &$26.04 \pm 1.94$ &$1.99 \pm 0.25$ &$63.82 \pm 7.72$ &$5.28 \pm 0.96$ &$10.89 \pm 1.30$ &$4.46$ &$1.59$ &$0.03$ &$0.19$\\
B2 0242+23 &$26.05 \pm 1.26$ &$2.10 \pm 0.18$ &$34.78 \pm 4.12$ &$4.95 \pm 0.69$ &$6.21 \pm 0.48$ &$1.93$ &$3.08$ &$0.69$ &$0.62$\\
4C +71.07 &$32.34 \pm 1.30$ &$2.24 \pm 0.16$ &$43.72 \pm 3.72$ &$4.38 \pm 0.40$ &$9.94 \pm 0.52$ &$23.20$ &$22.70$ &$0.14$ &$4.23$\\
PKS 2022+031 &$24.37 \pm 1.31$ &$2.18 \pm 0.17$ &$32.99 \pm 3.38$ &$11.10 \pm 1.17$ &$7.10 \pm 0.84$ &$0.12$ &$2.32$ &$0.96$ &$0.45$\\
MG2 J153938+2744 &$19.46 \pm 0.80$ &$2.12 \pm 0.13$ &$51.00 \pm 3.55$ &$8.96 \pm 0.96$ &$8.28 \pm 0.52$ &$0.30$ &$1.50$ &$8.67$ &$0.20$\\
S4 0917+44 &$27.14 \pm 0.81$ &$2.42 \pm 0.11$ &$69.94 \pm 6.35$ &$8.84 \pm 0.67$ &$5.32 \pm 0.26$ &$3.83$ &$7.48$ &$1.27$ &$0.91$\\
PMN J2135-5006 &$5.45 \pm 0.21$ &$2.85 \pm 0.06$ &$158.70 \pm 14.13$ &$355.10 \pm 48.94$ &$1.30 \pm 0.10$ &$0.66$ &$3.23$ &$2.85$ &$0.18$\\
OX 131 &$21.83 \pm 0.77$ &$1.60 \pm 0.12$ &$109.00 \pm 10.71$ &$13.76 \pm 0.93$ &$1.76 \pm 0.10$ &$0.41$ &$4.30$ &$0.04$ &$0.21$\\
PMN J1959-4246 &$20.98 \pm 0.84$ &$2.02 \pm 0.17$ &$57.41 \pm 4.50$ &$5.07 \pm 0.63$ &$7.88 \pm 0.71$ &$0.17$ &$2.54$ &$1.13$ &$0.34$\\
PKS 0446+11 &$23.19 \pm 1.12$ &$1.98 \pm 0.17$ &$16.67 \pm 2.26$ &$5.53 \pm 0.73$ &$5.34 \pm 0.46$ &$0.16$ &$4.36$ &$8.05$ &$1.42$\\
PKS 1329-049 &$22.29 \pm 0.87$ &$2.30 \pm 0.16$ &$69.16 \pm 8.12$ &$12.00 \pm 1.99$ &$3.62 \pm 0.34$ &$6.52$ &$8.00$ &$0.43$ &$0.87$\\
PMN J0134-3843 &$24.71 \pm 0.98$ &$2.54 \pm 0.19$ &$86.57 \pm 8.43$ &$4.61 \pm 0.61$ &$15.17 \pm 0.90$ &$9.01$ &$1.59$ &$1.40$ &$0.19$\\
87GB 080551.6+535010 &$26.16 \pm 1.11$ &$1.90 \pm 0.20$ &$59.01 \pm 5.16$ &$3.34 \pm 0.34$ &$4.58 \pm 0.40$ &$3.78$ &$3.21$ &$0.46$ &$0.46$\\
PKS B1043-291 &$26.98 \pm 1.61$ &$2.50 \pm 0.20$ &$56.45 \pm 4.78$ &$10.30 \pm 1.28$ &$9.58 \pm 1.15$ &$1.89$ &$1.80$ &$0.88$ &$0.25$\\
OM 127 &$22.03 \pm 1.16$ &$2.59 \pm 0.14$ &$49.91 \pm 5.54$ &$18.73 \pm 3.19$ &$7.31 \pm 0.66$ &$2.48$ &$2.56$ &$1.90$ &$0.42$\\
PKS 0227-369 &$20.02 \pm 0.68$ &$2.83 \pm 0.09$ &$81.14 \pm 8.13$ &$25.89 \pm 2.97$ &$5.32 \pm 0.50$ &$2.52$ &$3.59$ &$1.22$ &$0.41$\\
OF 200 &$15.55 \pm 0.59$ &$2.13 \pm 0.11$ &$38.66 \pm 1.92$ &$11.03 \pm 0.68$ &$5.45 \pm 0.24$ &$1.22$ &$2.14$ &$0.88$ &$0.35$\\
B3 0803+452 &$19.64 \pm 0.94$ &$2.28 \pm 0.25$ &$44.72 \pm 5.38$ &$3.17 \pm 0.43$ &$8.56 \pm 0.77$ &$1.35$ &$2.11$ &$3.03$ &$0.42$\\
4C +01.02 &$26.29 \pm 0.98$ &$1.88 \pm 0.13$ &$28.05 \pm 2.99$ &$5.55 \pm 0.52$ &$3.04 \pm 0.19$ &$2.08$ &$9.23$ &$5.25$ &$1.88$\\
PKS 1348+007 &$24.93 \pm 1.94$ &$1.70 \pm 0.21$ &$59.32 \pm 8.57$ &$7.87 \pm 0.81$ &$2.50 \pm 0.29$ &$0.04$ &$2.95$ &$2.19$ &$0.27$\\
SDSS J100326.63+020455.6 &$26.93 \pm 0.76$ &$2.03 \pm 0.08$ &$40.09 \pm 1.28$ &$7.69 \pm 0.39$ &$12.03 \pm 0.28$ &$1.12$ &$1.75$ &$0.13$ &$0.27$\\
PKS 0528+134 &$15.71 \pm 0.87$ &$2.15 \pm 0.10$ &$46.04 \pm 5.30$ &$20.93 \pm 2.60$ &$3.67 \pm 0.25$ &$0.80$ &$7.91$ &$0.90$ &$1.01$\\
TXS 0322+222 &$31.80 \pm 0.99$ &$2.12 \pm 0.25$ &$122.40 \pm 11.04$ &$2.38 \pm 0.22$ &$7.92 \pm 0.58$ &$4.45$ &$3.51$ &$0.51$ &$0.32$\\
4C +13.14 &$20.30 \pm 1.03$ &$1.72 \pm 0.14$ &$44.37 \pm 5.16$ &$4.10 \pm 0.43$ &$8.13 \pm 0.62$ &$11.00$ &$3.38$ &$1.21$ &$0.49$\\
PMN J0625-5438 &$12.52 \pm 0.70$ &$2.61 \pm 0.23$ &$200.50 \pm 27.17$ &$13.10 \pm 2.07$ &$5.12 \pm 0.55$ &$3.38$ &$2.38$ &$0.95$ &$0.12$\\
OX 110 &$24.62 \pm 0.87$ &$2.08 \pm 0.08$ & $29.37 \pm 2.40$ &$18.94 \pm 1.26$ &$4.89 \pm 0.23$ &$0.86$ &$1.06$ &$0.29$ &$0.19$\\
PKS 0549-575 &$27.49 \pm 2.14$ &$1.86 \pm 0.21$ & $37.63 \pm 4.32$ &$4.47 \pm 0.43$ &$6.42 \pm 0.69$ &$0.06$ &$3.22$ &$1.29$ &$0.52$\\
PKS B1412-096 &$5.24 \pm 0.17$ &$2.32 \pm 0.14$ & $199.30 \pm 17.68$ &$13.22 \pm 1.08$ &$3.17 \pm 0.26$ &$2.17$ &$2.20$ &$40.60$ &$0.10$\\

\hline
SDSS J145059.99+520111.7 &$ 40.41 \pm 0.35 $ &$ 1.71 \pm 0.01 $ &$ 1.41 \pm 0.03 $ &$ 153.30 \pm 1.30 $ &$ (1.08 \pm 0.01) \times 10^{-2} $ &$ - $ &$ 245.50 $ &$ 0.26 $ &$ 105.50 $\\
PMN J0124-0624 &$ 49.66 \pm 0.67 $ &$ 2.09 \pm 0.01 $ &$ 2.67 ±\pm 0.07 $ &$ 378.70 \pm 10.95 $ &$ (1.91\pm0.06)\times{10^{-3}} $ &$ - $ &$ 1126.00 $ &$ 0.22 $ &$ 1196.00 $\\
SDSS J105707.47+551032.2 &$ 28.94 \pm 0.43 $ &$ 2.15 \pm 0.01 $ &$ 1.08 \pm 0.03 $ &$ 2541.00 \pm 47.98 $ &$ (1.72 \pm 0.02) \times 10^{-2} $ &$ - $ &$ 59.21 $ &$ 1.53 $ &$ 159.80 $\\
1RXS J032342.6-011131 &$ 29.71 \pm 0.19 $ &$ 1.90 \pm 0.004 $ &$ 1.84 \pm 0.01 $ &$ 439.90 \pm 5.40 $ &$ (3.99 \pm 0.04) \times 10^{-2} $ &$ - $ &$ 40.14 $ &$ 3.92 $ &$ 27.82 $\\
NVSS J090226+205045 &$ 49.84 \pm 0.33 $ &$ 1.79 \pm 0.01 $ &$ 22.93 \pm 0.23 $ &$ 212.60 \pm 2.17 $ &$ (4.36\pm0.04)\times{10^{-3}}$ &$ - $ &$ 514.00 $ &$ 1.14 $ &$ 38.29 $\\
87GB 105148.6+222705 &$ 29.10 \pm 0.67 $ &$ 2.13 \pm 0.01 $ &$ 2.37 \pm 0.10 $ &$ 363.80 \pm 10.50 $ &$ (7.32 \pm 0.22) \times 10^{-3} $ &$ - $ &$ 936.70 $ &$ 0.93 $ &$ 1235.00 $\\
PKS 0437-454 &$ 44.39 \pm 1.43 $ &$ 1.94 \pm 0.01 $ &$ 1.34 \pm 0.06 $ &$ 171.20 \pm 5.92 $ &$ (1.65\pm0.07)\times10^{-3} $ &$ - $ &$ 3852.00 $ &$ 0.10 $ &$ 4502.00 $\\
\end{tabular}}
\end{table*}

\section{Conclusion}\label{concl}
High redshift blazars have consistently been the subject of intense investigation due to their crucial role in understanding the connection between jet formation, accretion processes, and black hole development in the early stages of black hole formation. In this paper, we investigate the multiwavelength properties of 79 blazars with redshifts ranging from $z = 2.0$ to $2.5$ using data accumulated by \textit{Fermi}, Swift XRT/UVOT, and NuSTAR.

In the \gray\ band, the flux and photon index of the selected sources span from $5.32 \times 10^{-10}$ to $3.40 \times 10^{-7}$ photon cm\(^{-2}\) s\(^{-1}\) and between $1.66$ and $3.15$, respectively, illustrating the diverse characteristics of the sources under consideration. Meanwhile, the luminosity ranges from \( (3.67 \pm 1.37) \times 10^{46} \) to \( (6.62 \pm 0.05) \times 10^{48} \) erg s\(^{-1}\), positioning some of these sources among the brightest blazars detected in the \gray\ band. Similarly, in the 0.3-10 keV band, the weakest sources exhibit fluxes on the order of \( (1.06 \pm 0.32) \times 10^{-13} \) erg cm\(^{-2}\)s\(^{-1}\), while the brighter ones reach \( (2.96 \pm 0.02) \times 10^{-11} \) erg cm\(^{-2}\)s\(^{-1}\), with the majority of the sources exhibiting a soft photon index (less than 2.0). This soft photon index extends into the hard X-ray band ($3-30$ keV), where the index ranges from $1.09$ to $1.67$. The comparison of \gray\ and X-ray fluxes reveals no clear correlation in time-averaged measurements. This is exemplified by sources such as 4C +71.07 and PKS 2149-306, which are bright in both \gray\ and X-ray bands, and 4C +01.02, which, despite its high \gray\ flux, exhibits only a moderate X-ray flux.

Using the adaptive binning methods for light curve, we found flux variations in 31 sources, with the most pronounced variability in 4C+01.02, 4C+71.07, PKS 0226-559, PKS 1329-049, PKS 2149-306, S3 0458-02, and S4 0917+44. The sources like PKS 2149-306, 4C+71.07, PKS 1329-049 and 4C+01.02 showed significant increases in \gray\ emission, with peak fluxes occasionally exceeding \(10^{-6}\: \text{photon\:cm}^{-2}\:\text{s}^{-1}\) while for the other sources show more modest increases in \gray\ flux: the \gray\ fluxes exceed \(10^{-7}\: \text{photon\:cm}^{-2}\:\text{s}^{-1}\) during flaring periods. Noticeably, the \gray\ luminosity of 4C+71.07, PKS 1329-049, and 4C+01.02 occasionally was above $10^{50}\:{\rm erg\:s^{-1}}$, putting them among the most luminous sources in the \gray\ band.

In addition to \gray\ variability, our analysis extended to X-ray and optical/UV bands. Despite the limited data, X-ray flux variability was evident for 4C +71.07, PKS 0226-559, PKS 1329-049, PKS 2149-306, S3 0458-02 and  S4 0917+44, while in optical/UV band clear flux variation is found for 4C+01.02, PKS 0226-559 and PKS 2149-306.

In this study, we employed a leptonic one-zone synchrotron and inverse Compton model, considering both internal and external photons, to interpret the multiwavelength SEDs of selected FSRQs and BL Lacs. The range of parameters estimated from the modeling provides a general view of the sources' emissions in an average state. The power-law index of the electrons, indicative of particle acceleration mechanisms, varied from 1.56 to 2.85. For FSRQs the highest energy emissions, as characterized by the cutoff Lorentz factor, varied broadly from 237.6 to \(3.53\times10^3\), reflecting the diverse characteristics of the sources studied. The total jet luminosity varied within a relatively narrow range from \(3.20\times10^{44}\) to \(6.51\times10^{45}\) erg s\(^{-1}\). The disk luminosity, estimated for sources with discernible thermal components in their SEDs, provided insights into the accretion processes. These disk luminosities, ranging dramatically from \(4.15\times10^{44}\) to \(3.97\times10^{47}\) erg s\(^{-1}\), highlighted the varying scales of accretion efficiency and energy conversion in these sources.

In summary, this comprehensive study of 79 high-redshift blazars illuminates the intricate dynamics and energetic processes at play in these distant and powerful objects. Our findings, encompassing a broad spectrum of multiwavelength observations, highlight the diverse characteristics and behaviors of blazars, underscoring their value in probing the physics of jet formation and accretion processes, and black hole development in the early Universe. The observed variations in flux and luminosity across different bands provide key insights into the particle acceleration and emission mechanisms. Future investigations, when a larger number of distant blazars are known, will allow a more systematic and statistical comparison of their emission properties, thereby unveiling the complexities of these energetic sources and providing a window into the HE universe and its evolution over cosmic time.
\section*{Acknowledgements}
We acknowledge the use of analysis tools and services from the Markarian Multiwavelength Data Center (https://mmdc.am), the Astrophysics Science Archive Research Center (HEASARC) platforms, as well as data from the Fermi-LAT, Swift and NuSTAR telescopes. This work was support by the Higher Education and Science Committee of the Republic of Armenia, in the frames of the research project No 21T-1C260.

\section*{Data Availability}
The data underlying this article will be shared on reasonable request to the corresponding author.



\bibliographystyle{mnras}
\bibliography{biblio} 








\bsp	
\label{lastpage}
\end{document}